\documentclass{article}

\usepackage[final]{neurips_2022}

\usepackage[utf8]{inputenc} 
\usepackage[T1]{fontenc}    
\usepackage{hyperref}       
\usepackage{url}            
\usepackage{booktabs}       
\usepackage{amsfonts}       
\usepackage{nicefrac}       
\usepackage{microtype}      
\usepackage{xcolor}         

\usepackage{graphicx}
\usepackage{subcaption}
\usepackage{amsmath}
\usepackage{amssymb}
\usepackage{placeins}
\usepackage{multirow}
\usepackage{pgffor}
\usepackage{enumitem}

\title{Bessel Equivariant Networks for Inversion of Transmission Effects in Multi-Mode Optical Fibres}

\author{%
  Joshua Mitton\thanks{https://github.com/JoshuaMitton} \\
  School of Computing Science \\
  \texttt{j.mitton.1@research.gla.ac.uk} \\
  \And
  Simon Peter Mekhail\\
  School of Physics \& Astronomy, \\
  \texttt{Simon.Mekhail@glasgow.ac.uk} \\
    \And
  Miles Padgett\\
  School of Physics \& Astronomy, \\
  \texttt{Miles.Padgett@glasgow.ac.uk} \\ 
    \And
  Daniele Faccio\\
  School of Physics \& Astronomy, \\
  \texttt{Daniele.Faccio@glasgow.ac.uk} \\
   \And
   Marco Aversa \\
   School of Computing Science \\
   \texttt{marco.aversa@glasgow.ac.uk} \\
   \And
   Roderick Murray-Smith \\
   School of Computing Science \\
   \texttt{roderick.murray-smith@glasgow.ac.uk} \\
   \AND
   {}\vspace{-0.5cm}\\
   University of Glasgow,
   Glasgow, Scotland, UK.\\
}

\newenvironment{enumerate*}%
{\begin{enumerate}%
\setlength{\itemsep}{0pt}%
\setlength{\parskip}{0pt}}%
{\end{enumerate}}

\begin{document}

\maketitle
\vspace{-0.6cm}

\begin{abstract}
  We develop a new type of model for solving the task of inverting the transmission effects of multi-mode optical fibres through the construction of an $\mathrm{SO}^{+}(2,1)$-equivariant neural network. This model takes advantage of the of the azimuthal correlations known to exist in fibre speckle patterns and naturally accounts for the difference in spatial arrangement between input and speckle patterns. In addition, we use a second post-processing network to remove circular artifacts, fill gaps, and sharpen the images, which is required due to the nature of optical fibre transmission. This two stage approach allows for the inspection of the predicted images produced by the more robust physically motivated equivariant model, which could be useful in a safety-critical application, or by the output of both models, which produces high quality images. Further, this model can scale to previously unachievable resolutions of imaging with multi-mode optical fibres and is demonstrated on $256 \times 256$ pixel images. This is a result of improving the trainable parameter requirement from $\mathcal{O}(N^4)$ to $\mathcal{O}(m)$, where $N$ is pixel size and $m$ is number of fibre modes. Finally, this model generalises to new images, outside of the set of training data classes, better than previous models.
\end{abstract}

\section{Introduction}
Multi-mode fibres (MMF) have many potential applications in medical imaging, cryptography, and communications. In the medical domain, the use of multi-mode fibre imaging has potential to create hair-thin endoscopes for imaging sensitive areas of the body. However, to achieve these applications, the fibre transmission properties must be compensated for to return a clear image \citep{stasio2017multimode}. A MMF has multiple different fibre modes, each of which propagates at a different velocity. This leads to an amplitude and phase mixing of the image as it propagates through the fibre \citep{mitschke2016fiber}. As a result, an input image creates a complex-valued  speckled pattern on the output of the MMF. The ability to accurately and in a scalable way learn to invert the transmission effects would unlock MMF imaging as a useful tool across a range of domains. This work concerns the use of a single multi-mode fibre and not fibre bundles.

Inverting a speckled image is challenging for multiple reasons. Firstly, the speckled images have a non-local relationship with respect to the original images. As a result, solely local patch-based models, such as convolutional neural networks, do not make sense as a solution without some dense mapping function. Therefore, the non-locality necessitates mapping the speckled images into a spatial arrangement similar to the original images before typical image-based deep learning techniques can be used, such as convolutions and pooling. In addition, the speckled images have circular correlation, which could be taken advantage of, although as noted by \cite{moran2018deep}, finding these requires solving the inversion, creating a chicken-and-egg problem. Finally, the fibre is equivalent to an unknown complex transmission matrix (TM) so, the inverse of this could be found using a complex-valued linear model, although this presents challenges in terms of memory requirements. A mapping between $350 \times 350$ original and speckled images would result in a TM with $350^4 \approx 15$ billion entries, requiring a linear model with as many parameters.

Previous work in inverting the transmission effects of MMFs has either required extensive experimentation to characterise the TM of the fibre \citep{vcivzmar2011shaping, vcivzmar2012exploiting, choi2012scanner, mahalati2013resolution, papadopoulos2012focusing, ploschner2015seeing, leite2021observing}, where the number of experimental measurements required for re-calibration was reduced by \citet{li2021compressively} by exploiting sparstiy in the TM; made use of dense linear models \citep{moran2018deep, fan2019deep, caramazza2019}; or made use of convolutional models \citep{borhani2018learning, rahmani2018multimode}. For the machine learning approaches to tackling the inversion task, those which make use of a dense linear model \citep{moran2018deep, fan2019deep, caramazza2019} can naturally account for the difference in spatial arrangement between speckled and original images, although they scale badly with the resolution of the images considered ($\mathcal{O}(N^4)$ for $N \times N$ resolution). On the other hand, in theory, the convolutional neural network (CNN) models \citep{borhani2018learning, rahmani2018multimode} improve upon the scalability issue, but in practice due to the need to approximate the transmission matrix and its non-local effects, the models require a large number of layers in order to be able to effectively map every pixel in the speckled image to every pixel in the original image, and hence in practice do not over come the scalability issues. All of these approaches are mostly expected to work for classes of objects that belong to the class that was used for training \citep{borhani2018learning}, with \cite{rahmani2018multimode} making initial steps towards general imaging and \cite{caramazza2019} demonstrating this for a more diverse testing dataset. We provide further details on each of the previous methods in Appendix~\ref{sec:relworkapp}.

In this work we present a model which naturally accounts for the difference in spatial arrangement between speckled and original images and scales more efficiently than previous methods to higher resolution images. We believe this is the first method to demonstrate an ability to invert $256 \times 256$ pixel speckled images into $256 \times 256$ pixel original images. Our approach also takes advantage of the circular correlations in the speckled images, 
 and improves upon previous general imaging results. Concerning the equivariance literature, we develop a model comprising of cylindrical harmonic basis functions, a basis set which has seen little attention in the equivariance literature, and make the connection between the transmission of light through a fibre and the group theoretic understanding used in developing equivariant neural networks. Our contributions are:
\begin{enumerate}
    \itemsep0em 
    \item A more data-efficient, scalable model to solve the inversion of MMF transmission effects.
    \item A model that provides better generalisation to out-of-training domain images.
    \item A connection between group theoretic equivariant neural networks and the inversion of MMF transmission effects, providing a new type of model to tackle the problem.
\end{enumerate}

\section{Background}
\label{sec:background}

\subsection{Multi-Mode Fibres}
\begin{figure}[htb]
  \centering
  
  \begin{subfigure}{0.19\linewidth}
    \includegraphics[width=\linewidth]{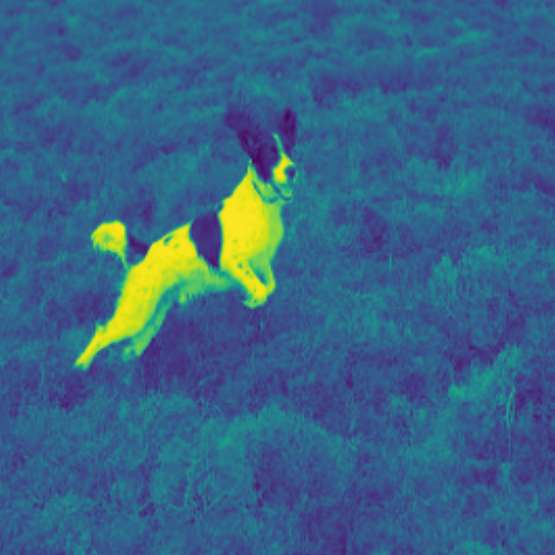}
    \centering
    (a) Original
  \end{subfigure}
  \begin{subfigure}{0.19\linewidth}
    \includegraphics[width=\linewidth]{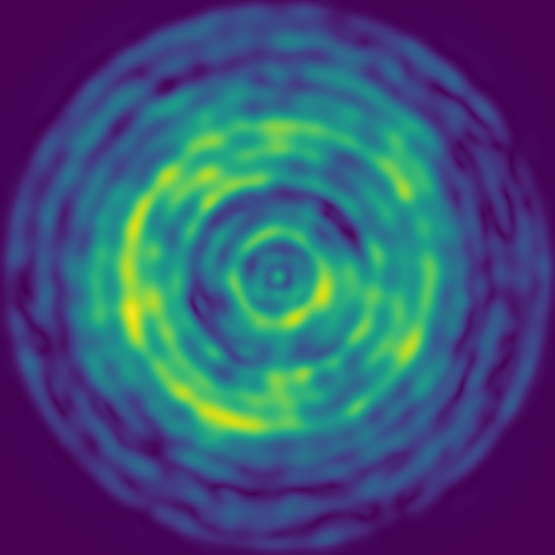}
    \centering
    (b) Speckled
  \end{subfigure}
  \begin{subfigure}{0.19\linewidth}
    \includegraphics[width=\linewidth]{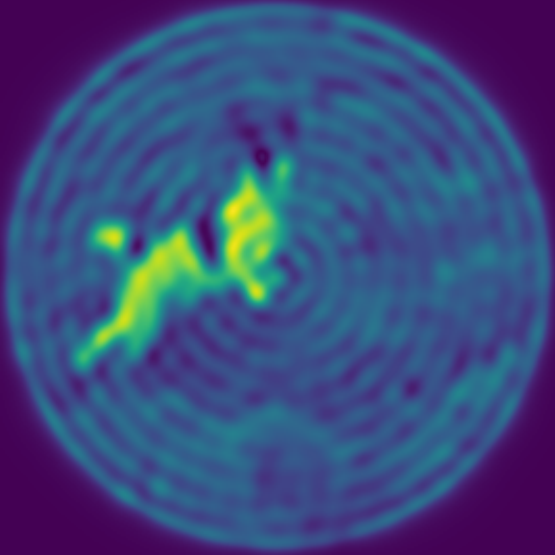}
    \centering
    (c) Inverted
  \end{subfigure}

  \caption{(a) Original image. (b) Speckled image from passing original images through a theoretical TM. (c) Inverted image created by passing speckled images through the inverted theoretical TM.}\vspace{-0.4cm}
  \label{fig:origspecinvmain}
\end{figure}

\looseness-1 Multi-mode fibres present a clear advantage over single-mode fibre bundles due to having 1-2 orders of magnitude greater density of modes than a fibre bundle \citep{choi2012scanner}. Despite this, the different propagation velocities of each mode, which result in the fibre producing scrambled images, presents a significant challenge. If each propagation mode and velocity was known the TM could be computed, providing a linear system that inverts the transmission effects, although in general this is not known. Further, mode specific losses and imperfect mapping between the input pixels and fibre modes can lead to information loss such that inverting the transmission does not yield the true original input image, as is demonstrated in Figure~\ref{fig:origspecinvmain}. Therefore, a model which correctly models the physics of inverting the transmission effects should produce the inverted image in Figure~\ref{fig:origspecinvmain} and producing the original image requires addition information to be learned. We discuss the propagation of light through optical fibres and how we construct theoretical TMs in Section~\ref{sec:theorytm} and further details on the inversion of the TM in Appendix~\ref{invTM}.

\subsection{Generation of Theoretical Transmission Matrices}
\label{sec:theorytm}

\begin{figure}[htb]
  \centering
  \begin{subfigure}{0.12\linewidth}
    \includegraphics[width=\linewidth]{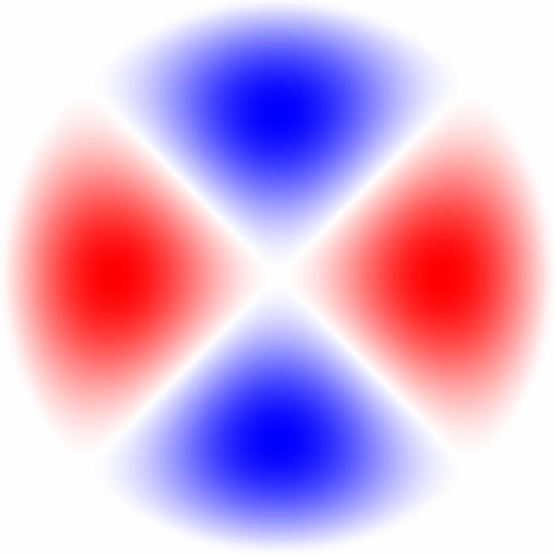}
  \end{subfigure}
  \begin{subfigure}{0.12\linewidth}
    \includegraphics[width=\linewidth]{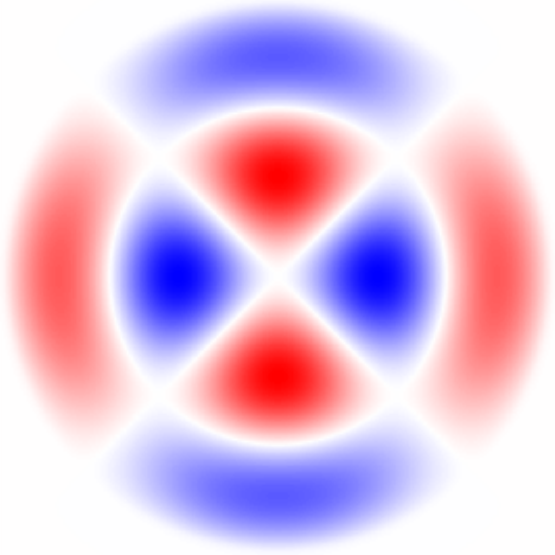}
  \end{subfigure}
  \begin{subfigure}{0.12\linewidth}
    \includegraphics[width=\linewidth]{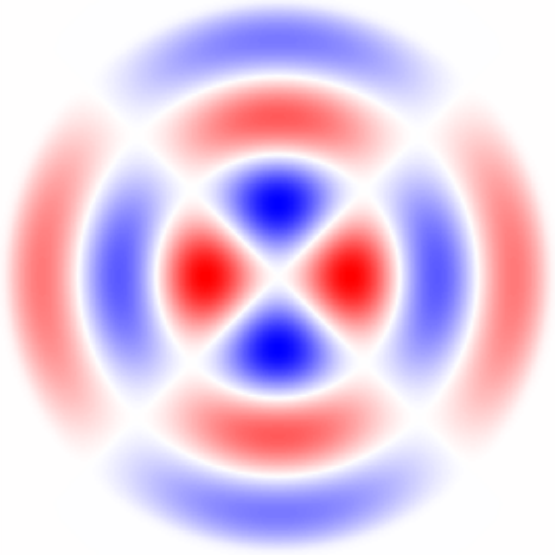}
  \end{subfigure}
  \begin{subfigure}{0.12\linewidth}
    \includegraphics[width=\linewidth]{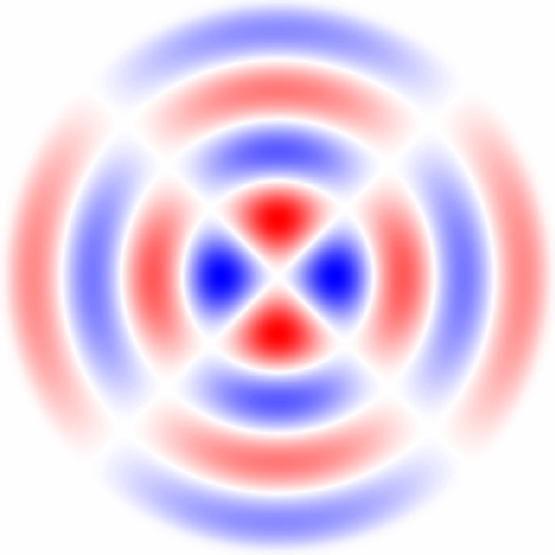}
  \end{subfigure}
  \begin{subfigure}{0.12\linewidth}
    \includegraphics[width=\linewidth]{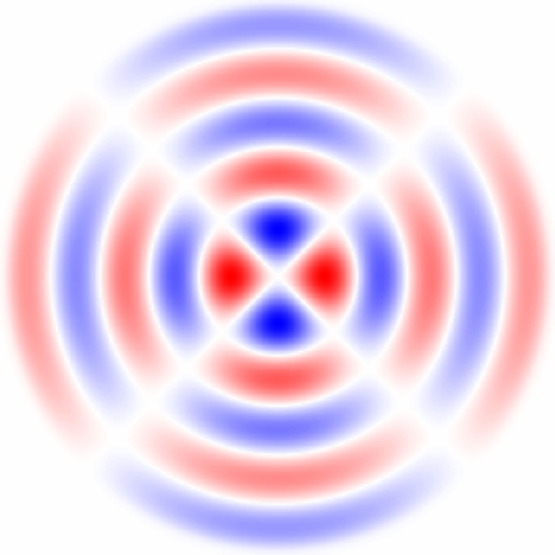}
  \end{subfigure}
  
  \begin{subfigure}{0.12\linewidth}
    \includegraphics[width=\linewidth]{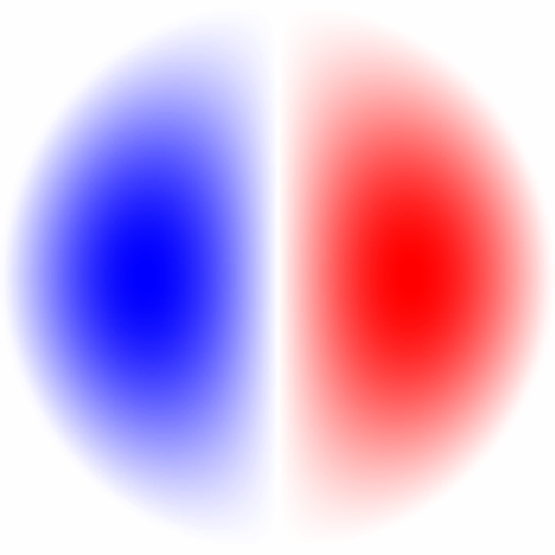}
  \end{subfigure}
  \begin{subfigure}{0.12\linewidth}
    \includegraphics[width=\linewidth]{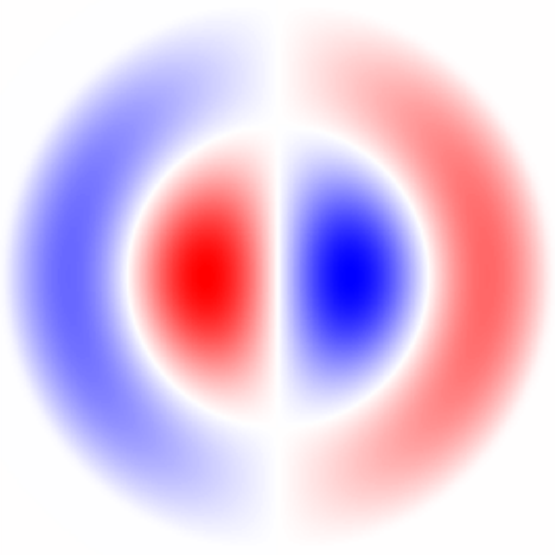}
  \end{subfigure}
  \begin{subfigure}{0.12\linewidth}
    \includegraphics[width=\linewidth]{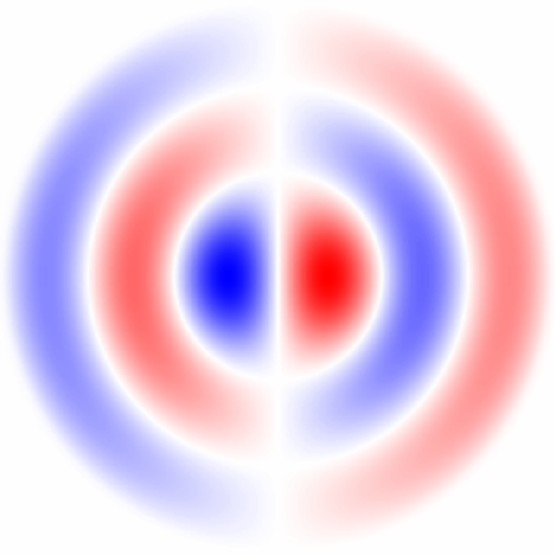}
  \end{subfigure}
  \begin{subfigure}{0.12\linewidth}
    \includegraphics[width=\linewidth]{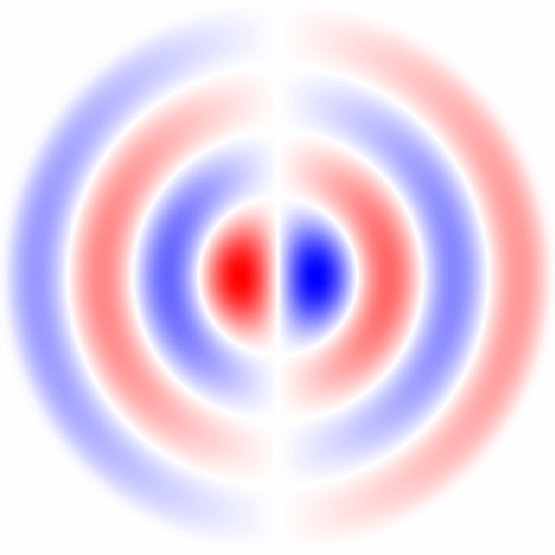}
  \end{subfigure}
  \begin{subfigure}{0.12\linewidth}
    \includegraphics[width=\linewidth]{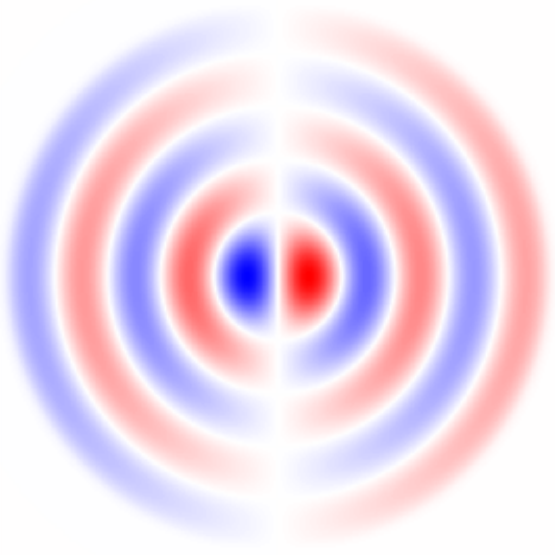}
  \end{subfigure}
  
  \begin{subfigure}{0.12\linewidth}
    \includegraphics[width=\linewidth]{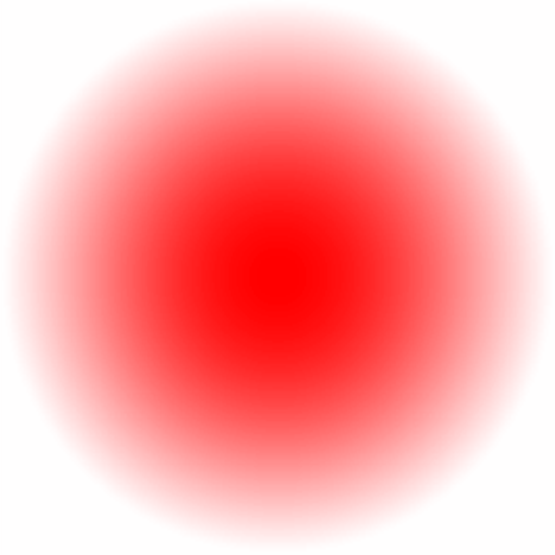}
    \centering
  \end{subfigure}
  \begin{subfigure}{0.12\linewidth}
    \includegraphics[width=\linewidth]{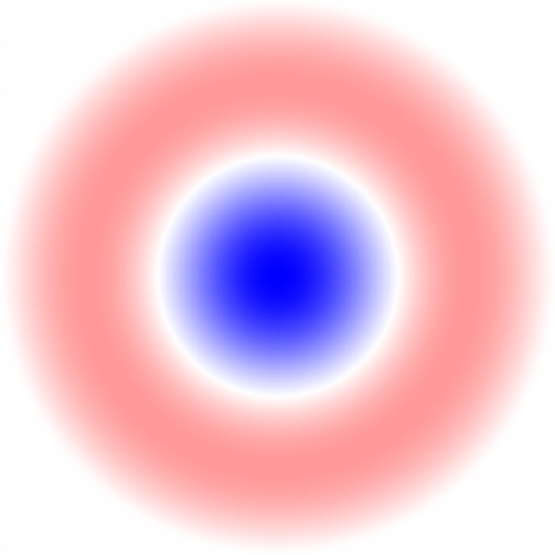}
    \centering
  \end{subfigure}
  \begin{subfigure}{0.12\linewidth}
    \includegraphics[width=\linewidth]{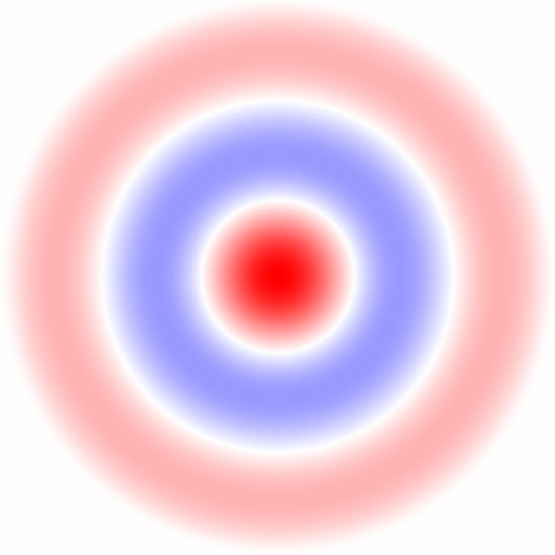}
    \centering
  \end{subfigure}
  \begin{subfigure}{0.12\linewidth}
    \includegraphics[width=\linewidth]{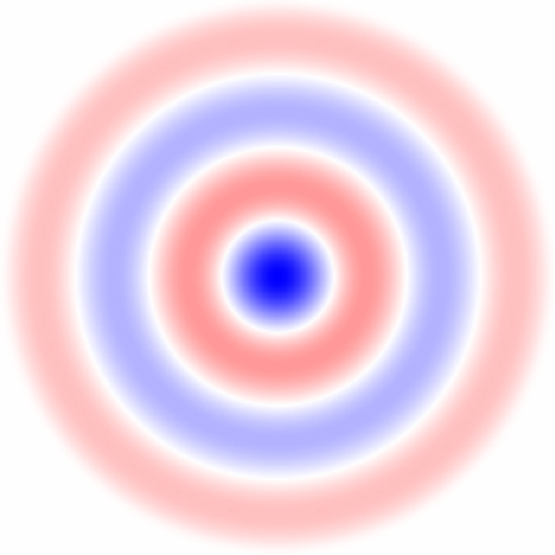}
    \centering
  \end{subfigure}
  \begin{subfigure}{0.12\linewidth}
    \includegraphics[width=\linewidth]{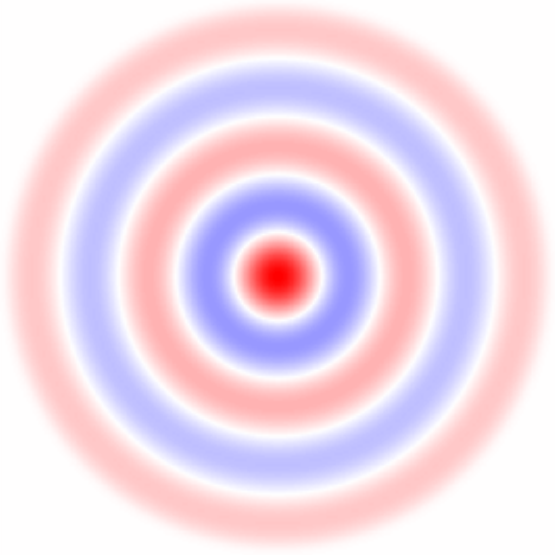}
    \centering
  \end{subfigure}
  \caption{Electric field amplitude profiles of the Bessel bases used within our model. Here red is positive, blue is negative and the colour saturation represents the electric field strength. These bases are often known as LP modes.}
  \label{fig:bases}
\end{figure}

Light propagation through a MMF is characterised by the transmission modes of the fibre. The fibres considered in this work have a step index profile with the refractive index within the core being constant and at a higher value than the cladding. Analytical solutions for the fibre modes can be determined by solving the Helmholtz equation in cylindrical coordinates (details in Appendix~\ref{sec:grouptheoryfibre}). The problem can therefore be expressed as an eigenvalue eigenfunction problem, where the eigenfunctions are the fibre modes and the eigenvalues are the propagation constants, $\beta$, of the modes. The solutions for the electric field within the fibre are comprised of Bessel functions of the first kind inside the core and modified Bessel functions of the second kind in the cladding as follows:
\begin{equation}
    f^{core}_l(r,\theta) = \frac{J_l(u(\beta) \cdot \frac{r}{R} )}{J_l(u(\beta))} e^{\pm i l \theta}, \quad\quad
    f^{clad}_l(r,\theta) = \frac{K_l (w(\beta) \cdot \frac{r}{R})}{K_l(w(\beta))}e^{\pm i l \theta},
    \label{eq:modifiedbesselsecond}
\end{equation}
where $r$ and $\theta$ are the radial and azimuthal coordinates, respectively, $l$ is the azimuthal index of the mode, $R$ is the fibre core radius, and $u$ and $w$ are normalised frequencies defined as follows: 
\begin{equation}
    u(\beta) = R\sqrt{k_0^2n_{core}^2-\beta^2_{lm}},
    \quad\quad
    w(\beta) = R\sqrt{\beta^2_{lm}-k_0^2n_{clad}^2},
    \label{eq:normalisedfrequencyw}
\end{equation}
where $n_{core}$ and $n_{clad}$ are the refractive indices of the core and cladding, respectively, $k_0$ is the vacuum wave number, and $m$ is the radial mode index.
Further details on the connection between propagation of light through MMFs and group theoretic equivariant neural networks are provided in Appendix~\ref{sec:grouptheoryfibre}. Taking the derivative of Equations~
\ref{eq:modifiedbesselsecond}, and equating the resulting functions asserts a smoothness condition on the electric field across the core-cladding boundary. Solving for this for all values of $\beta$ over all possible integer values of $l$ gives the propagation constants. These constants correspond to the fibre modes given by the now appropriately parameterised Equations~
\ref{eq:modifiedbesselsecond}. Examples of the mode fields are given in Figure~\ref{fig:bases}.

A diagonal fibre propagation matrix, $F$, can then be made from the eigenvalues, $\beta$, of the Helmholtz problem and the corresponding eigenfunctions can be recorded in a matrix, $M$, which maps the image space to the fibre mode, or group space. We can find the inverse mapping bases from the fibre or group space back to the image space, $M^{\dagger}$, by taking the conjugate transpose. These three components allow us to construct the transmission matrix as: $\mathrm{TM} = M F M^{\dagger}.$

\subsection{Equivariance}
Equivariance in neural networks concerns the study of symmetries and uses these to build an inductive bias into a model \citep{cohen2016group, bronstein2021geometric}. Equivariance as a property of a neural network guarantees that a transformation of the input data produces a predictable transformation of the predicted features \citep{worrall2017harmonic}. Formally, we say that, given two transformations $T$ and $T^{\prime}$ which are linear representations of a group $G$ and a network or layer $\Phi$, the network is equivariant if applying transform $T_{g}$ to an input $x$ and then passing it through the network $\Phi$ yields the same result as first passing $x$ through $\Phi$ and then transforming by $T_{g}^{\prime}$ \citep{cohen2016group}; that is,
\begin{equation}
    \Phi T_{g}(x) = T_{g}^{\prime}(\Phi(x)).
\end{equation}

Convolutional neural networks (CNNs) are an early example of equivariance on images, where by design they are translationally equivariant \citep{lecun1998gradient}. More recently, equivariance has been considered for other group actions than that of the translation group. The majority of works on images involve incorporating rotation or reflection equivariance into the model, where a variety of different group transformations have been considered \citep{cohen2016group, cohen2016steerable, weiler2019general, muruganso, wiersma2020cnns, de2020gauge,  franzen2021general}.

We are not interested in developing a group equivariant convolutional model here due to the non-similar spatial arrangements between the speckled and original image domains. Despite this, the concept of equivariance still applies, as the concept of learning a model from a fixed basis set, which guarantees symmetry properties are conserved, is relevant due to the nature of transmission through multi-mode optical fibres.

\subsection{Circular Harmonics}
Of the works that incorporate rotation equivariance into CNNs we are particularly interested in those using a continuous rotation group $\mathrm{SO}(2)$. In the case of discrete rotation groups a different feature space is associated to each rotation angle. Storing features for an infinite number of rotation angles is not computationally tractable. One option is to turn to a Fourier representation, whereby instead of choosing a discrete number of rotations a maximum frequency can be chosen. To use a Fourier representation for the weights of the model the domain $\mathbb{R}^{2}$ is decomposed into a radial profile and angular function. Solving for the kernel space of permissible filters that can be used within a CNN and ensuring that the model maintains rotation equivariance yields only the spectrally localised circular harmonics \citep{worrall2017harmonic, weiler2019general, franzen2021general}. Therefore a rotation equivariant CNN can be created by solving for the circular harmonics up to a certain rotational frequency, combining this with a radial profile function, and sampling a basis of resolution given by the CNN filter size. This yields a set of bases which can be linearly combined with a learnable weighting applied to each to form the kernel for the CNN.

\subsubsection{Cylindrical Harmonics}
While the circular harmonics have seen some attention in the deep learning community due to the use in constructing rotational equivariant CNNs, the cylindrical harmonics have seen less attention \citep{klicpera2020directional}. The cylindrical harmonics appear as a solution to Bessel functions for integer $\alpha$ and are therefore of interest for problems where information transformation is characterised by such functions. For example Bessel functions are used when solving wave or heat propagation. Bessel functions are solutions for different complex numbers $\alpha$ of Bessel's differential equation:
\begin{equation}
    x^{2} \frac{d^{2}y}{dx^{2}} + x \frac{dy}{dx} + (x^{2} - \alpha^{2})y = 0.
\end{equation}

For integer values of $\alpha$ the Bessel function solutions are a linearly independent set of functions expressed in cylindrical coordinates. Each function consists of the product of three functions. The radially dependent term is typically called the cylindrical harmonics. Further details on the connection between group theory and the Bessel basis functions is provided in Appendix~\ref{sec:grouptheoryfibre}.

\section{Our Method}

Our method comprises two models: a Bessel equivariant model and a convolutional post-processing model. The Bessel equivariant model uses a basis set comprising cylindrical harmonics and learns a complex mapping function to use the known symmetries in optical fibres. The Bessel equivariant model is constrained by the basis choice to produce circular images, which can have circular artifacts also seen in a true inversion, see Figure~\ref{fig:origspecinvmain}. The post-processing model is used to fill gaps in the corners, remove circular artifacts, and sharpen the images. The two stage approach is useful in safety-critical applications as users can inspect the output of both models, where the Bessel equivariant model output is less likely to depend on the training data due to the choice of inductive bias. 
 
\subsection{Bessel Basis Equivariant Model}

The core of our model has an inductive bias that better replicates the physics of the task of inverting the transmission effects of MMF imaging. To achieve this we utilise prior knowledge of MMFs and the modes with which an image propagates through the fibre. Similar to the successes of rotation equivariant neural networks, which make use of circular harmonics to achieve rotation equivariance, we utilise cylindrical harmonics as an inductive bias of the model. This is motivated by the propagation modes of the fibre being characterised by Bessel functions, where the radially dependent solution is given by the cylindrical harmonics. We make the connection to the group theoretic development of equivariant neural networks in Appendix~\ref{sec:grouptheoryfibre}, demonstrating the connection between the group $\mathrm{SO}^{+}(2,1)$, the light cone, and Bessel function solution to the wave equation.

The first stage of the model is the computation of two sets of basis functions. The first set transforms the input image into a function that lives on the group and the second transforms back into the image domain. The model's weight matrix linearly transforms the function mapped by the first basis set. The basis filters are computed in accordance with equations in Appendix~\ref{sec:theorytm}. 

These functions are sampled on a grid given by the size of the speckled images for the first basis set and the original images for the second basis set. An example of the bases produced is given in Figure~\ref{fig:bases}. The Bessel function bases are computed in a pre-training stage and therefore this step only has to be completed once when creating the model. Further, these bases are not trainable parameters and are not updated during training of the model. These basis functions are an alternative to those used in general rotation equivariant neural networks, which typically offer equivariance to the group $\mathrm{SO}(2)$, except they correctly model the symmetry group, $\mathrm{SO}^{+}(2,1)$ of the task of inverting transmission effects of a MMF due to the added time dimension.

\begin{figure}[htb]
  \centering
  \includegraphics[width=\textwidth]{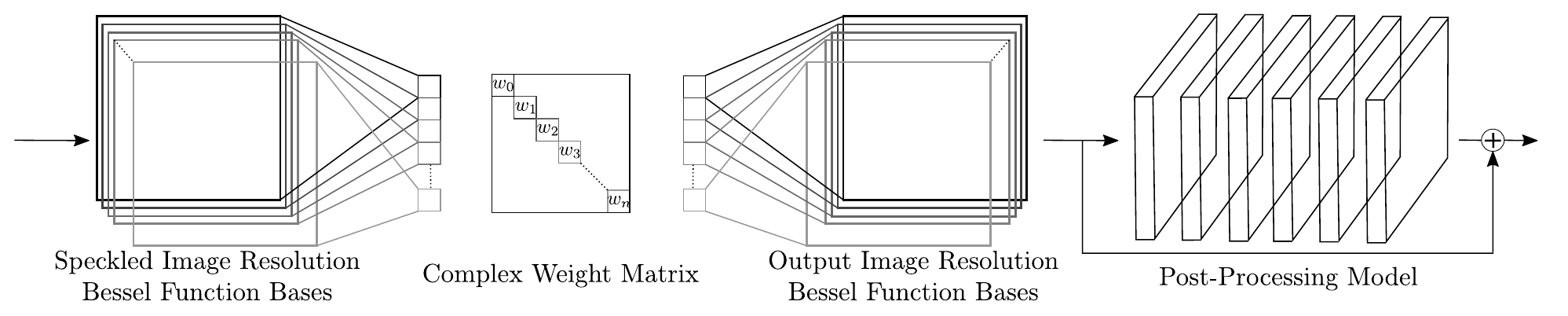}
  \caption{Our two stage model architecture comprising of a Bessel equivariant model and post-processing model. Input (left) is speckled images. Output (right) is predicted image. The speckled image resolution Bessel function bases transform the speckled image into a mode space of the group $\mathrm{SO}^{+}(2,1)$. The complex weight matrix is diagonal. The original image resolution Bessel function basis provides a mapping from the mode space to the output image space. The trainable parameters of the Bessel equivariant model are the diagonal complex weight matrix only. The post-processing model is a convolutional model.}\vspace{-0.4cm}
  \label{fig:equivariant_architecture}
\end{figure}

The model trainable parameters are complex-valued weights of size equal to the number of bases. The complex weight matrix is a diagonal matrix. The images transformed by the first basis set are multiplied by the weight matrix before being transformed by the second basis set to produce the predicted image. The overall model architecture is given in Figure~\ref{fig:equivariant_architecture}.

\subsection{Post-Processing Model}
\label{sec:ppmodel}
As demonstrated in Section~\ref{invTM} the limited number of modes means that the transmission effects of a MMF are typically not fully invertible leading to information loss. We would therefore not expect an equivariant model informed by the physics of the task to be able to fully invert the transmission effects. To overcome this we add a second model which is similar to a super-resolution model, except that it does not change the resolution and instead learns to remove circular artifacts, fill in gaps, and sharpen the images. It is a fully convolutional model comprising of convolutional layers and non-linearities. This model takes as input the output of our Bessel equivariant model, which we suspect will be similar to what is achievable by passing the speckled images through an inverted TM, and predicts the original images. This model is composed of eight convolutional layers with ReLU non-linearities. The motivation behind the explicit separation of the inversion and enhancement models is that in some applications, especially safety-critical applications, it is important to be able to see the results of an inversion conditioned on the observations alone, regardless of prior expectations.

\section{Experiments}

We investigate performance with both data generated with a theoretical TM and experimental data collected using a real MMF fibre. We provide details of the theoretical TM generation in Section~\ref{sec:theorytm}. We use the data from a real fibre, as described in \cite{moran2018deep}, to validate that our new approach can model the same data, despite using significantly fewer trainable model parameters. Further, we generate data using a theoretical TM as this allows us to flexibly adapt the parameters of the experiment and create new datasets to validate the new model design. Finally, the theoretical TM allows us to experiment with images with higher resolution than has been previously been considered. We provide details on the training of the models and datasets in Appendix~\ref{sec:trainingdets}.

\subsection{Real Multimode Fibre}
\label{sec:realfibre}

We compare our model to the complex-valued linear model developed by \cite{moran2018deep} by considering both MNIST and fMNIST data. For this we train and test separate models for the MNIST and fMNIST datasets. In addition, we compare multiple different versions of our model where the diagonal restriction of the mapping between bases is relaxed to allow for a block diagonal structure. We do this to account for manufacturing defects, sharp bends, dopant diffusion, elliptical cores which causes the diagonal mapping of the fibre matrix within the TM to be block diagonal \citep{carpenter2014maximally}. This relaxation allows for $x$-offsets above and below the main diagonal.

\begin{table}[b]
  \vspace{-0.4cm}
  \caption{Comparison of the loss values of each model trained with MNIST or fMNIST data.}
  \label{tab:reallosscomb}
  \centering
  \begin{tabular}{lcccc}
    \toprule
     & \multicolumn{2}{c}{MNIST} & \multicolumn{2}{c}{fMNIST}  \\
    Model & Train Loss & Test Loss & Train Loss & Test Loss  \\
    \midrule
    Complex Linear & 0.00396 & 0.00684 & 0.00509 & 0.01061 \\
    Bessel Equivariant Diag & 0.03004 & 0.03125 & 0.02903 & 0.03140 \\
    Bessel Equivariant Diag + Post Proc & 0.01317 & 0.01453 & 0.01609 & 0.01749 \\
    Bessel Equivariant 10 Off Diag + Post Proc & 0.00488 & 0.00576 & 0.00943 & 0.01162 \\
    Bessel Equivariant Full & 0.00300 & 0.00684 & 0.00548 & 0.01380 \\
    Bessel Equivariant Full + Post Proc & \textbf{0.00145} & \textbf{0.00378} & \textbf{0.00306} & \textbf{0.00964} \\
    \bottomrule
  \end{tabular}
\end{table}

\begin{figure}[htb]
  \centering

  \begin{subfigure}{0.16\linewidth}
    \includegraphics[width=\linewidth]{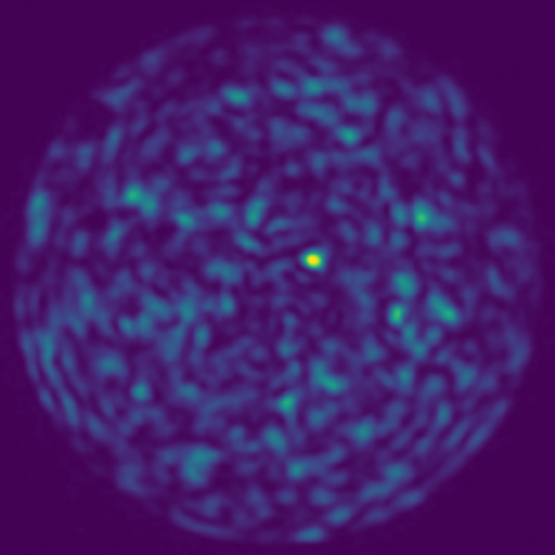}
  \end{subfigure}
  \begin{subfigure}{0.16\linewidth}
    \includegraphics[width=\linewidth]{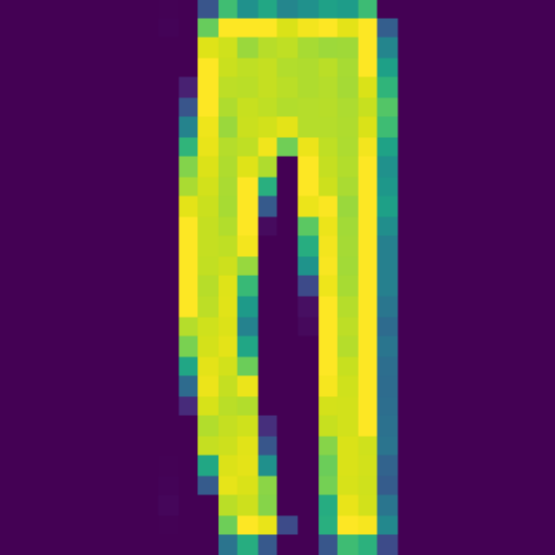}
  \end{subfigure}
  \begin{subfigure}{0.16\linewidth}
    \includegraphics[width=\linewidth]{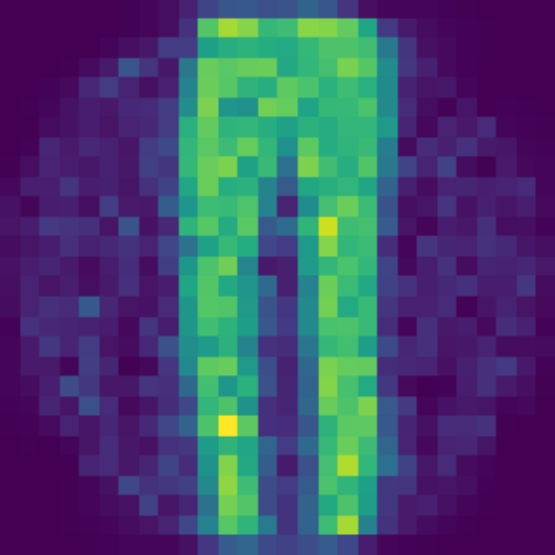}
  \end{subfigure}
  \begin{subfigure}{0.16\linewidth}
    \includegraphics[width=\linewidth]{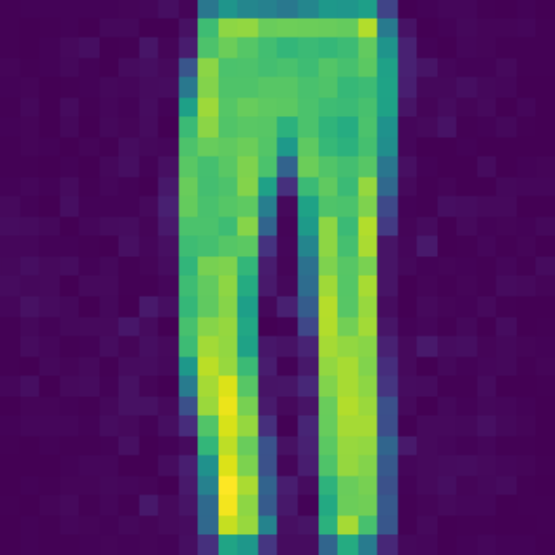}
  \end{subfigure}
  \begin{subfigure}{0.16\linewidth}
    \includegraphics[width=\linewidth]{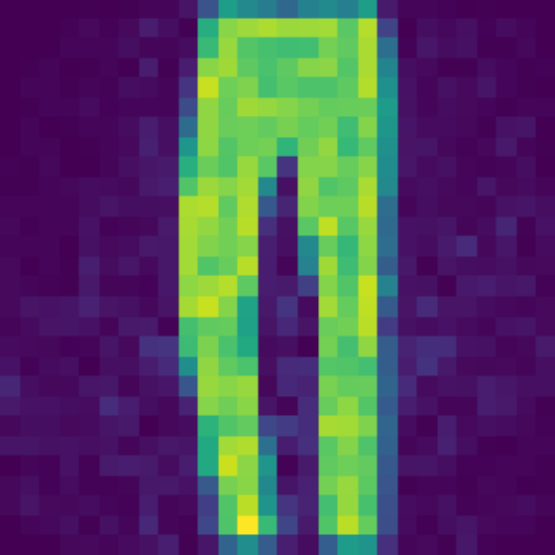}
  \end{subfigure}
  
  \begin{subfigure}{0.16\linewidth}
    \includegraphics[width=\linewidth]{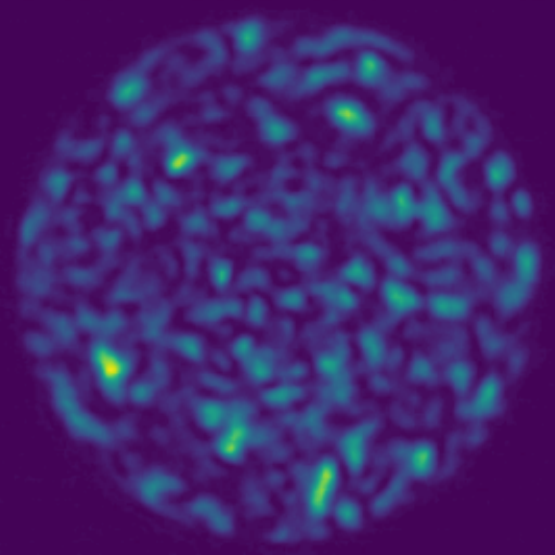}
    \centering
    (a) Input
  \end{subfigure}
  \begin{subfigure}{0.16\linewidth}
    \includegraphics[width=\linewidth]{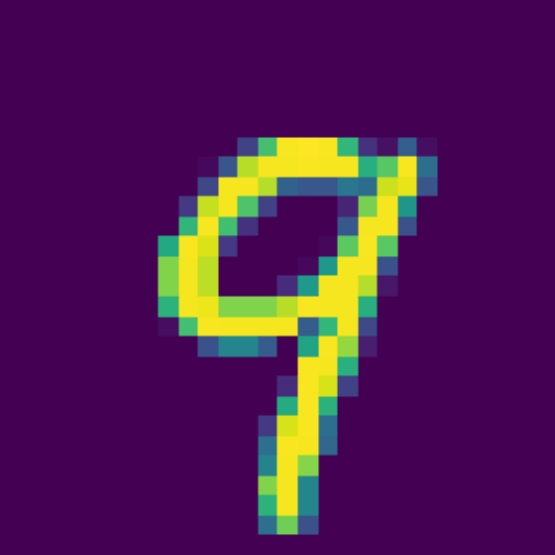}
    \centering
    (b) Target
  \end{subfigure}
  \begin{subfigure}{0.16\linewidth}
    \includegraphics[width=\linewidth]{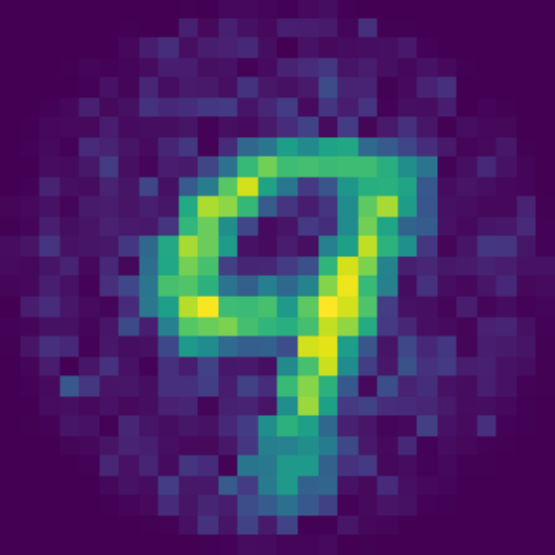}
    \centering
    (c) BEM
  \end{subfigure}
  \begin{subfigure}{0.16\linewidth}
    \includegraphics[width=\linewidth]{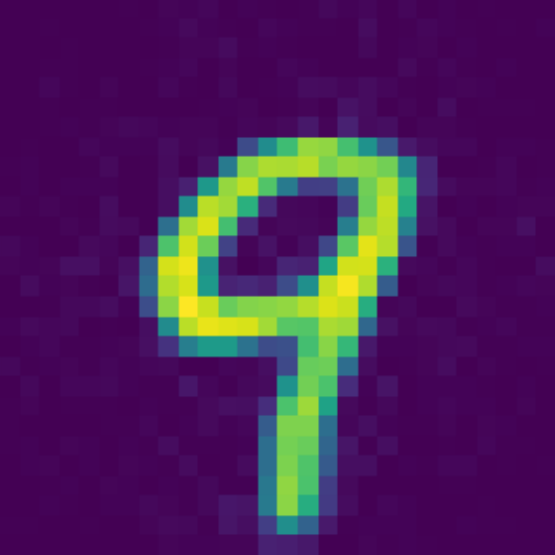}
    \centering
    (d) BEM+PP
  \end{subfigure}
  \begin{subfigure}{0.16\linewidth}
    \includegraphics[width=\linewidth]{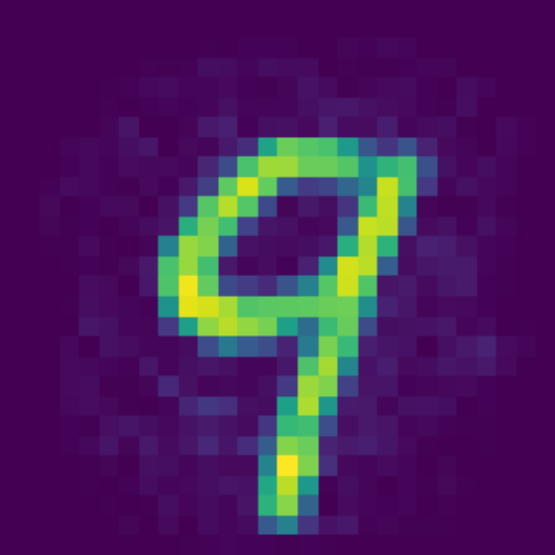}
    \centering
    (e) Complex
  \end{subfigure}
  \caption{Comparison of predicted images from inverting transmission effects of a MMF. The upper row is fMNIST data, lower row MNIST data. (a) Input speckled image, (b) the target original image to reconstruct, (c) Output of the Bessel equivariant model, (d) Output of the combination of Bessel equivariant and post-processing model, 
  and (e) the output of the complex-valued linear model.}\vspace{-0.4cm}
  \label{fig:realmnist}
\end{figure}

\looseness-1 Table~\ref{tab:reallosscomb} shows that our model with a full mapping matrix between bases, i.e. full relaxation of the diagonal constraint, outperforms all other models. In addition, with a relaxation to allow for a 10 block diagonal structure our model performs comparably. On the other hand, the complex linear model outperforms our Bessel equivariant model when the diagonal restriction on the mapping function between bases is enforced. Despite this, our model still provides clear and accurate reconstructions of the target images whilst using orders of magnitude fewer trainable parameters. Additional results are presented in Tables~\ref{tab:reallossmnistapp} and \ref{tab:reallossfmnistapp}. Further to a comparison of the loss values, we visually inspect the reconstruction quality in Figure~\ref{fig:realmnist}, and we provide further visualisations in Figures~\ref{fig:realmnistapp} and \ref{fig:realfmnistapp}. Thus despite our Bessel equivariant diagonal model achieving a larger loss value, the digit or item is still clearly visible in the prediction and is correctly sharpened to a realistic prediction of the original image by our post-processing model. The complex linear model can also be seen to predict a realistic looking output of the correct digit and object, although some noise does feature in the prediction. In addition, we visually compare each of our Bessel equivariant models with the diagonal restriction relaxed in Figures~\ref{fig:realMNISTmod} and \ref{fig:realfmnistmod}, which demonstrates that as the diagonal restriction is relaxed the model can better predict the original images with less noise in the prediction from the Bessel equivariant model.

We also compare the number of trainable parameters within the model in Table~\ref{tab:realparamsmod}. This highlights that our model requires two orders of magnitude fewer parameters that the complex linear model when using the diagonal Bessel equivariant model. Therefore, our model has the potential to scale to higher resolution images where the complex linear model will run out of GPU memory. As the diagonal mapping between Bessel bases is relaxed to include $10$-block diagonal structure the model still requires two order of magnitude fewer parameters. In the most flexible version of our model the number of trainable parameters is comparable with the complex Linear model, although we have demonstrated that this level of flexibility is not required to achieve good image reconstruction.

\begin{table}[htb]
  \vspace{-0.4cm}
  \caption{Comparison of \# trainable parameters in each model trained with MNIST or fMNIST data.}
  \label{tab:realparamsmod}
  \centering
  \begin{tabular}{lc}
    \toprule
    Model & Number of Trainable Parameters (Millions)  \\
    \midrule
    Complex Linear & 78.826 \\
    Bessel Equivariant Diag + Post Proc & 0.500 \\
    Bessel Equivariant 10 Off Diag + Post Proc & 0.617 \\
    Bessel Equivariant Full + Post Proc & 43.108 \\
    \bottomrule \vspace{-0.0cm}
  \end{tabular}
\end{table}

\subsection{Theoretical TM}
\label{sec:theoryfibre}

We compare our model to the complex linear model developed by \cite{moran2018deep} with data created using a theoretical TM. We create multiple datasets one using $28 \times 28$ fMNIST images, with $180 \times180$ speckled images, another with MNIST images, of the same size, and one with a subset of images from the ImageNet dataset \citep{deng2009imagenet} where we fix the resolution to be $256 \times 256$ of both the images and speckled images. Using  fMNIST and MNIST with a theory TM is useful for developing understanding, due to the ease of creating different experiments and datasets, and for testing the generalisability of the model between the two datasets. We include the ImageNet-based dataset as these are higher resolution images containing more challenging information than fMNIST. To the best of our knowledge, demonstrating the ability to invert transmission effects of such high resolution images has not been previously achievable with a machine learning based approach.

\subsubsection{Generalisability of Models}

We provide a comparison of the predictions of a complex valued linear model \citep{moran2018deep}, our equivariant model only, and our equivariant model coupled with a post-processing model trained on fMNIST images and tested on both fMNIST and MNIST to analyse to what extent the models can generalise to a new dataset. Table~\ref{tab:fmnistloss} presents the loss values for both the training and testing of each model. Considering the fMNIST data this shows that the complex linear and Bessel equivariant models achieve similar loss values. Although this is a result of the Bessel equivariant model under-performing due to not being able to reconstruct pixel values close to the edge of the image as a consequence of the circular nature of the Bessel bases functions, while the complex linear model under-performs due to failing to reconstruct higher frequency information and over-fitting to the general clothing categories. Finally, when our Bessel equivariant model is combined with the post-processing model it outperforms all other models. The ability to generalise to a new data is assessed through the MNIST dataset. This shows that our Bessel equivariant model significantly out-performs the complex linear model in generalising to a new dataset.

\begin{table}[t]
  \vspace{-0.0cm}
  \caption{Comparison of the loss values of each model trained with fMNIST data.}
  \label{tab:fmnistloss}
  \centering
  \begin{tabular}{lccc}
    \toprule
          & \multicolumn{2}{c}{fMNIST} & MNIST \\
    Model & Train Loss & Test Loss     & Test Loss  \\
    \midrule
    Complex Linear                 & 0.0149 & 0.0146 & 0.0363 \\
    Bessel Equivariant             & 0.0141 & 0.0139 & \textbf{0.0026} \\
    Bessel Equivariant + Post Proc & \textbf{0.0032} & \textbf{0.0032} & 0.0028 \\
    \bottomrule
  \end{tabular}
\end{table}

\begin{figure}[b]
  \centering
  
  
  \begin{subfigure}{0.16\linewidth}
    \includegraphics[width=\linewidth]{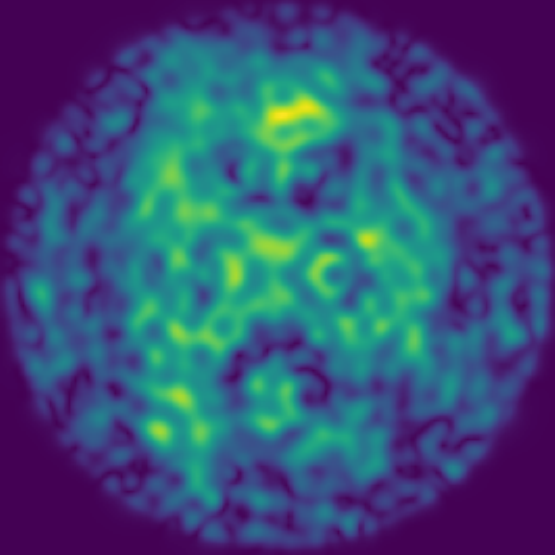}
    \centering
    (a) Input
  \end{subfigure}
  \begin{subfigure}{0.16\linewidth}
    \includegraphics[width=\linewidth]{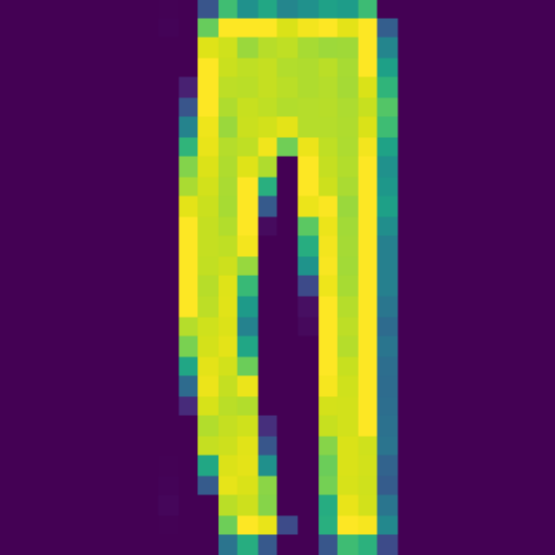}
    \centering
    (b) Target
  \end{subfigure}
  \begin{subfigure}{0.16\linewidth}
    \includegraphics[width=\linewidth]{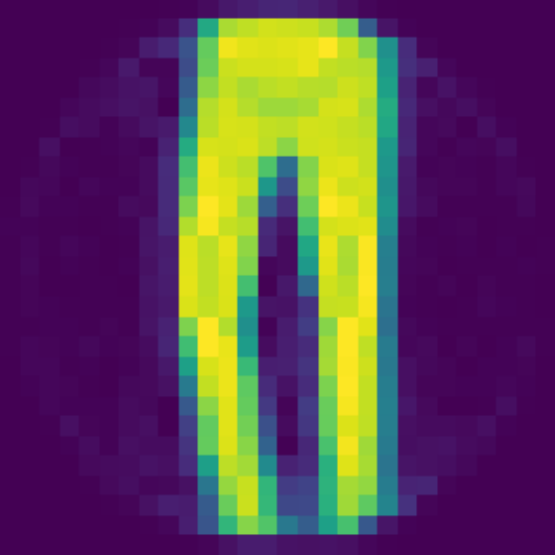}
    \centering
    (c) BEM
  \end{subfigure}
  \begin{subfigure}{0.16\linewidth}
    \includegraphics[width=\linewidth]{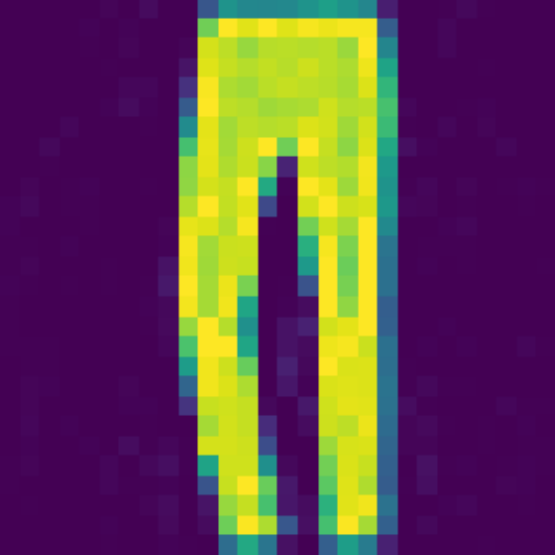}
    \centering
    (d) BEM + PP
  \end{subfigure}
  \begin{subfigure}{0.16\linewidth}
    \includegraphics[width=\linewidth]{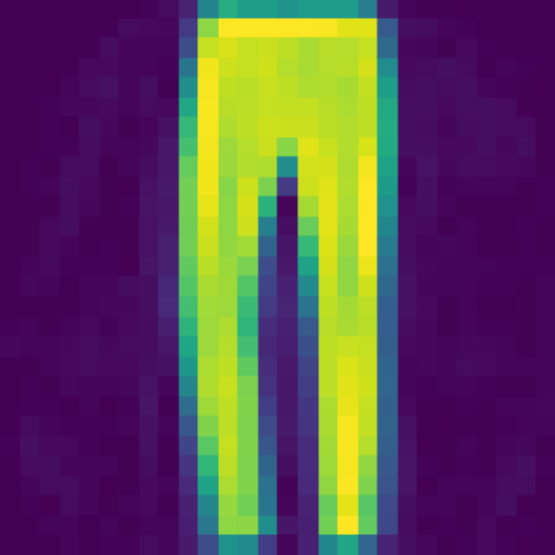}
    \centering
    (e) Complex
  \end{subfigure}
  \caption{Comparison of predicted images from inverting transmission effects of a MMF. (a) Input speckled image, (b) Target original image to reconstruct, (c) Output of the Bessel equivariant model, (d) Output of the Bessel equivariant and post-processing model, 
  and (e) Output of the complex valued linear model.}
  \label{fig:fmnistplots}
  \vspace{-0.2cm}
\end{figure}

 Figure~\ref{fig:fmnistplots} shows reconstructions produced by the three models, with further reconstructions provided in Figure~\ref{fig:theoryfmnistapp}. This demonstrates visually that our model, even without the post-processing part, is able to better reconstruct the original images. The post-processing model refines the output of the equivariant model and fills in gaps outside the fibre. Despite it being possible to estimate the general category of clothing from the outputs of the complex linear model, the higher frequency information has been lost, i.e. the pattern on the jumper or the curve in the trouser leg. On the other hand, our model is able to reconstruct this higher frequency information.

\begin{figure}[htb]
  \centering
  \begin{subfigure}{0.16\linewidth}
    \includegraphics[width=\linewidth]{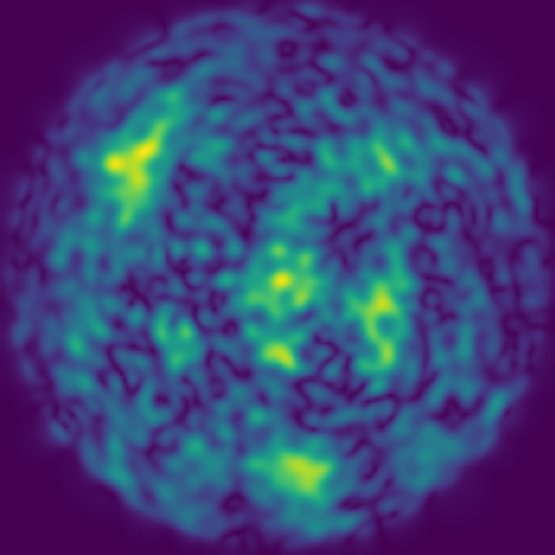}
    \centering
    (a) Input
  \end{subfigure}
  \begin{subfigure}{0.16\linewidth}
    \includegraphics[width=\linewidth]{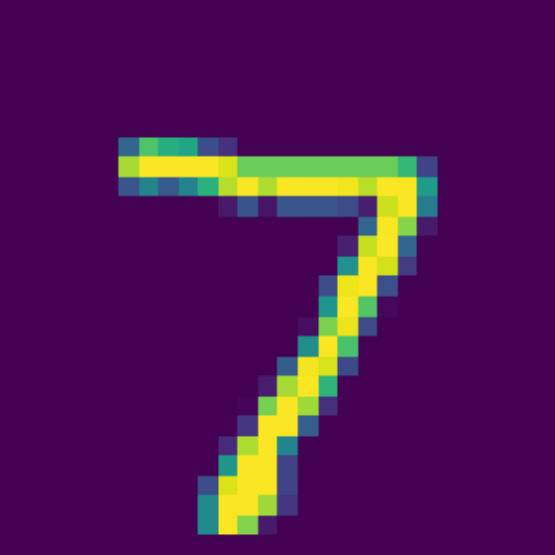}
    \centering
    (b) Target
  \end{subfigure}
  \begin{subfigure}{0.16\linewidth}
    \includegraphics[width=\linewidth]{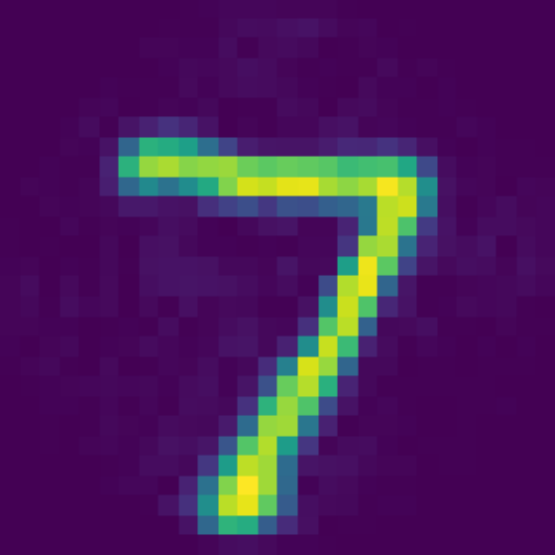}
    \centering
    (c) BEM
  \end{subfigure}
  \begin{subfigure}{0.16\linewidth}
    \includegraphics[width=\linewidth]{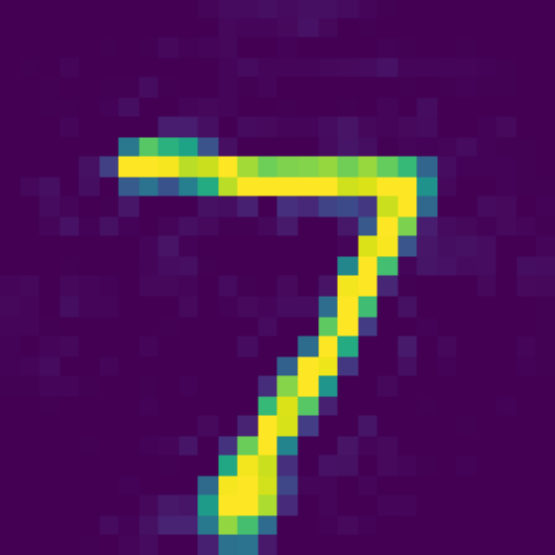}
    \centering
    (d) BEM + PP
  \end{subfigure}
  \begin{subfigure}{0.16\linewidth}
    \includegraphics[width=\linewidth]{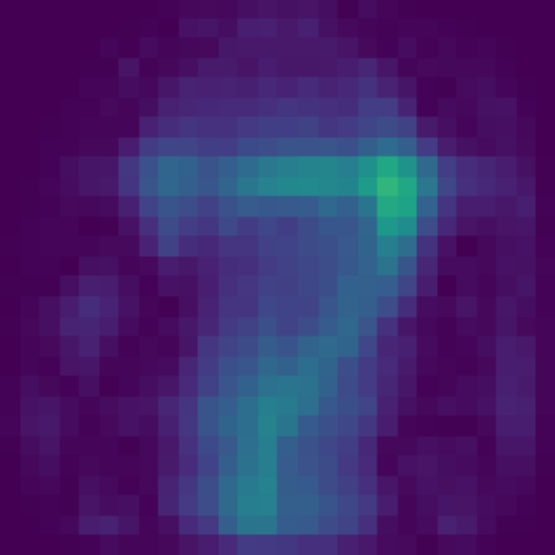}
    \centering
    (e) Complex
  \end{subfigure}

  \caption{Comparison of predicted images from inverting transmission effects of a MMF. (a) Input speckled images, (b) Target original images, (c) Output of the Bessel equivariant model, (d) Output of Bessel equivariant and post-processing model, 
  and (e) Output of the complex valued linear model.}
  \vspace{-0.4cm}
  \label{fig:mnistplots}
\end{figure}

Previous works have demonstrated some ability to generalise to new data classes \citep{rahmani2018multimode, caramazza2019}, although in each case the results are not perfect. Here, we compare the different methods trained on fMNIST and then tested on MNIST data. Figure~\ref{fig:mnistplots} demonstrates that our Bessel equivariant model generalises well to a new data domain, with further reconstructions provided in Figure~\ref{fig:theorymnistapp}. On the other hand, the complex linear model somewhat predicts the correct digits. In this out of training domain our post-processing model does not add any value to the original Bessel equivariant model. This highlights the benefit of our two-stage modelling approach as we have one robust model and a second model to sharpen the images, as a result if a user believes the situation could be unusual they could reliably use the output of our Bessel equivariant model and not the post-processing part. Further to achieving strong generalisation results, we also consider the effect of noise in the speckled images in Appendix~\ref{sec:noisefmnist}.

Figure~\ref{fig:scalability_shared} shows how the number of trainable model parameters scale. Even for $28\times28$ images our model requires two orders of magnitude fewer trainable parameters than the complex linear model.


\subsubsection{Scaling to Larger Images -- ImageNet}

We now experiment with a subset of images from the ImageNet dataset, where we fix the resolution to be $256 \times 256$. Inverting the transmission effects of such higher resolution images has not been previously achievable.  We only compare our equivariant model with our equivariant model coupled with a fully convolutional post-processing model. The complex-valued linear model cannot fit in the GPU memory of an A6000 with 48.7GiB of memory with this resolution of image.

\begin{figure}[htb]
  \centering
  
  
  \begin{subfigure}{0.24\linewidth}
    \includegraphics[width=\linewidth]{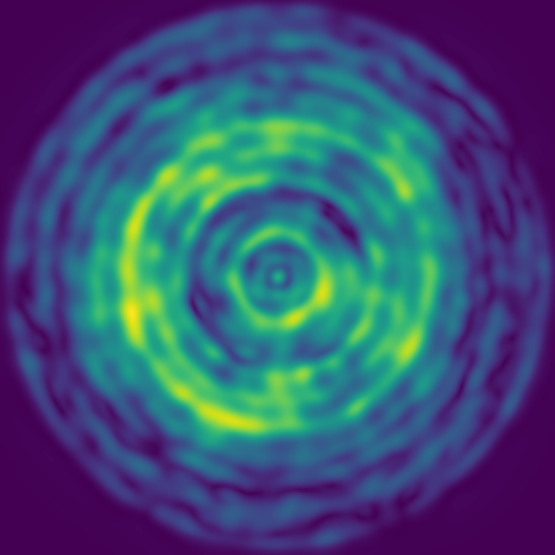}
    \centering
    (a) Input
  \end{subfigure}
  \begin{subfigure}{0.24\linewidth}
    \includegraphics[width=\linewidth]{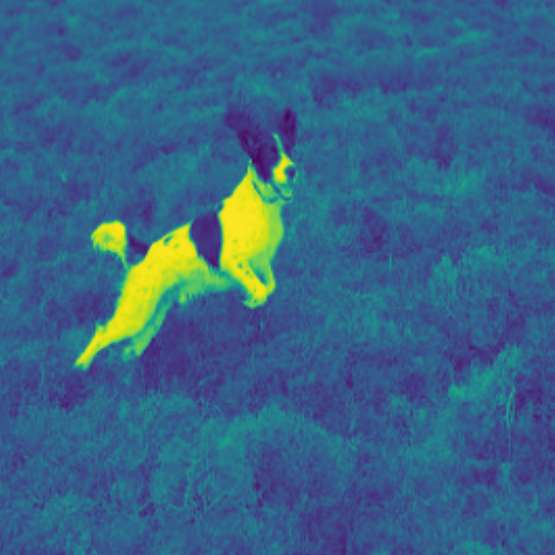}
    \centering
    (b) Target
  \end{subfigure}
  \begin{subfigure}{0.24\linewidth}
    \includegraphics[width=\linewidth]{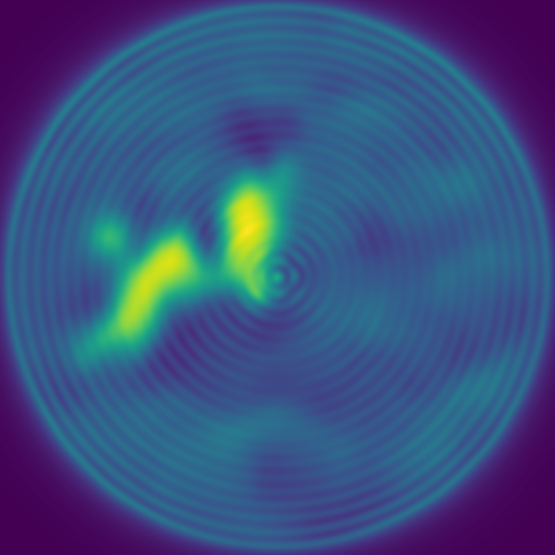}
    \centering
    (c) BEM
  \end{subfigure}
  \begin{subfigure}{0.24\linewidth}
    \includegraphics[width=\linewidth]{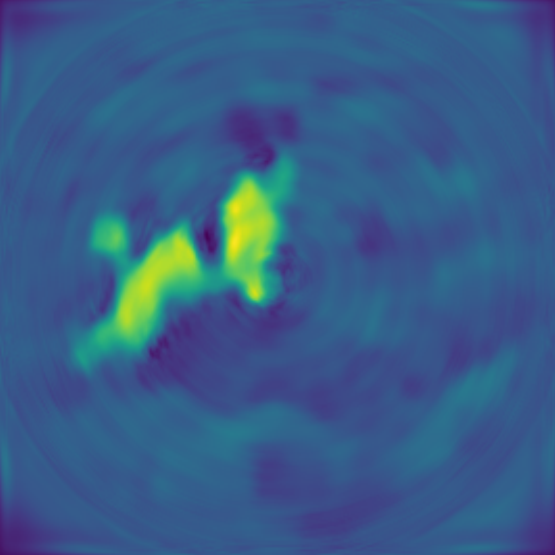}
    \centering
    (d) BEM + PP
  \end{subfigure}
  \caption{Comparison of predicted images from inverting transmission effects of a MMF using high resolution ImageNet data. (a) Input speckled image, (b) Target original image to reconstruct, (c) Output of Bessel equivariant model  (d) Output of Bessel equivariant and post-processing model.}
  \vspace{-0.2cm}
  \label{fig:imagenet}
\end{figure}

We analyse our model visually, with Figure~\ref{fig:imagenet} showing that our Bessel equivariant model produces a reconstruction from which one can determine the type of dog and its activity, although the reconstruction does not capture all the high frequency information, and information in the corners is lost due to the circular nature of the fibre. Further reconstructions are provided in Figure~\ref{fig:theoryimagenetapp}. When we combine the two models some of the artifacts of the Bessel equivariant model are removed as the post-processing model fills in information towards the corners and sharpens the image.

\begin{figure}[htb]
    \centering
    \includegraphics[width=0.96\linewidth]{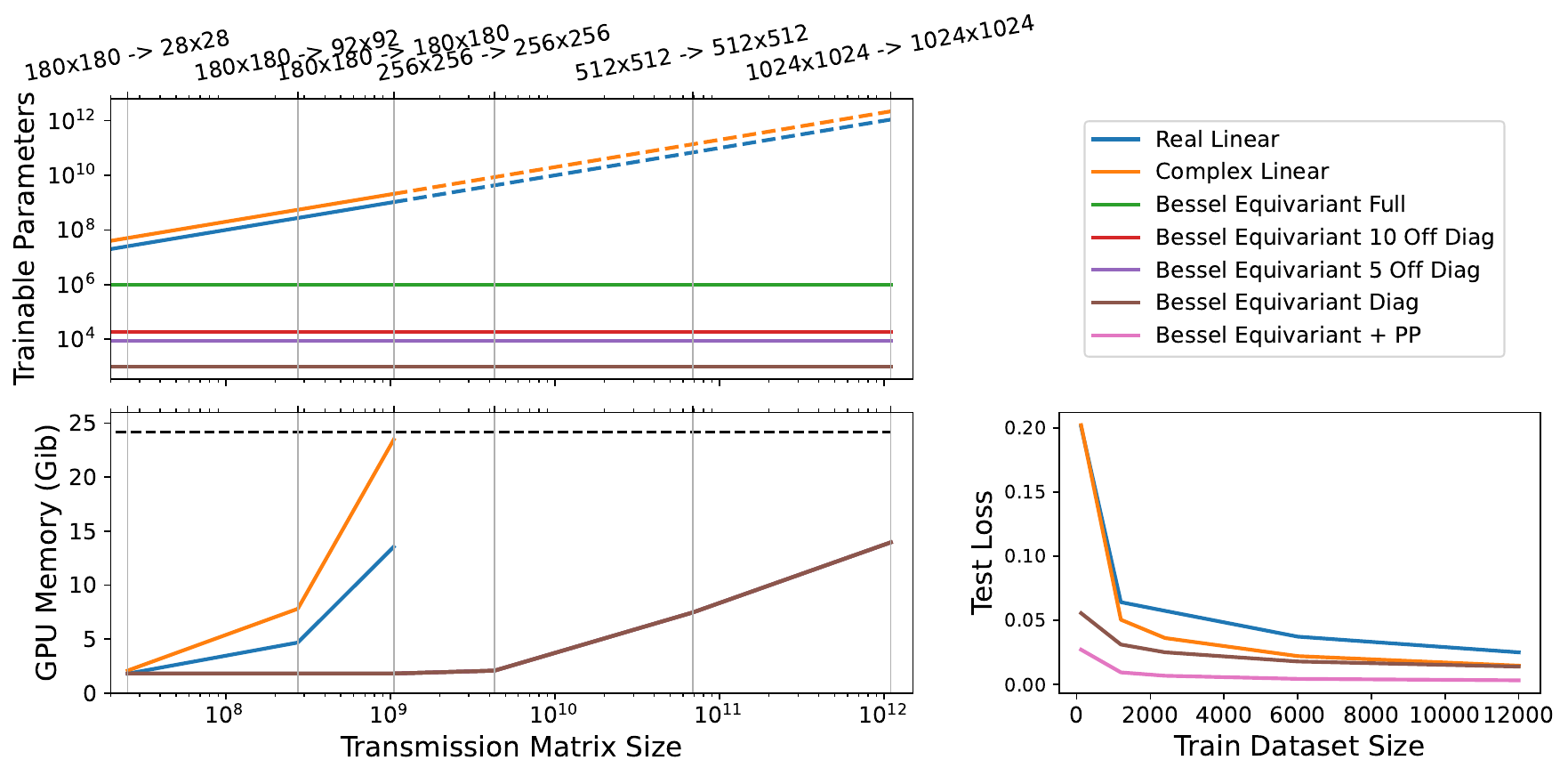}
    \caption{(Upper Left) Comparison of number of trainable parameters v.\ TM size (dashed line indicates model cannot in practice be built due to large memory requirements). (Lower Left) Comparison of required GPU memory against TM size. Vertical grey lines indicate a given image size -- our model can process $1024\times1024$ images on a single consumer-level Titan RTX GPU. (Lower Right) Comparison of reducing the size of the training dataset. All plots for a 1000 mode fibre.}
    \label{fig:scalability_shared}
    \vspace{-0.4cm}
\end{figure}

\begin{table}[b]
  \vspace{-0.6cm}
  \caption{Comparison of the loss values of each model trained with ImageNet data.}
  \label{tab:imagenetloss}
  \centering
  \begin{tabular}{lll}
    \toprule
    Model & Train Loss & Test Loss  \\
    \midrule
    Bessel Equivariant & 0.0541 & 0.0574 \\
    Bessel Equivariant + Post Proc & \textbf{0.0161} & \textbf{0.0159} \\
    \bottomrule\vspace{-0.4cm}
  \end{tabular}
\end{table}
 We present the loss values for both models in Table~\ref{tab:imagenetloss}. This shows a similar result as for fMNIST, that our Bessel equivariant model can solve the task well, but requires the post-processing model to sharpen the image and fill in gaps in the corners.
We compare the models' memory requirements in Figure~\ref{fig:scalability_shared}. Note how fewer trainable parameters translate into significantly less memory use than the complex linear model. This effect is seen more drastically when scaling to high resolution images, where the complex linear model runs out of memory on a $24.2$GiB Titan RTX for original and speckled images of resolution $180 \times 180$ pixels. Our model, in contrast, has been tested on $256 \times 256$ pixel images, requiring only $2.1$GiB, and can scale to megapixel images.

Finally, we explore the impact on each model of reducing the size of the training dataset. Requiring a reduced dataset size minimises the time taken in a lab collecting data. Figure~\ref{fig:scalability_shared} demonstrates that our model outperforms the complex linear model with $10 \times$-less training data, reducing the need to collect $12000$ samples to $1200$. The reconstructions are visualised in Appendix~\ref{app:redtrain}.

\section{Conclusions}
We present a new type of model for solving the task of inverting the transmission effects of a MMF through developing a $\mathrm{SO}^{+}(2,1)$-equivariant neural network and combining this with a post-processing network. This improves upon complex linear models through incorporating a useful physically informed inductive bias into the equivariant network. The equivariant network is shown to perform well on new image classes, providing a general model for fibre inversion. The post-processing network is specific to an image domain, as it generalises to new image classes as well as a general convolutional network, although it improves the quality of returned images. The use of a theoretical TM allows us to compare an `ideal' inverse with the output of previous models based on learning a full transmission matrix. This suggests that previous learned transmission matrices were combining elements of the actual transmission matrix with elements of the post-processing network, which would affect their ability to generalise to new images. We also anticipate that in interactive safety-critical applications users might want the ability to switch between modes, to be sure that the evidence was there, and not overly influenced by the priors in the training data.

This new model significantly improves the ability to scale to higher resolution images by improving the scaling law from $\mathcal{O}(N^4)$ to $\mathcal{O}(m)$, where $N$ is pixel size and $m$ is the number of fibre modes. We provide a comparison between our model and prior works on both data created using a real fibre and a theoretical TM, demonstrating in both cases that our model solves the task while using significantly fewer trainable parameters than the complex and real linear models. In addition, we demonstrate results on high resolution $256 \times 256$ images, which has previously been unachievable due to the growth of parameters with previous models. Furthermore, we demonstrate the ability of our model to generalise to new data classes outside of the training data, outperforming prior works. The dramatic reduction in the number of parameters for each fibre configuration opens the way for future models which can learn mappings for high-resolution images, from a wider set of perturbed fibre poses and combine these using architectures such as VAEs.

\begin{ack}
JM is supported by a University of Glasgow Lord Kelvin Adam Smith
Studentship. RM-S, MP, DF and SPM are grateful for EPSRC support through grants EP/T00097X/1  and
RM-S  for EP/R018634/1. MA is funded by dotPhoton and a UofG scholarship. 
DF  acknowledges funding through the Royal Academy of Engineering Chair in Emerging Technologies programme.
\end{ack}
\bibliography{neurips_2022}
\bibliographystyle{neurips_2022}

\clearpage
\section*{Checklist}

\begin{enumerate}

\item For all authors...
\begin{enumerate}
  \item Do the main claims made in the abstract and introduction accurately reflect the paper's contributions and scope?
    \answerYes{}
  \item Did you describe the limitations of your work?
    \answerYes{See Sections~\ref{sec:ppmodel} and \ref{invTM}}.
  \item Did you discuss any potential negative societal impacts of your work?
    \answerYes{} See Section~\ref{sec:negsocimpact}.
  \item Have you read the ethics review guidelines and ensured that your paper conforms to them?
    \answerYes{}
\end{enumerate}

\item If you are including theoretical results...
\begin{enumerate}
  \item Did you state the full set of assumptions of all theoretical results?
    \answerYes{} See Sections~\ref{sec:background}, \ref{sec:theorytm}, and \ref{invTM}.
  \item Did you include complete proofs of all theoretical results?
    \answerYes{} See Sections~\ref{invTM} and \ref{sec:grouptheoryfibre}. 
\end{enumerate}

\item If you ran experiments...
\begin{enumerate}
  \item Did you include the code, data, and instructions needed to reproduce the main experimental results (either in the supplemental material or as a URL)?
    \answerYes{}
  \item Did you specify all the training details (e.g., data splits, hyperparameters, how they were chosen)?
    \answerYes{} See Section~\ref{sec:trainingdets}.
  \item Did you report error bars (e.g., with respect to the random seed after running experiments multiple times)?
    \answerNo{} Due to compute constraints we did not run with multiple random seeds, although we do note that training was very stable and so the results will be reproducible.
  \item Did you include the total amount of compute and the type of resources used (e.g., type of GPUs, internal cluster, or cloud provider)?
    \answerYes{} See Section~\ref{sec:trainingdets}.
\end{enumerate}

\item If you are using existing assets (e.g., code, data, models) or curating/releasing new assets...
\begin{enumerate}
  \item If your work uses existing assets, did you cite the creators?
    \answerYes{} See Section~\ref{sec:realfibre}.
  \item Did you mention the license of the assets?
    \answerYes{} See Section~\ref{sec:trainingdets}.
  \item Did you include any new assets either in the supplemental material or as a URL?
    \answerYes{}
  \item Did you discuss whether and how consent was obtained from people whose data you're using/curating?
    \answerYes{} Consent provided by the authors who collected the data.
  \item Did you discuss whether the data you are using/curating contains personally identifiable information or offensive content?
    \answerYes{} It does not contain any sensitive information.
\end{enumerate}

\item If you used crowdsourcing or conducted research with human subjects...
\begin{enumerate}
  \item Did you include the full text of instructions given to participants and screenshots, if applicable?
    \answerNA{}
  \item Did you describe any potential participant risks, with links to Institutional Review Board (IRB) approvals, if applicable?
    \answerNA{}
  \item Did you include the estimated hourly wage paid to participants and the total amount spent on participant compensation?
    \answerNA{}
\end{enumerate}

\end{enumerate}


\clearpage
\appendix

\section{Appendix}

\subsection{Group Theoretic Understanding of Optical Fibre Transmission Modes}
\label{sec:grouptheoryfibre}

When a light beam propagates in free space or in a transparent homogeneous medium, its transverse intensity profile generally changes. Despite this, there exist certain distributions that do not change intensity profile as they traverse. These fixed profiles are the transmission modes of the space.

The development of group equivariant networks exploits a similar principle, where these networks are constructed under the more general principle of finding basis functions called irreducible representations of some group. Some examples of this principle include: for the circle $S^{1}$ or line $\mathbb{R}$ the irreducible representations are given by complex exponentials $\mathrm{exp}(in\theta)$, for the group $\mathrm{SO}(2)$ the irreducible representations are given by the circular harmonics, for the group $\mathrm{SO}(3)$ the irreducible representations are given by the Wigner D-functions, and for $S^{2}$ the irreducible representations are given by the spherical harmonics. Thus a function on the group can be composed as a linear combination of the corresponding irreducible representations. On the other hand, when learning a function it is possible to learn the weightings of each of the irreducible representations of the group to learn that function. Building a model such that its feature space comprises the irreducible representations of the group provides a method of constructing a model that is equivariant to the underlying group of the representations.

Here we seek to show a connection between the known properties of optical fibres and group equivariant networks. We start by providing some details on the propagation of light through a fibre. The propagation of light is governed by the time-independent (Helmholtz) equation, which in cylindrical coordinates is given by Equation~\ref{eq:wavecyl}.
\begin{equation}
    \frac{\partial^{2} E}{\partial r^{2}} + \frac{1}{r} \frac{\partial E}{\partial r} + \frac{1}{r^{2}} \frac{\partial^{2} E}{\partial \phi^{2}} + q^{2}E = 0.
    \label{eq:wavecyl}
\end{equation}

The standard approach to solving the above equation is to use the separation-of-variables procedure, which assumes a solution of the form given in Equation~\ref{eq:wavesolvform}.
\begin{equation}
    E_{z} = AF_{1}(r) F_{2}(\phi).
    \label{eq:wavesolvform}
\end{equation}


Due to the circular symmetry of the fibre, each component must not change when the coordinate $\phi$ is increased by a multiple of $2 \pi$. Therefore, we make the following assumption:
\begin{equation}
    F_{2} (\phi) = e^{i \nu \phi},
\end{equation}
where $\nu \in \mathbb{Z}$. Substituting into Equation~\ref{eq:wavecyl} yields a wave equation of the following form:
\begin{equation}
    \frac{\partial^{2} F_{1}}{\partial r^{2}} + \frac{1}{r} \frac{\partial F_{1}}{\partial r} + \left( q^{2} - \frac{\nu ^{2}}{r^{2}} \right ) F_{1} = 0,
    \label{eq:besselfnc}
\end{equation}
which is a differential equation for Bessel functions. Solving this both inside the core of the fibre and in the cladding of the fibre provides two solutions. In the core of the fibre, as $r \rightarrow \infty$ the guided modes must remain finite. Thus for $r \le a$ for core radius $a$ the solution is a Bessel function of the first kind:
\begin{equation}
    E_{z} (r \le a) = A J_{\nu} (ur) e^{i \nu \phi} e^{i (\omega t - \beta z)}
\end{equation}
and outside of the core the solution is a modified Bessel function of the second kind:
\begin{equation}
    E_{z} (r \ge a) = C K_{\nu} (wr) e^{i \nu \phi} e^{i (\omega t - \beta z)}.
\end{equation}
Solving these equations provides the transmission modes of the fibre, i.e. the non-changing distributions.

Next we consider the propagation of light through a fibre from a group theoretic perspective. The indefinite special orthogonal group $\mathrm{SO}(2,1)$ is the group considering two spatial dimensions and one time dimension, and can be realised as:
\begin{equation}
    SO(2,1) = \{ X \in \mathrm{Mat}_{3} (\mathbb{R}) | X^{t} \nu X = \nu , \mathrm{det}(X) = 1 \}
\end{equation}
where,
\begin{equation}
    \nu = \begin{pmatrix}
           1 & 0 & 0\\
           0 & 1 & 0\\
           0 & 0 & -1
           \end{pmatrix}.
\end{equation}

To identify the Lie algebra $\mathfrak{so}(2,1)$ we use the tangent space $T_{1} SO(2,1)$ to $\mathrm{SO}(2,1)$ at the identity $1$. We then choose a curve $a : L \rightarrow SO(21)$ such that $a^{\prime}(0) = A$ Then, the characterising equation of $A$ gives:
\begin{equation}
    a(t)^{T} \nu a(t) = \nu.
\end{equation}
Taking the derivative with respect to $t$ gives:
\begin{equation}
    a^{\prime}(t)^{T} \nu a(t) + a(t)^{T} \nu a^{\prime}(t) = 0.
\end{equation}
Then, evaluating the expression at $t=0$ gives:
\begin{equation}
    A^{T} \nu + \nu A = 0.
    \label{eq:liealgcurve0}
\end{equation}

Now we check the linear conditions determined by the above characterisation to then write out the Lie algebra, with $\nu$ having a natural block decomposition. Therefore, for a general element $A \in \mathfrak{so}(2,1)$, given as:
\begin{equation}
    A = \begin{pmatrix}
           W & x \\
           y^{T} & z
           \end{pmatrix},
\end{equation}
where $W \in M(2, \mathbb{R})$, $x, y \in \mathbb{R}^{2}$, and $z \in \mathbb{R}$. In this block decomposition $\nu = \begin{pmatrix} \mathbb{I}_{2} & 0 \\ 0 & -1 \end{pmatrix}$. Then, Equation~\ref{eq:liealgcurve0} becomes:
\begin{equation}
    \begin{pmatrix}
           W^{T} & y \\
           x^{T} & z
           \end{pmatrix}
    \begin{pmatrix}
           \mathbb{I}_{2} & 0 \\
           0 & -1
           \end{pmatrix}
    +
    \begin{pmatrix}
           \mathbb{I}_{2} & 0 \\
           0 & -1
           \end{pmatrix}
    \begin{pmatrix}
           W & x \\
           y^{T} & z
           \end{pmatrix}
    =
    \begin{pmatrix}
           0 & 0 \\
           0 & 0
           \end{pmatrix}.
\end{equation}
Therefore, the following conditions are imposed:
\begin{eqnarray*}
    W^{T} &=& -W\\
    y &=& x\\
    z &=& 0.
\end{eqnarray*}
Hence, the Lie algebra $\mathfrak{so}(2,1)$ is given by:
\begin{equation}
    \mathfrak{so}(2,1) = \left \{ 
    \begin{pmatrix}
           W & x \\
           x^{T} & 0
           \end{pmatrix}
    : W^{T} = -W
    \right \}
    =
    \left \{ 
    \begin{pmatrix}
           0     & -w    & x_{1} \\
           w     & 0     & x_{2} \\
           x_{1} & x_{2} &  0
           \end{pmatrix}
    \right \}.
\end{equation}

As a result we can see that the condition on $W$ is such that $W \in \mathfrak{so}(2, \mathbb{R})$. Therefore, the condition on $W$ characterises the rotational symmetry of the group. Given the goal is make the connection between the group theoretic understanding of group equivariant neural networks and solution to the wave equation in cylindrical coordinates, this is promising as the solutions to Bessel functions have rotational symmetries. 

Further, the group $\mathrm{SO}(2,1)$ is that of two spatial dimensions and one time dimension. The connection between the three spatial dimensional case and the view as a light cone was made in the development of Minkowski spacetime. Here you imagine a cone where the time axis runs from the point of a cone through the centre of the plane drawn on the open end and the two spatial axes form a plane which intersects the cone and is parallel to the plane drawn on the open end. This is known as the future light cone and is an interpretation of how light spreads out after an event occurs. The group actions of the group $\mathrm{SO}^{+}(2,1)$, which is the group $\mathrm{SO}(2,1)$ with the requirements that $t>0$, act on this space and can be viewed as rotations of the three dimensional Euclidean sphere. A connection can be drawn between this view and the fact that non compact generators of $\mathrm{SO}(n,1)$ differ from corresponding matrix elements of the same generators of $\mathrm{SO}(n+1)$, the group of rotations in $n+1$-dimensional space, by a factor of $\sqrt(-1)$ \citep{wong1974unitary}. Therefore, the connection can be made between the group actions of $\mathrm{SO}^{+}(2,1)$ and the light cone view of light propagation.

A final connection can be made between the light cone and the wave equation, given the connection between the group action and the light cone. Returning to the wave equation we note that it describes waves travelling with frequency independent speed. The character of the solution is different in odd and even dimensional spaces. In an odd dimensional space a disturbance propagates on the light cone and vanishes elsewhere. On the other hand, in an even dimensional space a disturbance propagates inside the entire light cone. Therefore, in an even dimensional space a disturbed medium never returns to rest. This phenomena is known as geometric dispersion. Here we are interested in even dimensional space as we have two spatial dimensions governing transmission through the fibre along with a third time dimension. We therefore expect the propagation of light through the fibre to be understood by propagation inside the entire light cone. Given the wave equation:
\begin{equation}
    u_{tt} = c^{2}(u_{xx} + u_{yy}).
\end{equation}

A solution is possible by considering the three-dimensional theory, if we regard $u$ as a function in three  dimensions and that the third dimension is independent. So, if we require:
\begin{eqnarray*}
    u_{0}(0,x,y) &=& 0\\
    u_{t}(0,x,y) &=& \phi(x,y),
\end{eqnarray*}
then the three-dimensional solution equation becomes:
\begin{equation}
    u(t,x,y) = tM_{ct} [ \phi ] = \frac{t}{4 \pi} \iint_{S} \phi (x + ct\alpha, y + ct\beta) d\omega,
\end{equation}
where $\alpha$ and $\beta$ are the first two coordinates on the unit sphere and $d\omega$ is the area element on the sphere. The integral can be written as a double integral over disc $D$ with centre $x,y$ and radius $ct$:
\begin{equation}
    u(t,x,y) = \frac{1}{2 \pi ct} \iint_{D} \frac{\phi(x+\xi, y+\nu)}{\sqrt{(ct)^{2} - \xi^{2} - \nu^{1}}} d\xi d\nu.
\end{equation}

It becomes clear that the solution does not only depend on the data on the light cone where:
\begin{equation}
    (x - \xi)^{2} + (y - \nu)^{2} = c^{2}t^{2},
\end{equation}
but on the entire data inside the light cone. Therefore, it can be seen how the solution to the wave equation yields the interpretation of the light cone in the precise way that was yielded by the analysis of the group $\mathrm{SO}^{+}(2,1)$.

This completes the connections between the group action of $\mathrm{SO}^{+}(2,1)$ and the solution to the Bessel functions that govern the transmission modes within optical fibres. Now we can compose a model in a similar way to how circular harmonics are used to construct rotation equivariant neural networks under the group $\mathrm{SO}(2)$, but with utilising the basis functions found by solving the Bessel function in Equation~\ref{eq:besselfnc} to compose a network which is equivariant to the group $\mathrm{SO}^{+}(2,1)$. This is useful as it positions the model with respect to other group equivariant neural networks and provides a model with suitable inductive bias for the task of learning a function approximation to transmission through optical fibres.

\subsection{Related Work}
\label{sec:relworkapp}
One method for inverting the transmission effects is to find the underlying TM of the multi-mode fibre. In general this is not known, although in principle the TM can be found by acquiring the output amplitude and phase relative to each mode \citep{vcivzmar2011shaping, vcivzmar2012exploiting}. \cite{choi2012scanner} present an approach to construct the TM in a scanner-free method based on measurement of amplitude and phase of the output. Although the method requires 500 measurement repetitions at different incidence angles. \cite{mahalati2013resolution} develop a method in which the number of resolvable image features approached four times the number of spatial modes. \cite{papadopoulos2012focusing} develop a digital phase conjugation technique to restore images without the requirement of calculating the full TM. Despite this, the solution of finding the TM, or part of it, has to be repeated for every different fibre, for every different length, under each different bending scenario, and for each different temperature, reducing the practicality of this solution. \cite{ploschner2015seeing} developed a procedure that could also incorporate bending through a precise characterisation of the fibre and a theoretical model. 

Another approach to inverting the transmission effects is to use machine learning to solve the inverse problem. \cite{moran2018deep} and \cite{ caramazza2019} approached solving the inverse problem through the use of either a Real and Complex linear model with a Hadamard layer to model the power drop off of the spatial light modulator. At its core this approach requires the use of a linear model, which maps from speckled images to the original images and scales as $\mathcal{O}(N_{s}^{2}N_{o}^{2})$, where $N_{s}$ is the resolution of the speckled image and $N_{o}$ is the resolution of the original image. This is a very memory-expensive operation and restricts the scalability of the approach to relatively low resolution images. Further, \cite{borhani2018learning} use a convolutional U-Net model to invert the transmission effects, which although using spatially localised convolutions is more memory efficient than a linear model, the speckled images are not in a similar spatial arrangement. An indication of this is likely in the fact their model requires 14 hidden layers to learn the inversion of $32 \times 32$ resolution images. \cite{fan2019deep} use a convolutional neural network to invert transmission effects, where the convolutions and pooling reduce the speckled resolution down before a dense linear layer predicts the original image. As a result the model will be sensitive to the learned down-sampling of the convolutions and suffer from similar scalability issues to \cite{moran2018deep} due to the inclusion of a dense linear layer. Further, \cite{rahmani2018multimode} also use a convolutional model and despite not having a dense linear layer, the model does comprise $22$ convolutional layers, so, for low resolution, $28 \times 28$, images it is not surprising the model can overcome the mapping between two image domains with non-similar spatial arrangements as a mapping from each speckled pixel to each original image pixel is possible. There is therefore no evidence that the convolutional based models truly improve the scalability issue as these models are very large considering the resolution of images. Finally, of the machine learning based methods for inverting the transmission effects \cite{rahmani2018multimode} make the first attempt to demonstrate generalisation outside of the training domain, although the results demonstrate limited evidence of true generalisation due to the choice of images featuring limited high frequency features and low quality reconstructions, while \cite{caramazza2019} demonstrate stronger generalisation due to the testing data choice, but there is still scope to improve the reconstruction quality outside of the training domain.

\subsection{Training Details}
\label{sec:trainingdets}

Throughout the paper we make use of three different datasets that are commonly used in machine learning, namely MNIST \citep{lecun1998gradient}, fMNIST \citep{xiao2017fashion}, and ImageNet \citep{deng2009imagenet}. Where MNIST and fMNIST are also commonly used in the study of inverting transmission effects of MMFs due to their low resolutions. MNIST images are $28\times28$ images of handwritten digits between $0$ and $9$. fMNIST images are $28\times28$ images of items of clothing, such as trousers, jumpers, or shoes. ImageNet is a large scale dataset of higher resolution images which we size to be $256x256$. We only use a subset of the imageNet dataset to demonstrate the ability to invert such high resolution images. For each of these datasets we split into a training and testing dataset. The training dataset is used for training, while the testing dataset is not used during training and we use this to test the model after training. We use the testing dataset to generate all the predicted images throughout the paper.

We train each model for 200 epochs with the stochastic gradient descent algorithm. We use the mean squared error as a loss function between the original images and the predicted images. We do not use any regularisation on the model weights. For the Real and Complex linear models and the Bessel equivariant model we train each model for 200 epochs. For the post-processing model, as it is an addition to the Bessel equivariant model, we train this model for 200 epochs after the Bessel equivariant model has already been trained for 200 epochs and the model weights frozen. For the training, all models are trained on 1 Titan RTX GPU with 24GiB of memory. We did attempt training the complex linear model on a A6000 GPU with 48GB of memory on ImageNet to check if this was possible, but it required more memory.

We utilise two different sources of data throughout the experimental section. The first of these is collected using a theoretical TM for which we have both the amplitude and phase information for the speckled patterns. In addition, we also have access to the original and inverted images, where the inverted images are created by first passing an image through the TM and then back through its inverse. The second source of data is that collected by \cite{moran2018deep}, which is data collected in the lab using a real fibre. For this data we only have the amplitude information for the speckled patterns. In addition, we also have the original images which have constant phase. This data has no licence and it accessible on the projects GitHub page. 
\newpage
\subsection{Invertibility of TMs}
\label{invTM}

\begin{figure}[htb]
  \centering
  \begin{subfigure}{0.19\linewidth}
    \includegraphics[width=\linewidth]{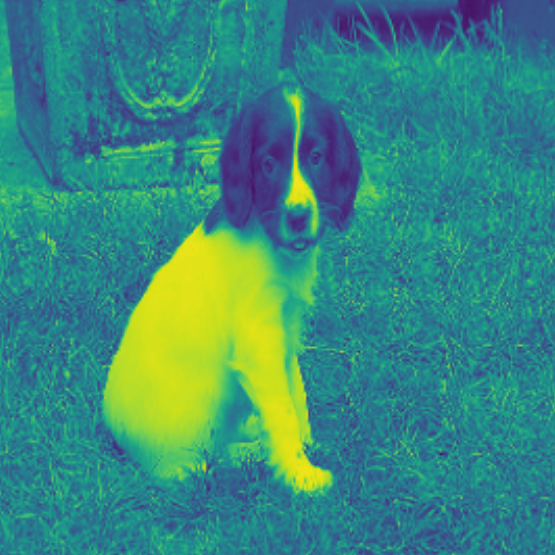}
  \end{subfigure}
  \begin{subfigure}{0.19\linewidth}
    \includegraphics[width=\linewidth]{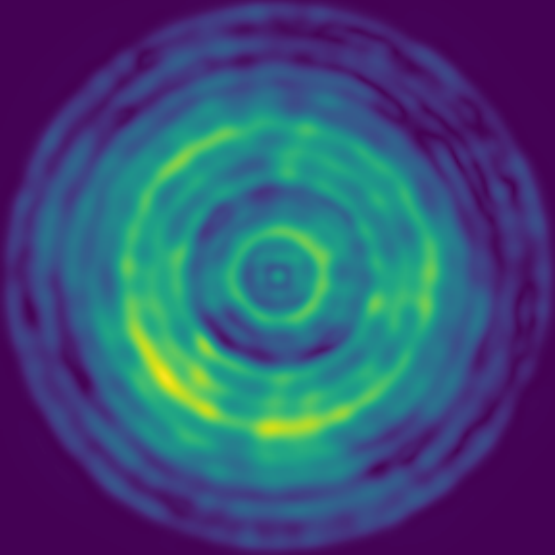}
  \end{subfigure}
  \begin{subfigure}{0.19\linewidth}
    \includegraphics[width=\linewidth]{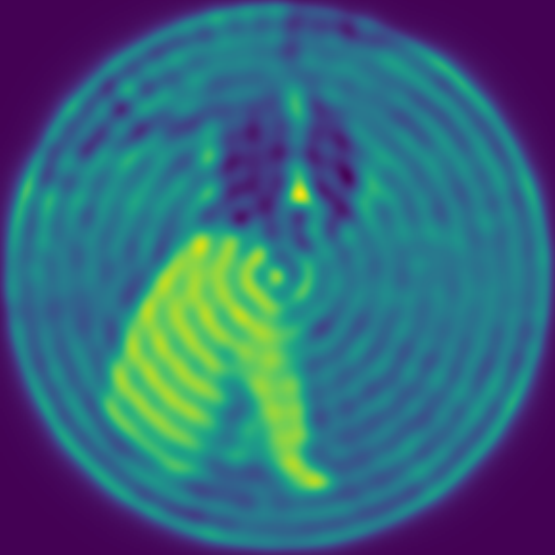}
  \end{subfigure}
  
  \begin{subfigure}{0.19\linewidth}
    \includegraphics[width=\linewidth]{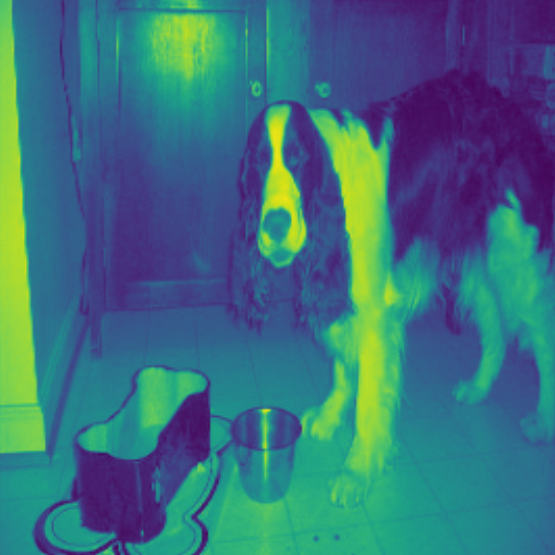}
  \end{subfigure}
  \begin{subfigure}{0.19\linewidth}
    \includegraphics[width=\linewidth]{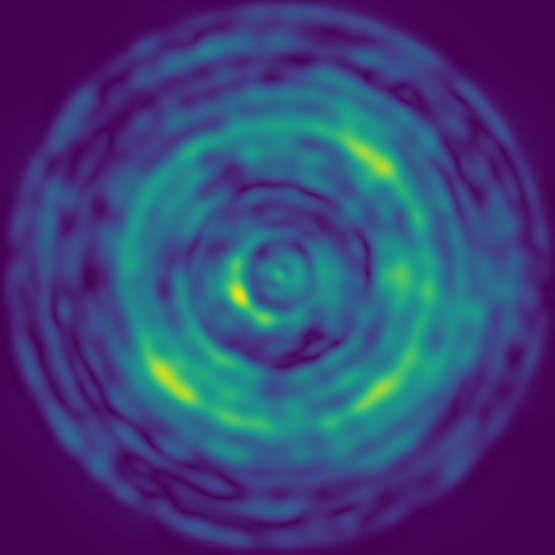}
  \end{subfigure}
  \begin{subfigure}{0.19\linewidth}
    \includegraphics[width=\linewidth]{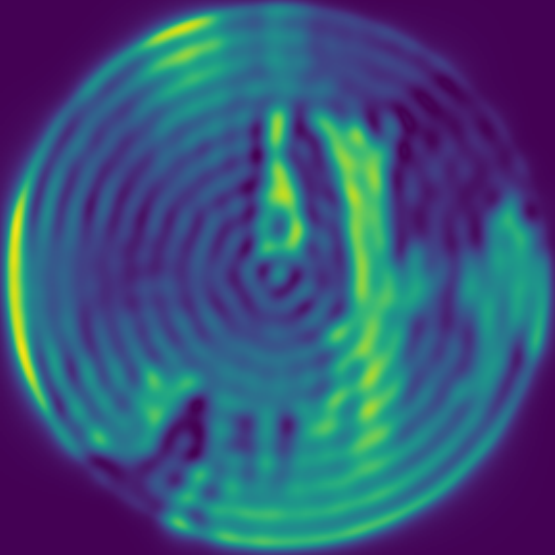}
  \end{subfigure}
  
  \begin{subfigure}{0.19\linewidth}
    \includegraphics[width=\linewidth]{results/imagenet/orig_speck_inv/original_2.pdf}
    \centering
    (a) Original
  \end{subfigure}
  \begin{subfigure}{0.19\linewidth}
    \includegraphics[width=\linewidth]{results/imagenet/orig_speck_inv/speckled_2.pdf}
    \centering
    (b) Speckled
  \end{subfigure}
  \begin{subfigure}{0.19\linewidth}
    \includegraphics[width=\linewidth]{results/imagenet/orig_speck_inv/inverse_2.pdf}
    \centering
    (c) Inverted
  \end{subfigure}

  \caption{(a) The original image. (b) The speckled image created by passing the original images through a theoretical TM. (c) The inverted image created by passing the speckled images through the inverted theoretical TM.}
  \label{fig:origspecinv}
\end{figure}

Given a TM, which models how an image propagates through a fiber, it is possible to invert the matrix and get the corresponding mapping back from the speckled image space to the original image space. We provide details of the construction of TMs in Appendix~\ref{sec:theorytm}. Utilising this we can inspect information loss due to the limited number of modes of the fibre by passing the image through the TM and back through the inverse to create an inverted image. Therefore, if we knew the TM this inverted image would be the recoverable information. We show an example of the original image, the speckled image, and the inverted image in Figure~\ref{fig:origspecinv}.

This is useful as it allows us to separate the task of inverting the transmission effects of a fibre into that which could be reconstructed by understanding the physics of the TM and that which requires generating due to being lost information. We believe this could therefore be viewed as two tasks: (1) a task of inverting the transmission effects which would have the aim of generating the inverted images and (2) a task, similar to a super-resolution task, of predicting the original images from the output of the first task.

\FloatBarrier

\subsection{Results for Accounting for Losses in a Real System}

As we noted in Section~\ref{sec:realfibre} the assumption that the mapping between Bessel bases is diagonal holds in theory, although in practice an imperfect system could lead to a necessity to relax this assumption. Here we explore this for the MNIST data collected using a real fibre by building three different versions of our model (1) the strong assumption of a diagonal matrix, (2) relaxing the assumption to allow five elements either side of the diagonal to be populated, (3) relaxing the assumption to allow ten elements either side of the diagonal to be populated, and (4) one with a full matrix mapping between bases. If the fibre and experimental set-up was in an ideal setting we would expect model (1) to produce results as strong as those demonstrated in Section~\ref{sec:theoryfibre} where data was created with a theoretical TM. On the other hand, model (2) allowing 5 elements to be populated either side of the diagonal would allow the model to capture some of the block diagonal structure seen in \cite{carpenter2014maximally}, although this would still occlude the mapping between modes further apart in our mapping space. Similarly, (3) allows for capturing more of the block diagonal structure by allowing 10 elements to be populated either side of the diagonal. Finally, model (4) allows the greatest flexibility and has the potential to capture deviations from theory as seen in practice, although, this model does not take advantage of the known diagonal structure that this mapping function has and is therefore over-parameterised.

\begin{figure}[htb]
  \centering
  \begin{subfigure}{0.11\linewidth}
    \includegraphics[width=\linewidth]{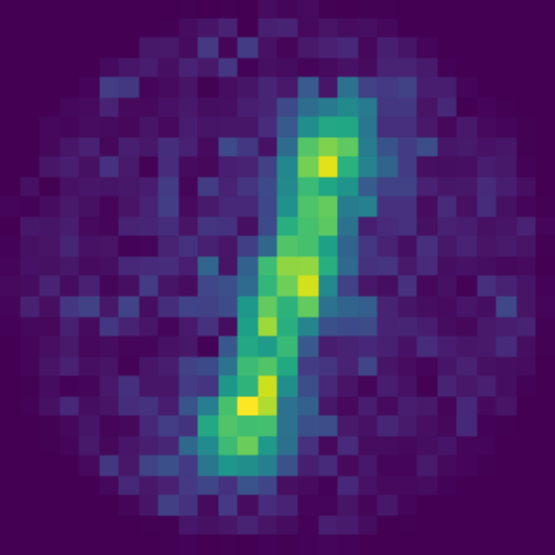}
  \end{subfigure}
  \begin{subfigure}{0.11\linewidth}
    \includegraphics[width=\linewidth]{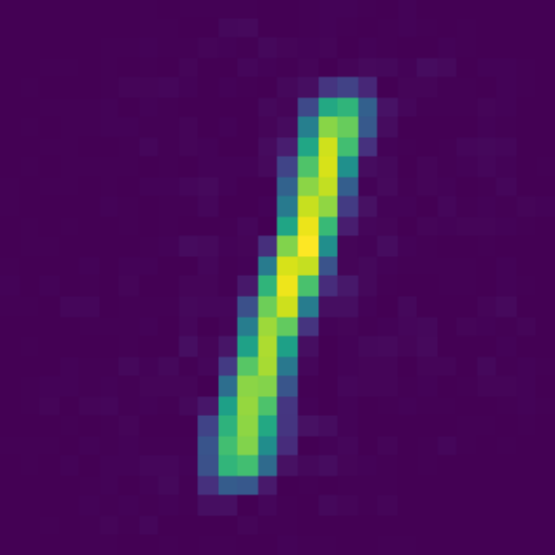}
  \end{subfigure}
  \begin{subfigure}{0.11\linewidth}
    \includegraphics[width=\linewidth]{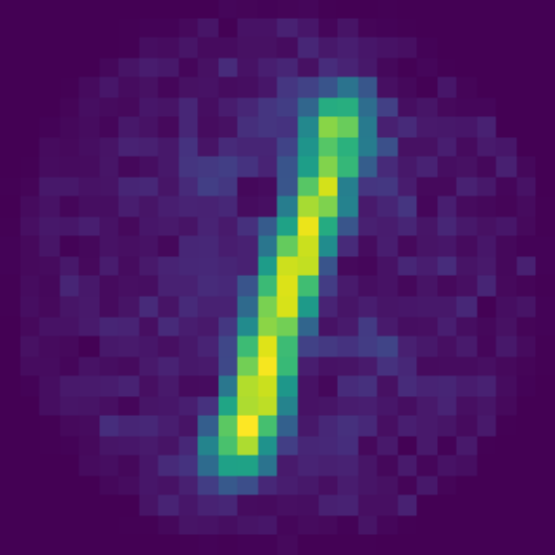}
  \end{subfigure}
  \begin{subfigure}{0.11\linewidth}
    \includegraphics[width=\linewidth]{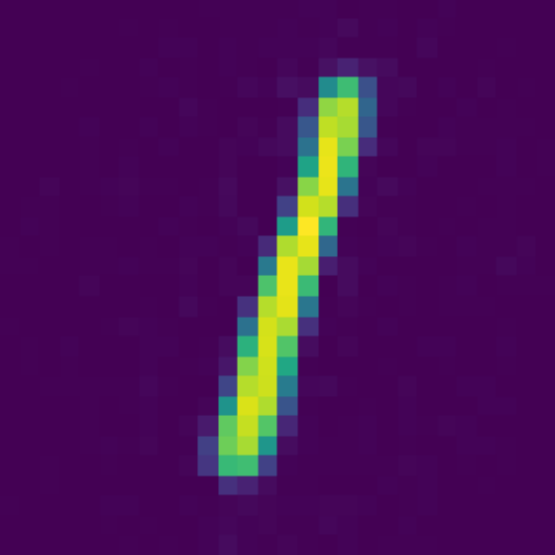}
  \end{subfigure}
  \begin{subfigure}{0.11\linewidth}
    \includegraphics[width=\linewidth]{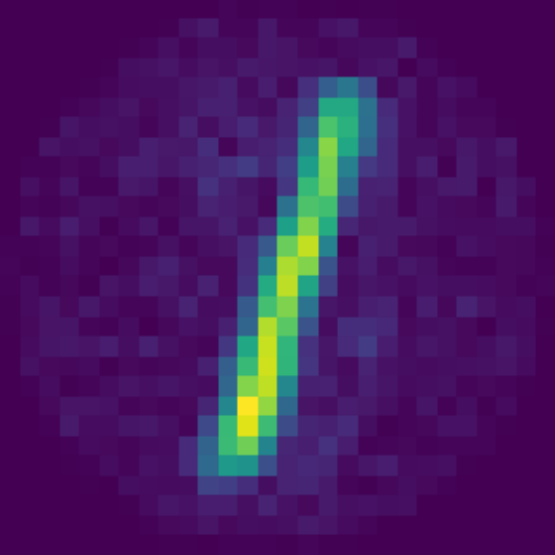}
  \end{subfigure}
  \begin{subfigure}{0.11\linewidth}
    \includegraphics[width=\linewidth]{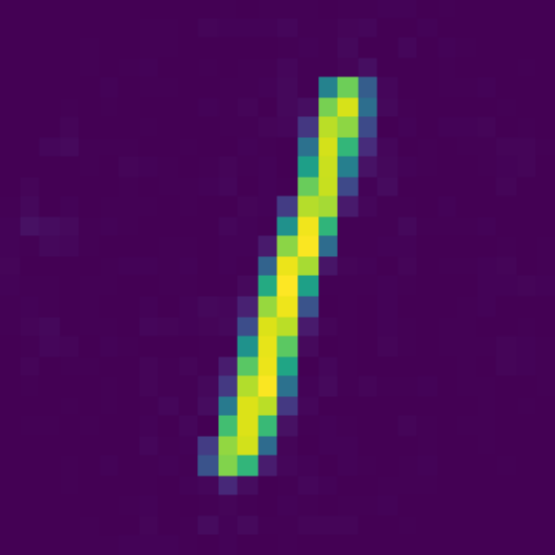}
  \end{subfigure}
  \begin{subfigure}{0.11\linewidth}
    \includegraphics[width=\linewidth]{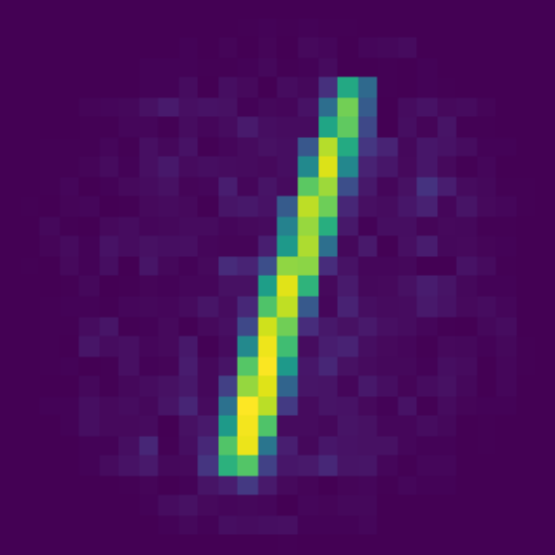}
  \end{subfigure}
  \begin{subfigure}{0.11\linewidth}
    \includegraphics[width=\linewidth]{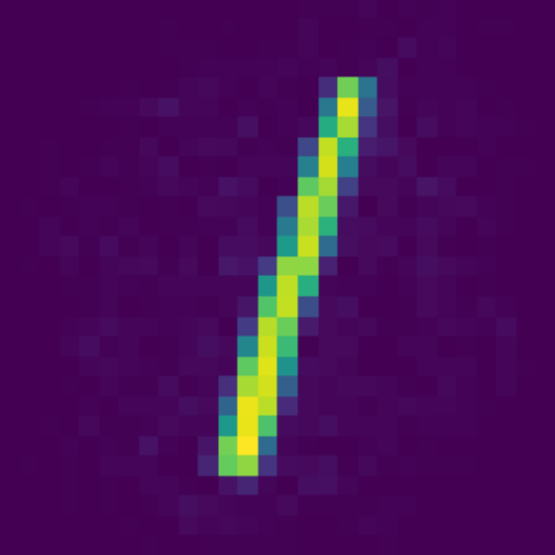}
  \end{subfigure}
  
  \begin{subfigure}{0.11\linewidth}
    \includegraphics[width=\linewidth]{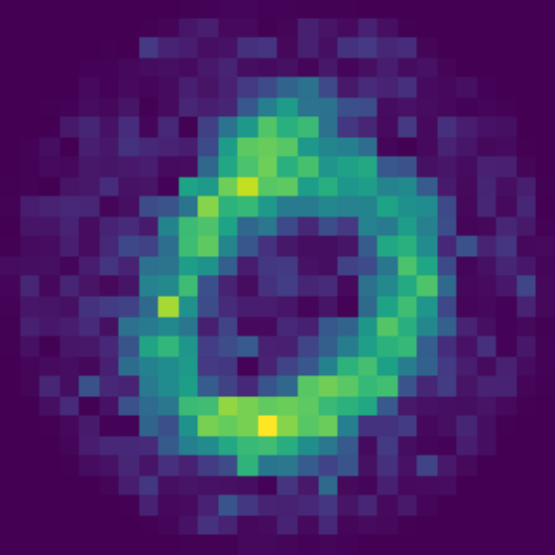}
  \end{subfigure}
  \begin{subfigure}{0.11\linewidth}
    \includegraphics[width=\linewidth]{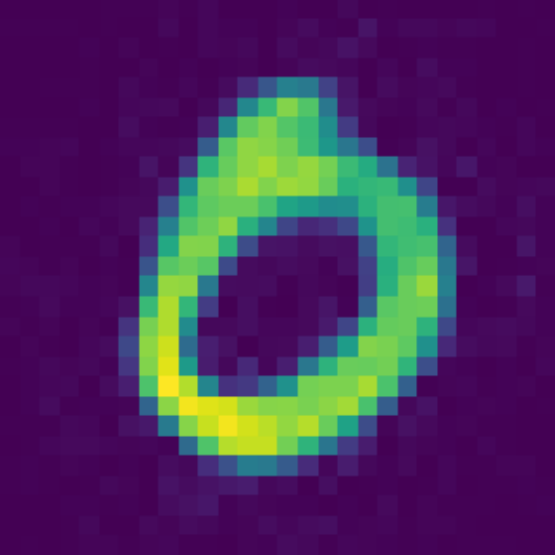}
  \end{subfigure}
  \begin{subfigure}{0.11\linewidth}
    \includegraphics[width=\linewidth]{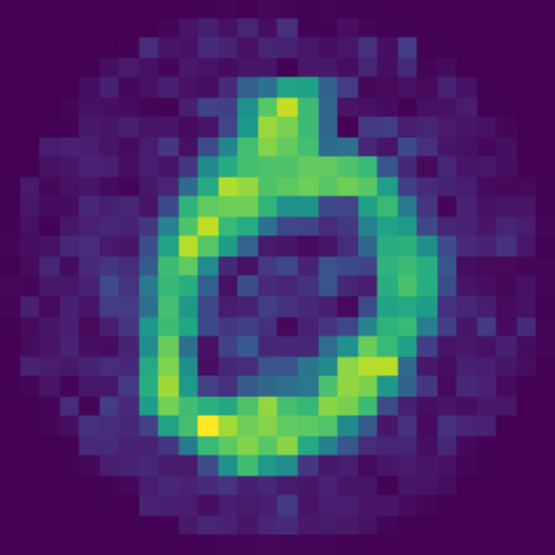}
  \end{subfigure}
  \begin{subfigure}{0.11\linewidth}
    \includegraphics[width=\linewidth]{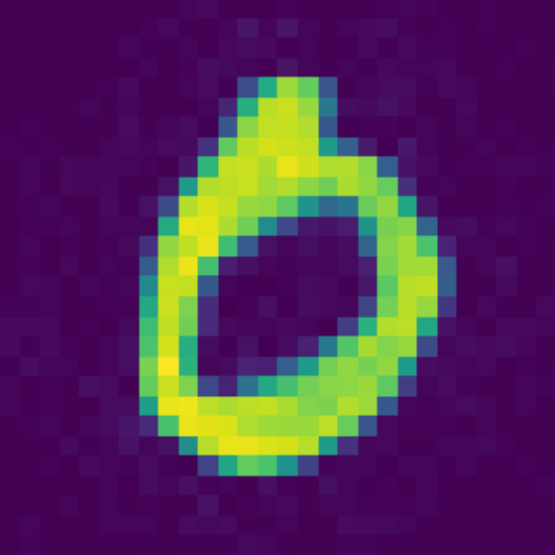}
  \end{subfigure}
  \begin{subfigure}{0.11\linewidth}
    \includegraphics[width=\linewidth]{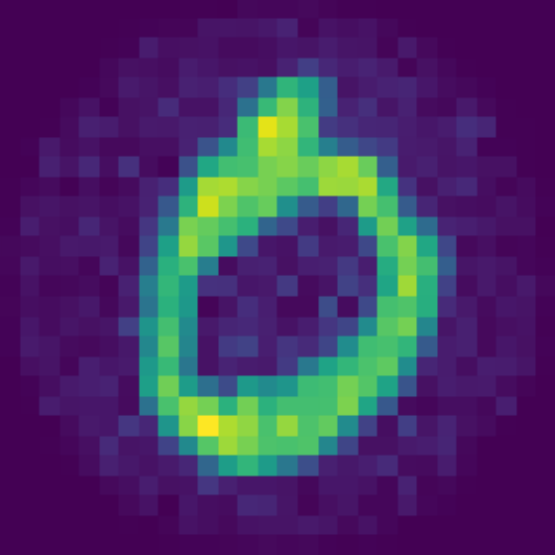}
  \end{subfigure}
  \begin{subfigure}{0.11\linewidth}
    \includegraphics[width=\linewidth]{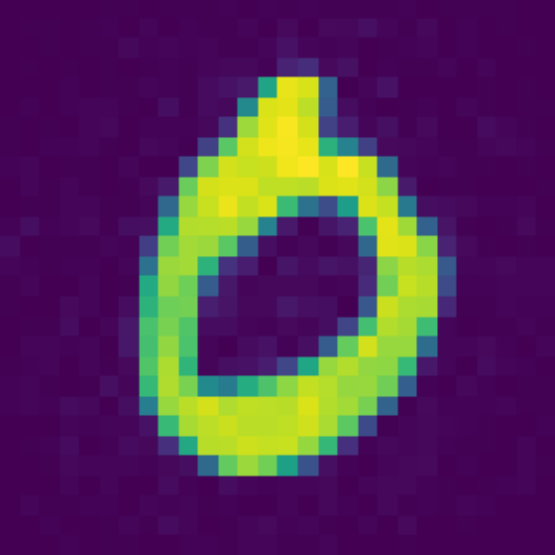}
  \end{subfigure}
  \begin{subfigure}{0.11\linewidth}
    \includegraphics[width=\linewidth]{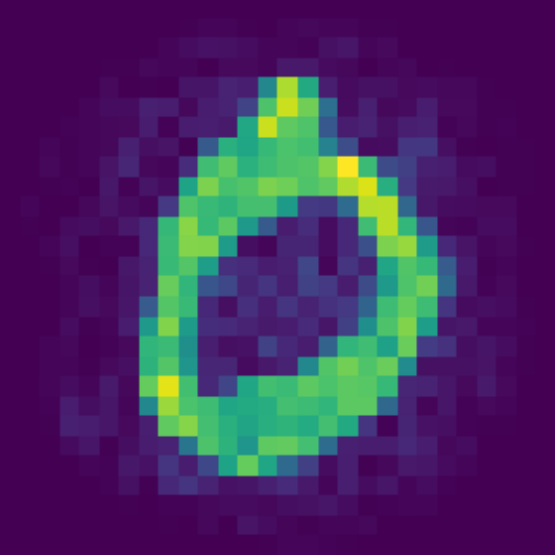}
  \end{subfigure}
  \begin{subfigure}{0.11\linewidth}
    \includegraphics[width=\linewidth]{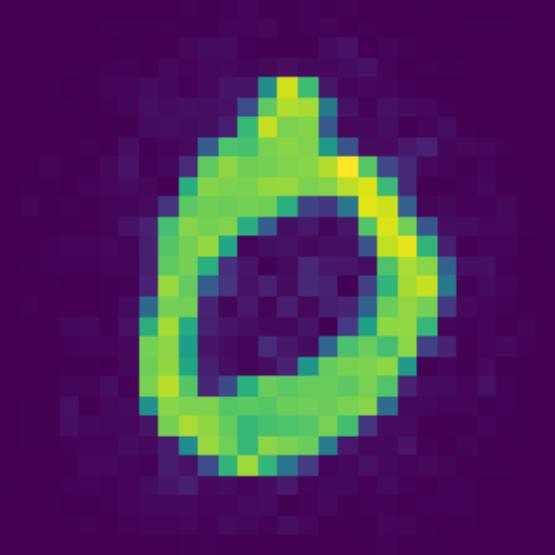}
  \end{subfigure}
  
  \begin{subfigure}{0.11\linewidth}
    \includegraphics[width=\linewidth]{results/real/figs1/TM_bases_2.pdf}
    \centering
    (a) 
  \end{subfigure}
  \begin{subfigure}{0.11\linewidth}
    \includegraphics[width=\linewidth]{results/real/figs1/TM_bases_SR_2.pdf}
    \centering
    (b)
  \end{subfigure}
  \begin{subfigure}{0.11\linewidth}
    \includegraphics[width=\linewidth]{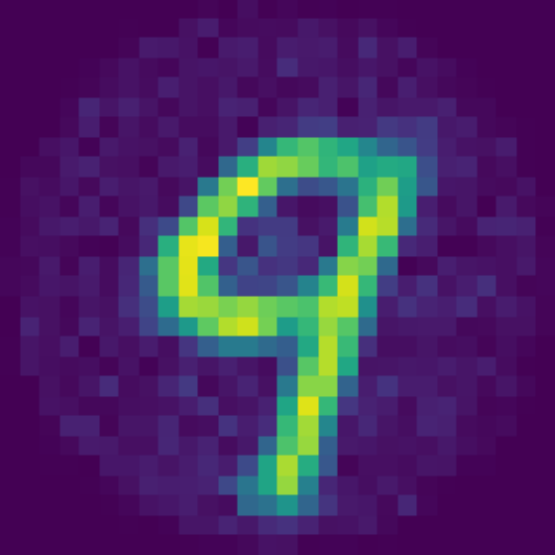}
    \centering
    (c)
  \end{subfigure}
  \begin{subfigure}{0.11\linewidth}
    \includegraphics[width=\linewidth]{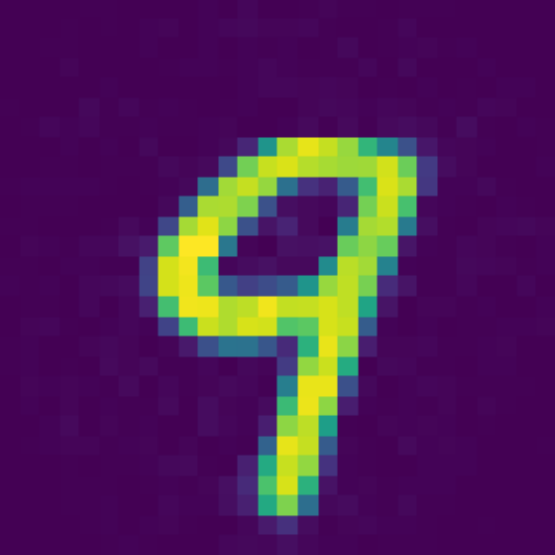}
    \centering
    (d)
  \end{subfigure}
  \begin{subfigure}{0.11\linewidth}
    \includegraphics[width=\linewidth]{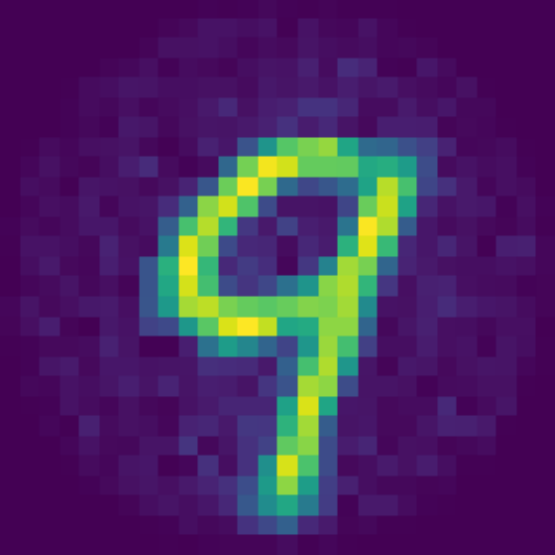}
    \centering
    (e)
  \end{subfigure}
  \begin{subfigure}{0.11\linewidth}
    \includegraphics[width=\linewidth]{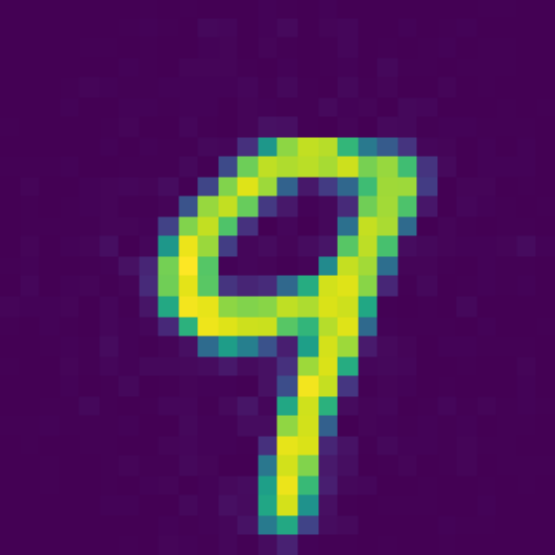}
    \centering
    (f)
  \end{subfigure}
  \begin{subfigure}{0.11\linewidth}
    \includegraphics[width=\linewidth]{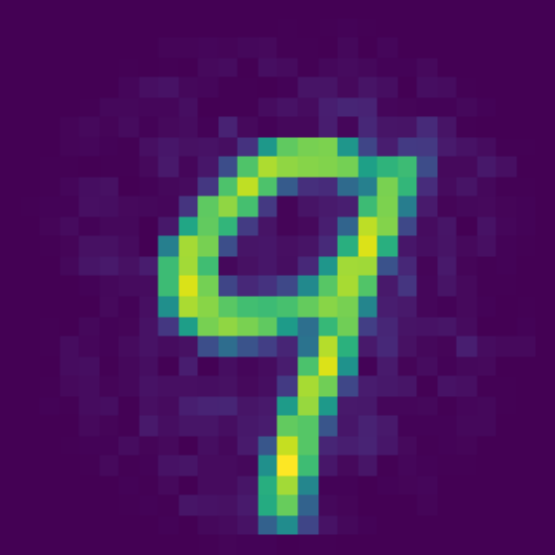}
    \centering
    (g)
  \end{subfigure}
  \begin{subfigure}{0.11\linewidth}
    \includegraphics[width=\linewidth]{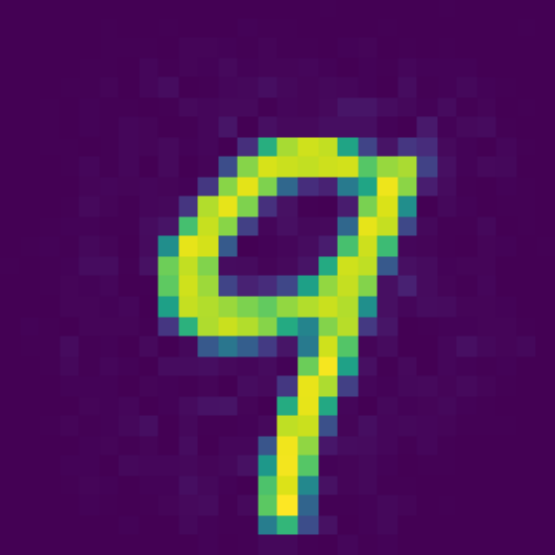}
    \centering
    (h)
  \end{subfigure}
  \caption{Comparison of predicted images from inverting transmission effects of a MMF with varying relaxations on the assumption the fibre has a diagonal fibre propagation matrix. (a) the output of the Bessel equivariant model with a diagonal propagation matrix assumption between Bessel bases, (b) the output of the combination of Bessel equivariant with a diagonal propagation matrix assumption between Bessel bases and post-processing model, (c) the output of the Bessel equivariant model with 5 diagonal offsets in the propagation matrix between Bessel bases, (d) the output of the Bessel equivariant model with 5 diagonal offsets in the propagation matrix between Bessel bases and post-processing model, (e) the output of the Bessel equivariant model with 10 diagonal offsets in the propagation matrix between Bessel bases, (f) the output of the Bessel equivariant model with 10 diagonal offsets in the propagation matrix between Bessel bases and post-processing model, (g) the output of the Bessel equivariant model with a full propagation matrix between Bessel bases, and (h) the output of the Bessel equivariant model with a full propagation matrix between Bessel bases and post-processing model.}
  \label{fig:realMNISTmod}
\end{figure}

Figure~\ref{fig:realMNISTmod} demonstrates that as the diagonal assumption of the mapping function between Bessel bases is relaxed, our Bessel equivariant model produced images closer to that original image through capturing more high frequency detail and predicting less noise. Our model with five elements either side of the diagonal populated with trainable parameters achieves a comparable result to the Real and Complex linear models, and our most constrained model is able to generate digits that like the correct digit. This is a promising result as it demonstrates that our model allows for a design choice of complexity, with the option to trade off performance for a reduced memory requirement whilst maintaining accurate results even in the lowest memory configuration; this is something that is not possible with the Real or Complex linear models. Choosing a model with a low number of off-diagonal trainable elements, we believe, is the best trade off between considering losses and imperfections of the real-world while benefiting from the known sparsity of the mapping function between Bessel bases.

\begin{figure}[htb]
  \centering
  \begin{subfigure}{0.11\linewidth}
    \includegraphics[width=\linewidth]{results/real/fmnist/TM_bases_0.pdf}
  \end{subfigure}
  \begin{subfigure}{0.11\linewidth}
    \includegraphics[width=\linewidth]{results/real/fmnist/TM_bases_SR_0.pdf}
  \end{subfigure}
  \begin{subfigure}{0.11\linewidth}
    \includegraphics[width=\linewidth]{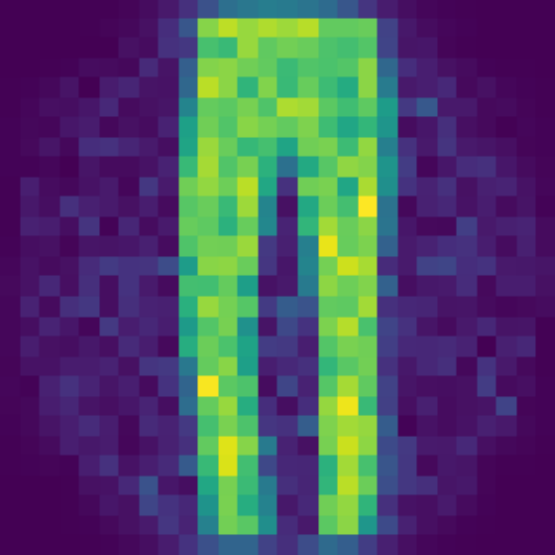}
  \end{subfigure}
  \begin{subfigure}{0.11\linewidth}
    \includegraphics[width=\linewidth]{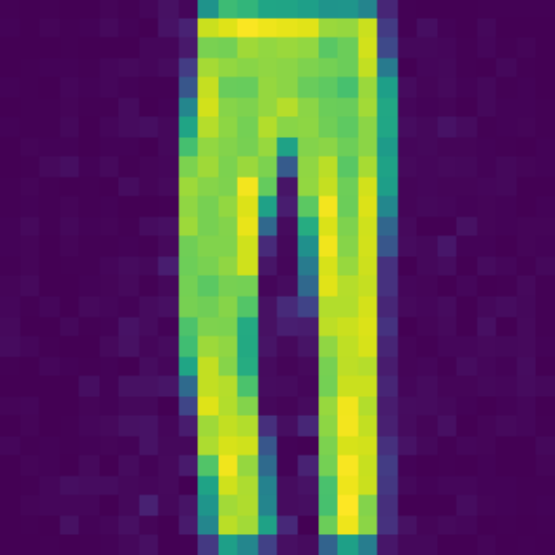}
  \end{subfigure}
  \begin{subfigure}{0.11\linewidth}
    \includegraphics[width=\linewidth]{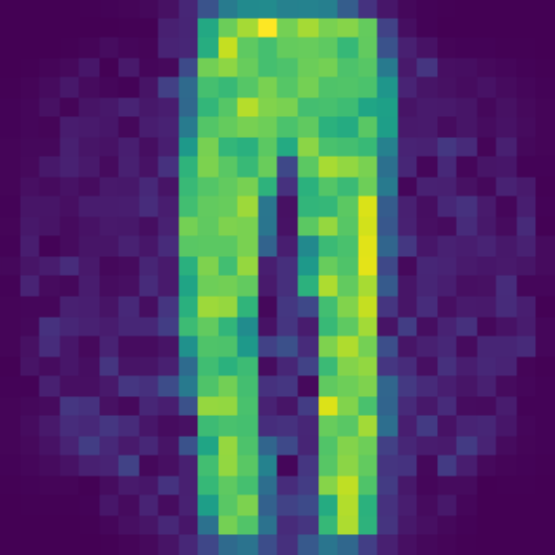}
  \end{subfigure}
  \begin{subfigure}{0.11\linewidth}
    \includegraphics[width=\linewidth]{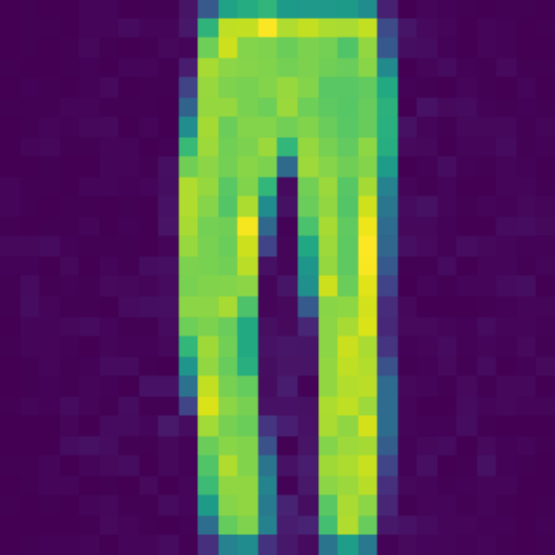}
  \end{subfigure}
  \begin{subfigure}{0.11\linewidth}
    \includegraphics[width=\linewidth]{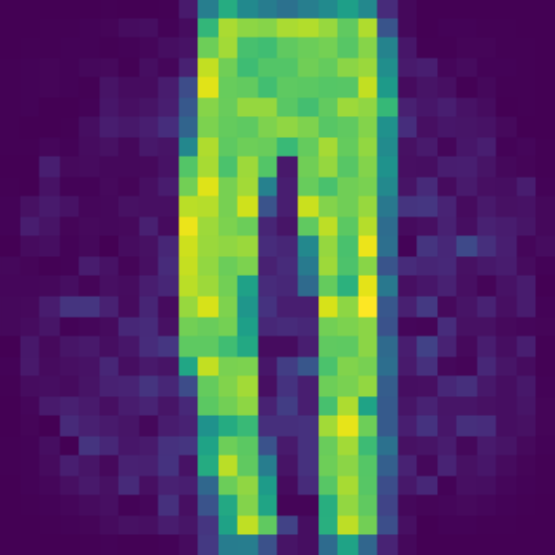}
  \end{subfigure}
  \begin{subfigure}{0.11\linewidth}
    \includegraphics[width=\linewidth]{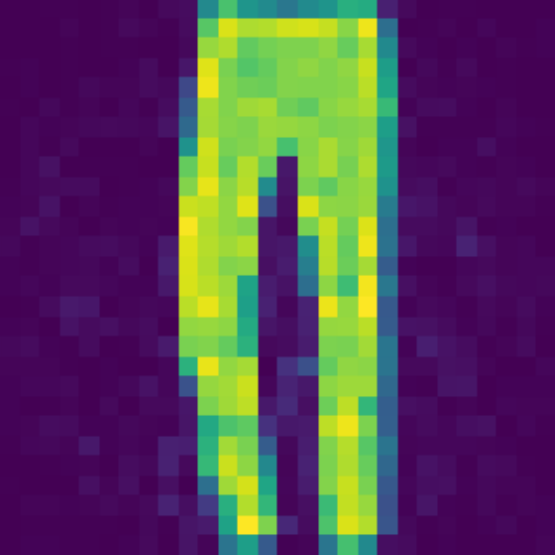}
  \end{subfigure}
  
  \begin{subfigure}{0.11\linewidth}
    \includegraphics[width=\linewidth]{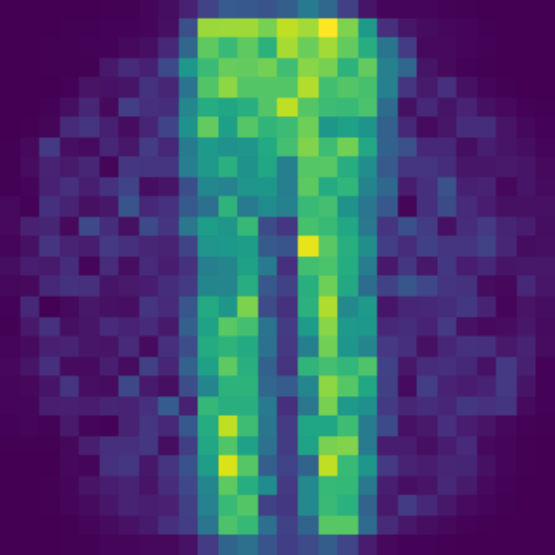}
  \end{subfigure}
  \begin{subfigure}{0.11\linewidth}
    \includegraphics[width=\linewidth]{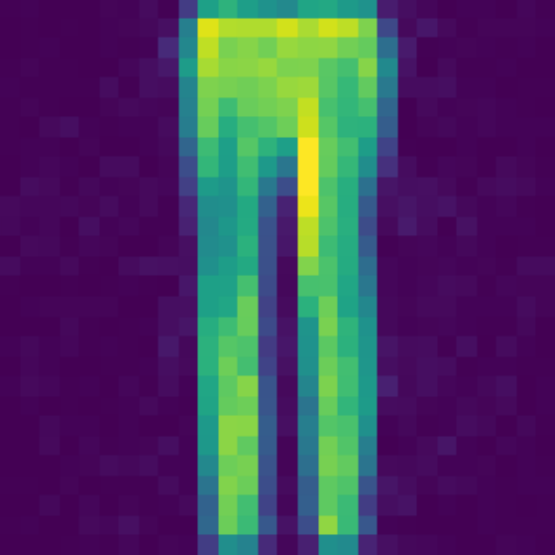}
  \end{subfigure}
  \begin{subfigure}{0.11\linewidth}
    \includegraphics[width=\linewidth]{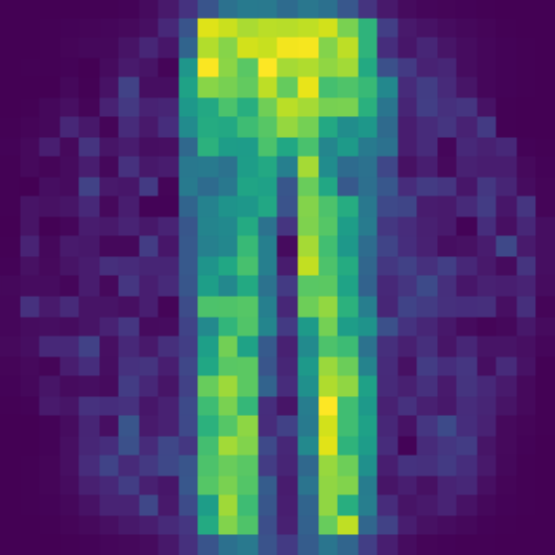}
  \end{subfigure}
  \begin{subfigure}{0.11\linewidth}
    \includegraphics[width=\linewidth]{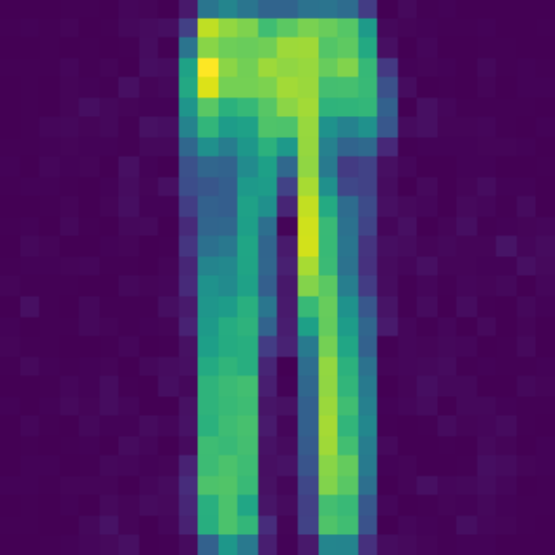}
  \end{subfigure}
  \begin{subfigure}{0.11\linewidth}
    \includegraphics[width=\linewidth]{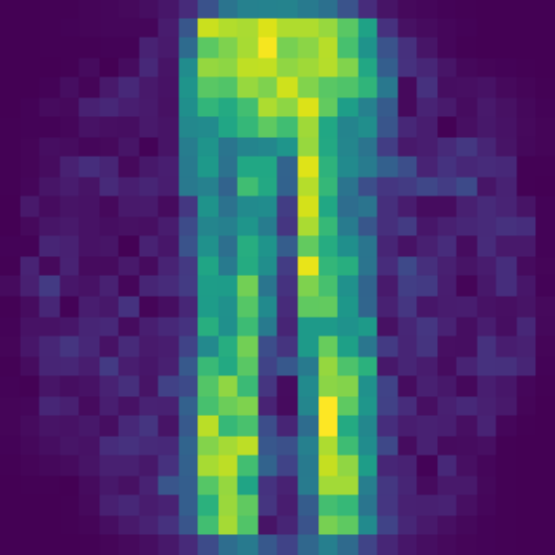}
  \end{subfigure}
  \begin{subfigure}{0.11\linewidth}
    \includegraphics[width=\linewidth]{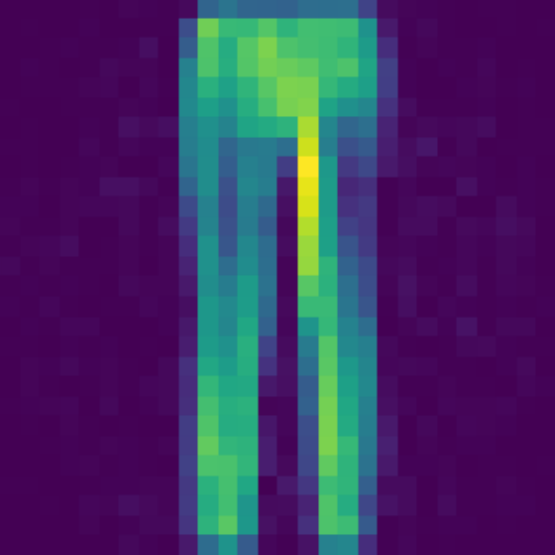}
  \end{subfigure}
  \begin{subfigure}{0.11\linewidth}
    \includegraphics[width=\linewidth]{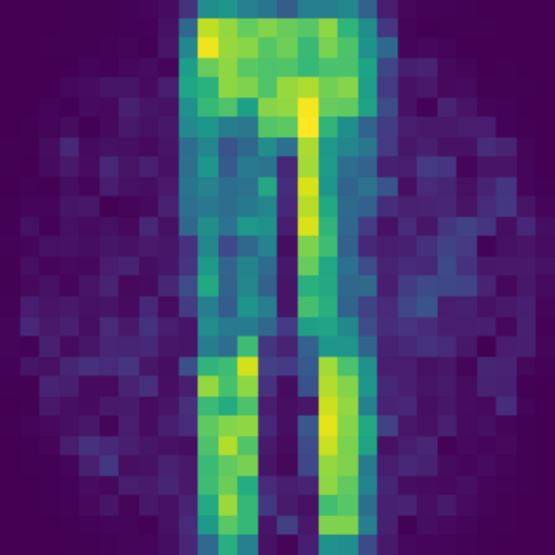}
  \end{subfigure}
  \begin{subfigure}{0.11\linewidth}
    \includegraphics[width=\linewidth]{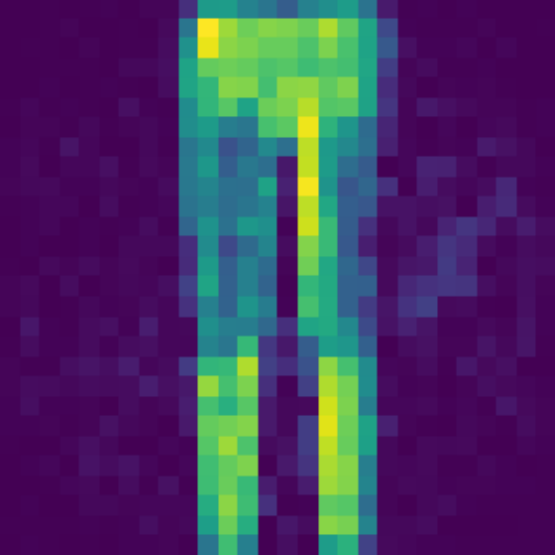}
  \end{subfigure}

  \begin{subfigure}{0.11\linewidth}
    \includegraphics[width=\linewidth]{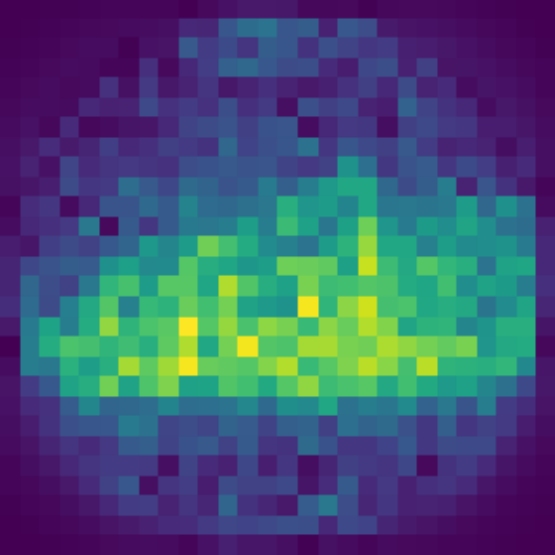}
    \centering
    (a)
  \end{subfigure}
  \begin{subfigure}{0.11\linewidth}
    \includegraphics[width=\linewidth]{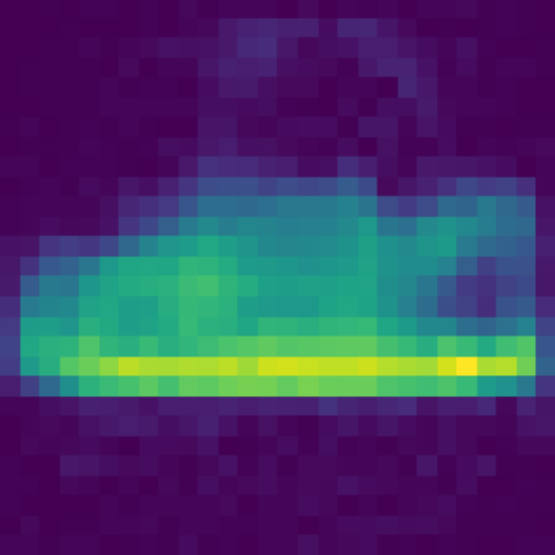}
    \centering
    (b)
  \end{subfigure}
  \begin{subfigure}{0.11\linewidth}
    \includegraphics[width=\linewidth]{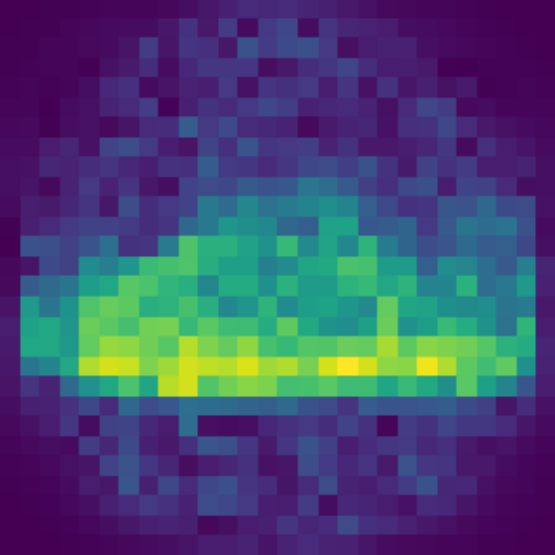}
    \centering
    (c)
  \end{subfigure}
  \begin{subfigure}{0.11\linewidth}
    \includegraphics[width=\linewidth]{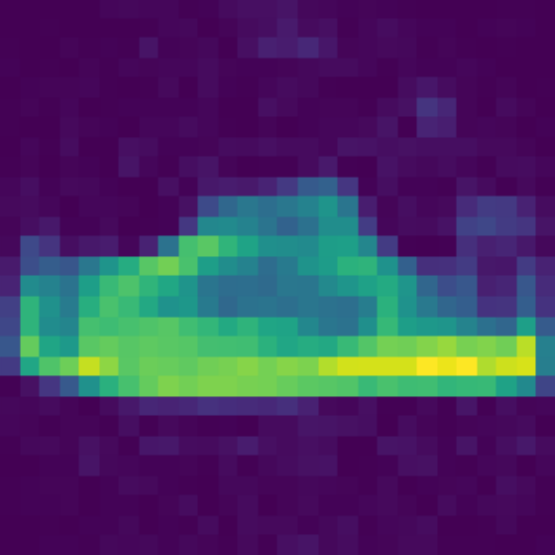}
    \centering
    (d)
  \end{subfigure}
  \begin{subfigure}{0.11\linewidth}
    \includegraphics[width=\linewidth]{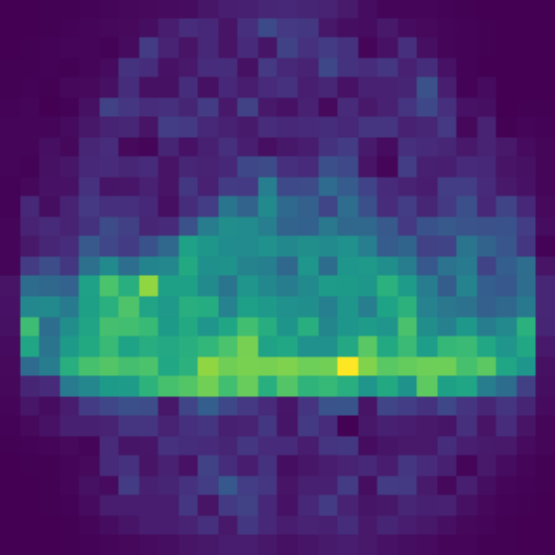}
    \centering
    (e)
  \end{subfigure}
  \begin{subfigure}{0.11\linewidth}
    \includegraphics[width=\linewidth]{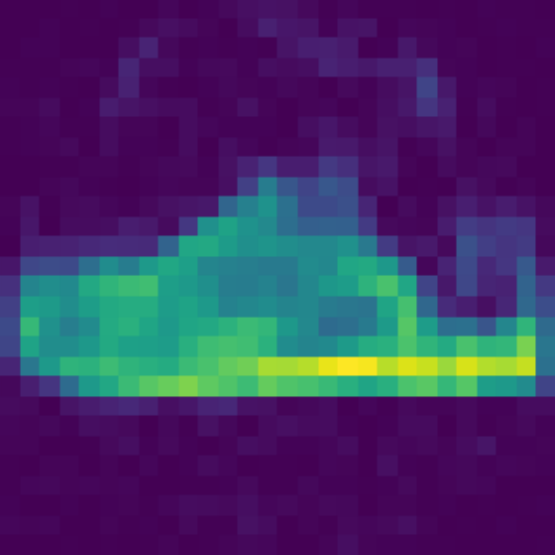}
    \centering
    (f)
  \end{subfigure}
  \begin{subfigure}{0.11\linewidth}
    \includegraphics[width=\linewidth]{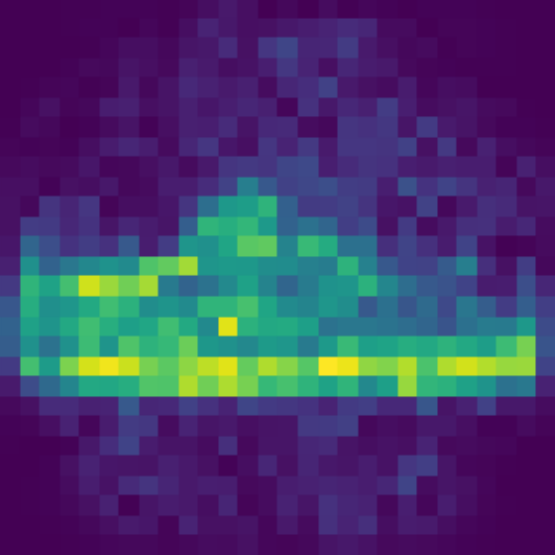}
    \centering
    (g)
  \end{subfigure}
  \begin{subfigure}{0.11\linewidth}
    \includegraphics[width=\linewidth]{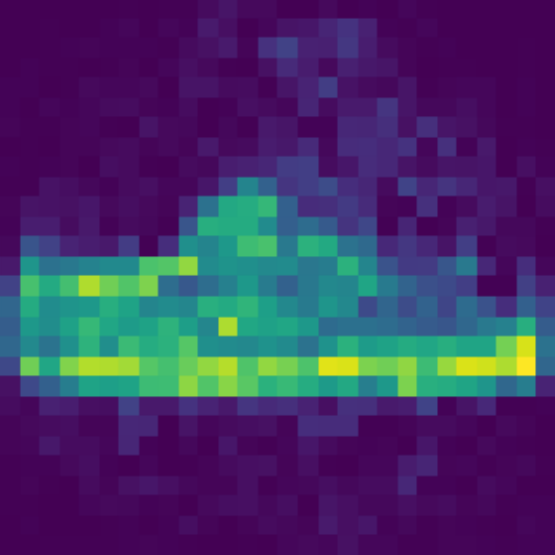}
    \centering
    (h)
  \end{subfigure}
  \caption{Comparison of predicted images from inverting transmission effects of a MMF with varying relaxations on the assumption the fibre has a diagonal fibre propagation matrix. (a) the output of the Bessel equivariant model with a diagonal propagation matrix assumption between Bessel bases, (b) the output of the combination of Bessel equivariant with a diagonal propagation matrix assumption between Bessel bases and post-processing model, (c) the output of the Bessel equivariant model with 5 diagonal offsets in the propagation matrix between Bessel bases, (d) the output of the Bessel equivariant model with 5 diagonal offsets in the propagation matrix between Bessel bases and post-processing model, (e) the output of the Bessel equivariant model with 10 diagonal offsets in the propagation matrix between Bessel bases, (f) the output of the Bessel equivariant model with 10 diagonal offsets in the propagation matrix between Bessel bases and post-processing model, (g) the output of the Bessel equivariant model with a full propagation matrix between Bessel bases, and (h) the output of the Bessel equivariant model with a full propagation matrix between Bessel bases and post-processing model.}
  \label{fig:realfmnistmod}
\end{figure}

\FloatBarrier

\subsection{Effects of Noise on Speckled Images}
\label{sec:noisefmnist}

\begin{figure}[htb]
  \centering
  \begin{subfigure}{0.16\linewidth}
    \includegraphics[width=\linewidth]{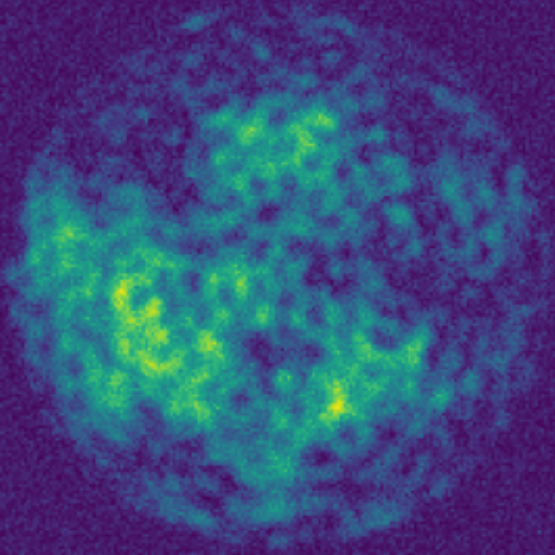}
  \end{subfigure}
  \begin{subfigure}{0.16\linewidth}
    \includegraphics[width=\linewidth]{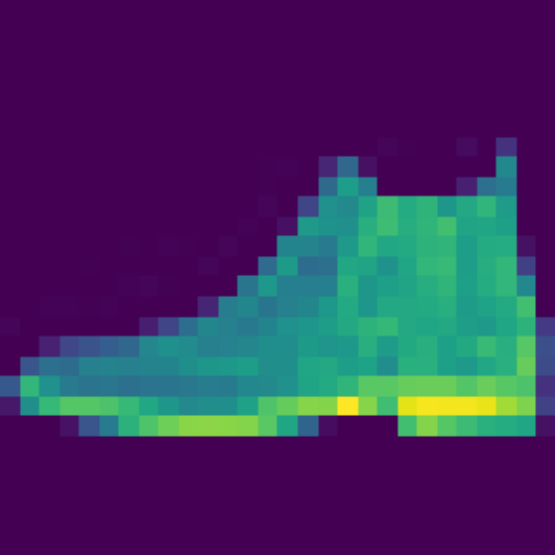}
  \end{subfigure}
  \begin{subfigure}{0.16\linewidth}
    \includegraphics[width=\linewidth]{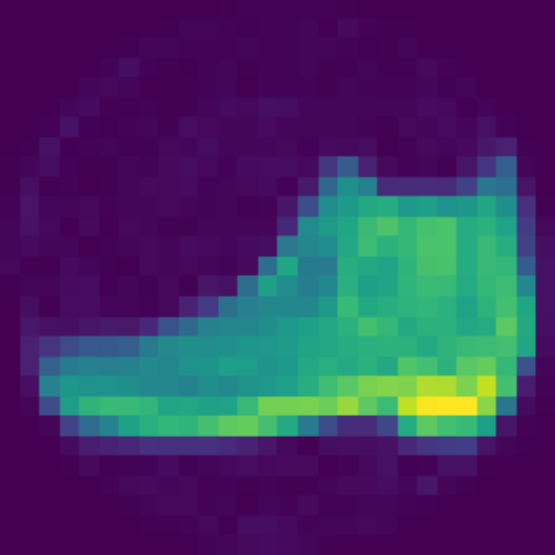}
  \end{subfigure}
  \begin{subfigure}{0.16\linewidth}
    \includegraphics[width=\linewidth]{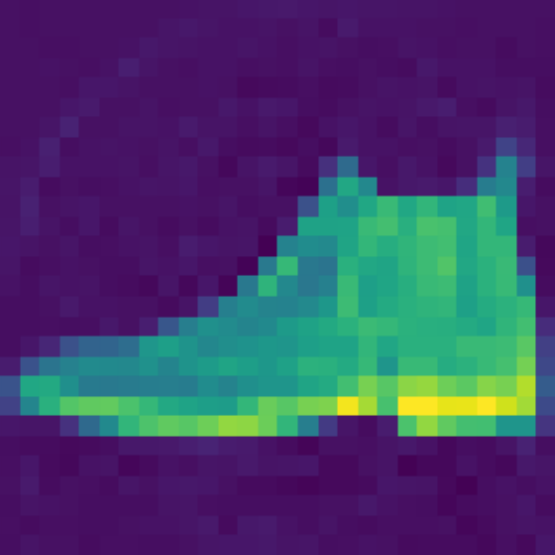}
  \end{subfigure}
  
  \begin{subfigure}{0.16\linewidth}
    \includegraphics[width=\linewidth]{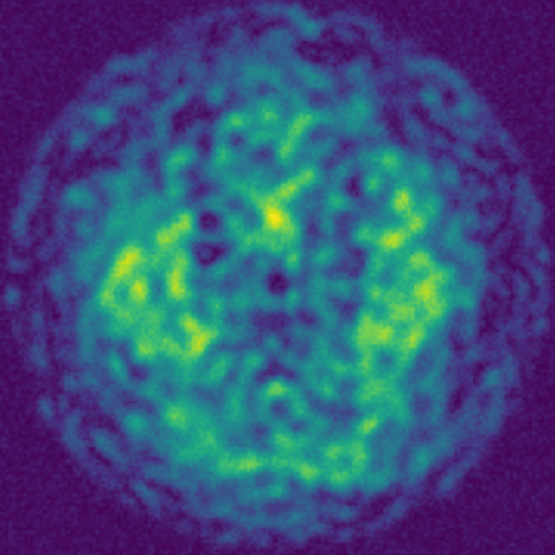}
  \end{subfigure}
  \begin{subfigure}{0.16\linewidth}
    \includegraphics[width=\linewidth]{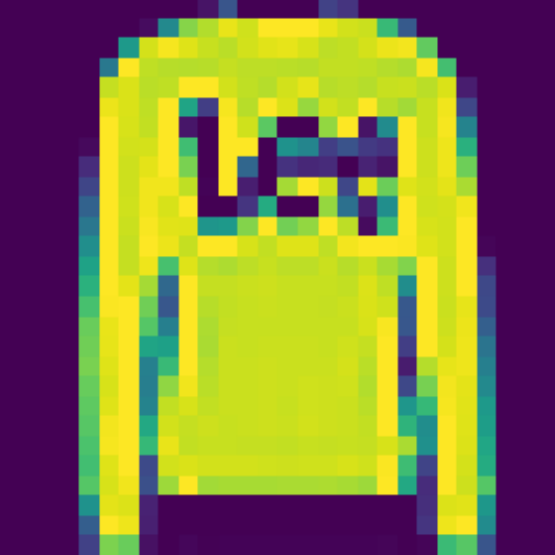}
  \end{subfigure}
  \begin{subfigure}{0.16\linewidth}
    \includegraphics[width=\linewidth]{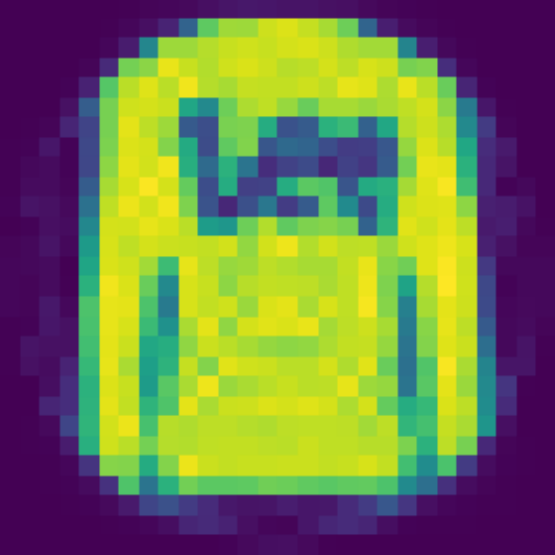}
  \end{subfigure}
  \begin{subfigure}{0.16\linewidth}
    \includegraphics[width=\linewidth]{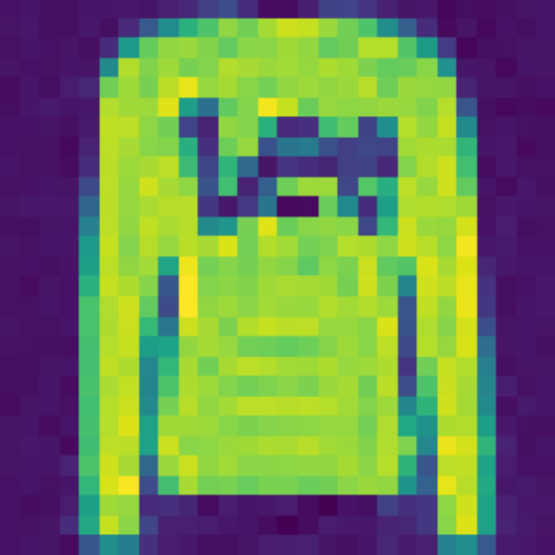}
  \end{subfigure}
  
  \begin{subfigure}{0.16\linewidth}
    \includegraphics[width=\linewidth]{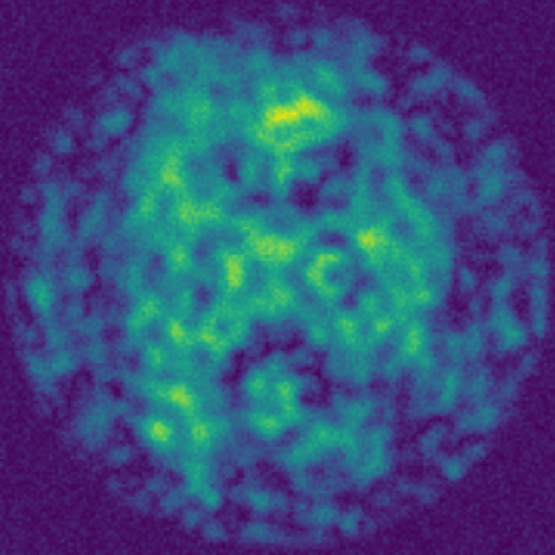}
    \centering
    (a) Input
  \end{subfigure}
  \begin{subfigure}{0.16\linewidth}
    \includegraphics[width=\linewidth]{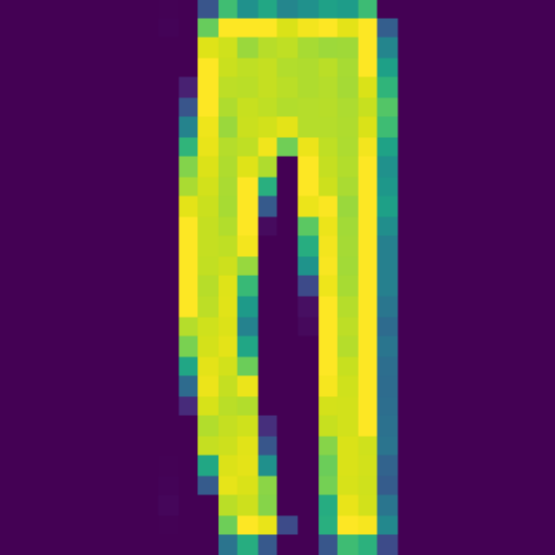}
    \centering
    (b) Target
  \end{subfigure}
  \begin{subfigure}{0.16\linewidth}
    \includegraphics[width=\linewidth]{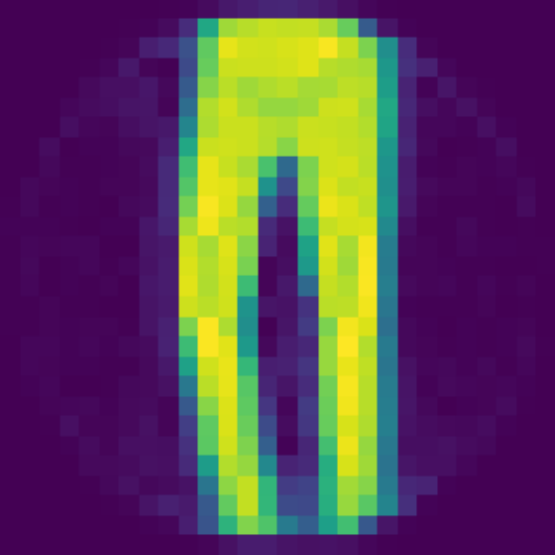}
    \centering
    (c) BEM
  \end{subfigure}
  \begin{subfigure}{0.16\linewidth}
    \includegraphics[width=\linewidth]{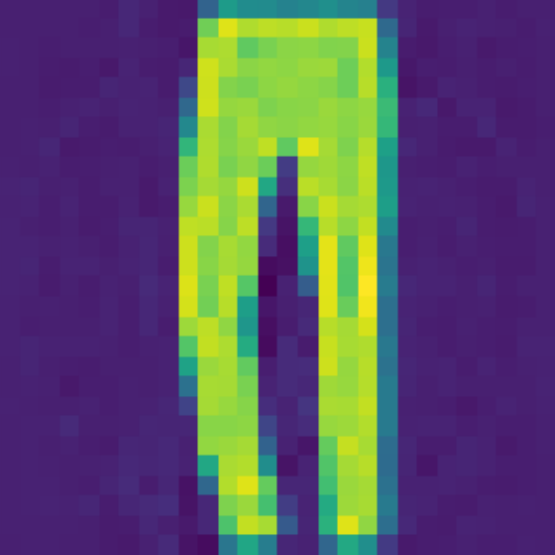}
    \centering
    (d) BEM + PP
  \end{subfigure}
  \caption{Comparison of predicted images from inverting transmission effects of a MMF using FMnist data created with a theoretical TM when the speckled images are saturated at $90\%$ of their maximum value and Gaussian noise with $0.01$ standard deviation added. (a) The input noisy speckled image, (b) the target original image to reconstruct, (c) the output of the Bessel equivariant model, and (d) the output of the combination of Bessel equivariant and post-processing model.}
  \label{fig:fmnist001noise}
\end{figure}

\begin{figure}[htb]
  \centering
  \begin{subfigure}{0.16\linewidth}
    \includegraphics[width=\linewidth]{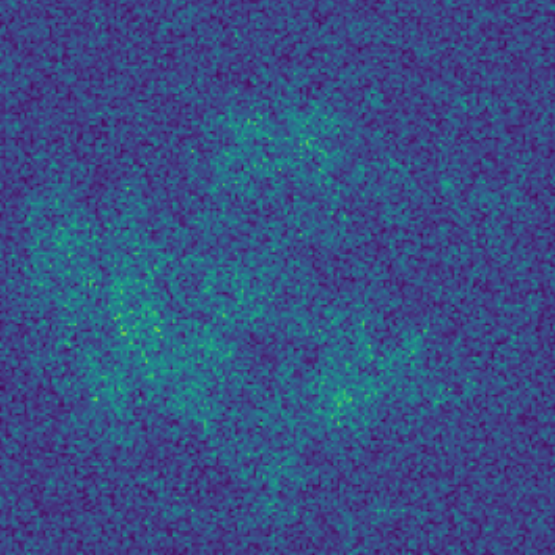}
  \end{subfigure}
  \begin{subfigure}{0.16\linewidth}
    \includegraphics[width=\linewidth]{results/fmnist/noise/original_0.pdf}
  \end{subfigure}
  \begin{subfigure}{0.16\linewidth}
    \includegraphics[width=\linewidth]{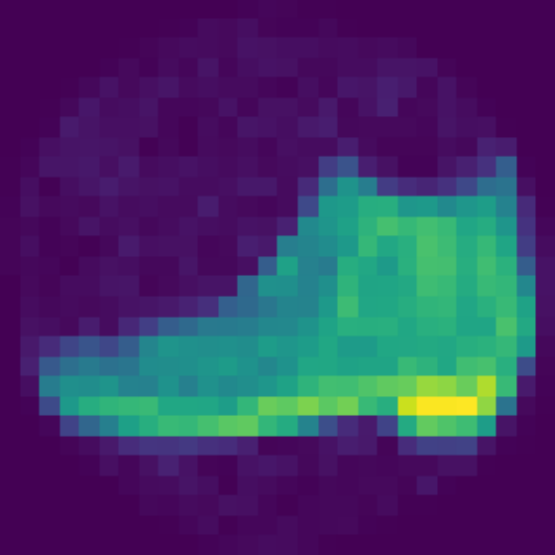}
  \end{subfigure}
  \begin{subfigure}{0.16\linewidth}
    \includegraphics[width=\linewidth]{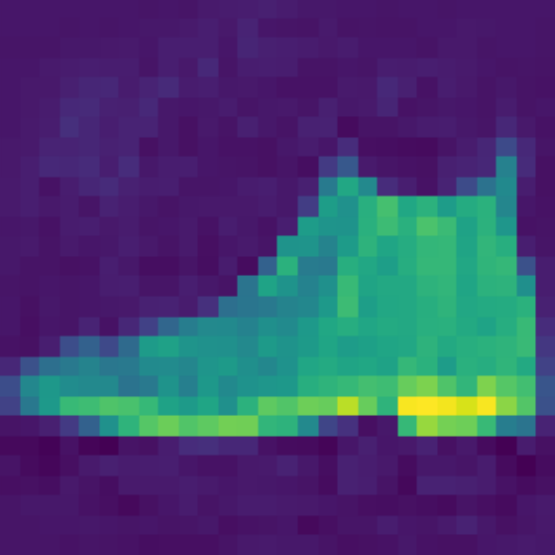}
  \end{subfigure}
  
  \begin{subfigure}{0.16\linewidth}
    \includegraphics[width=\linewidth]{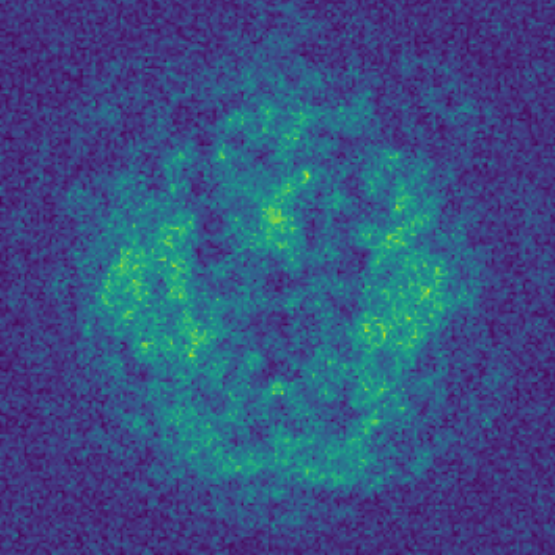}
  \end{subfigure}
  \begin{subfigure}{0.16\linewidth}
    \includegraphics[width=\linewidth]{results/fmnist/noise/original_1.pdf}
  \end{subfigure}
  \begin{subfigure}{0.16\linewidth}
    \includegraphics[width=\linewidth]{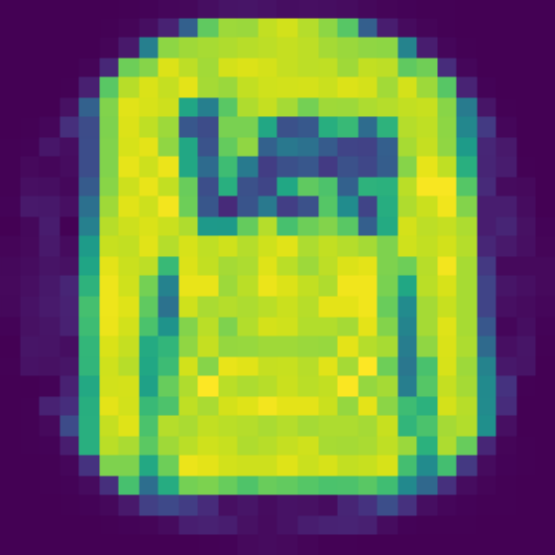}
  \end{subfigure}
  \begin{subfigure}{0.16\linewidth}
    \includegraphics[width=\linewidth]{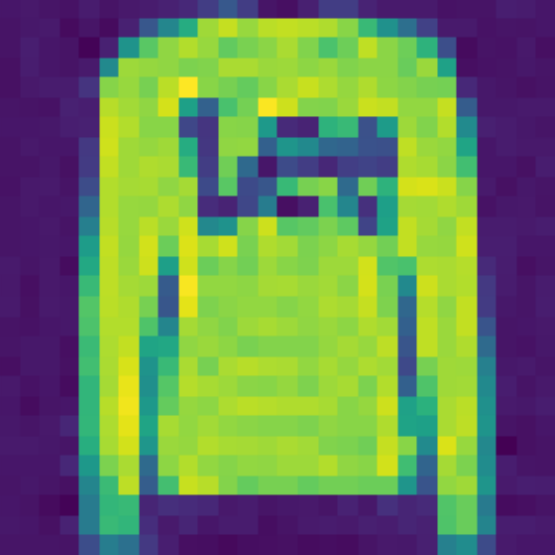}
  \end{subfigure}
  
  \begin{subfigure}{0.16\linewidth}
    \includegraphics[width=\linewidth]{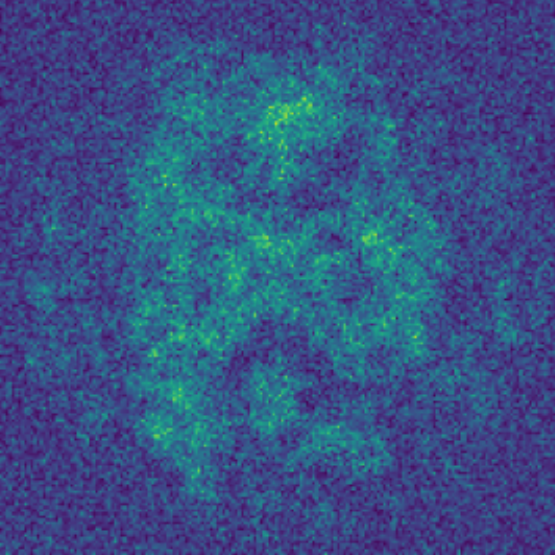}
    \centering
    (a) Input
  \end{subfigure}
  \begin{subfigure}{0.16\linewidth}
    \includegraphics[width=\linewidth]{results/fmnist/noise/original_2.pdf}
    \centering
    (b) Target
  \end{subfigure}
  \begin{subfigure}{0.16\linewidth}
    \includegraphics[width=\linewidth]{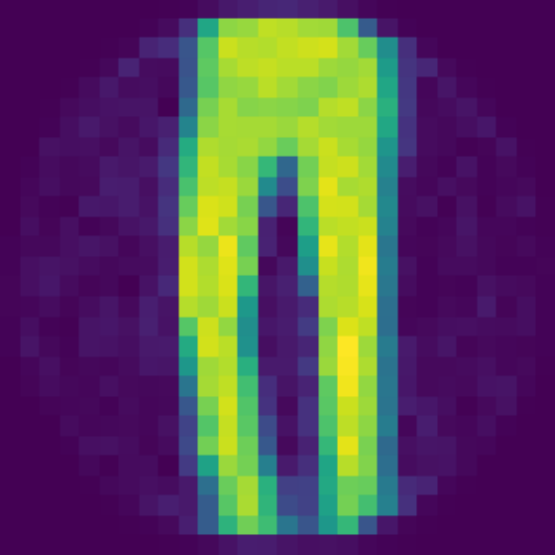}
    \centering
    (c) BEM
  \end{subfigure}
  \begin{subfigure}{0.16\linewidth}
    \includegraphics[width=\linewidth]{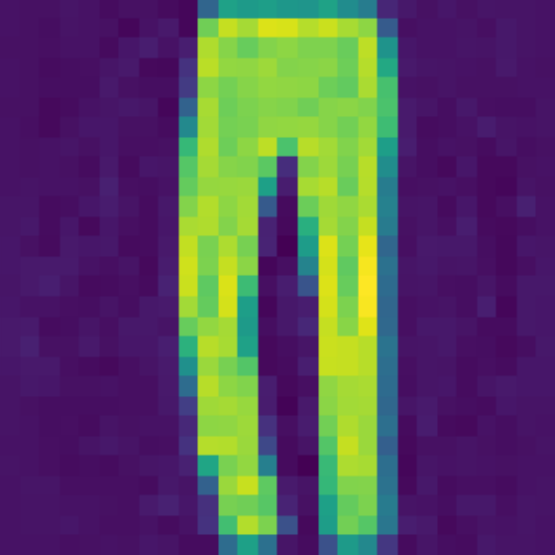}
    \centering
    (d)BEM + PP
  \end{subfigure}
  \caption{Comparison of predicted images from inverting transmission effects of a MMF using FMnist data created with a theoretical TM when the speckled images are saturated at $90\%$ of their maximum value and Gaussian noise with $0.05$ standard deviation added. (a) The input noisy speckled image, (b) the target original image to reconstruct, (c) the output of the Bessel equivariant model, and (d) the output of the combination of Bessel equivariant and post-processing model.}
  \label{fig:fmnist005noise}
\end{figure}

\begin{figure}[htb]
  \centering
  \begin{subfigure}{0.16\linewidth}
    \includegraphics[width=\linewidth]{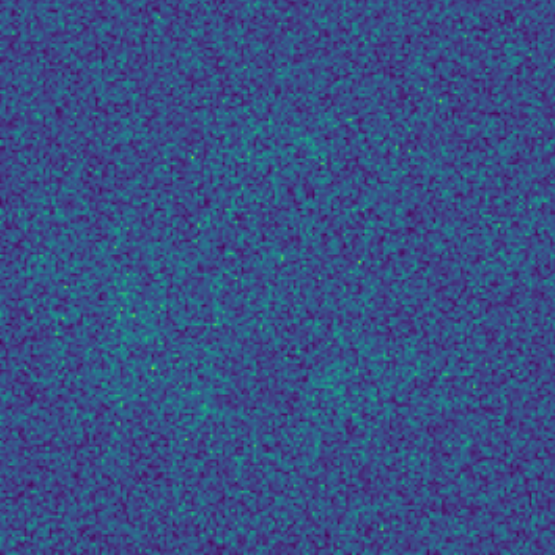}
  \end{subfigure}
  \begin{subfigure}{0.16\linewidth}
    \includegraphics[width=\linewidth]{results/fmnist/noise/original_0.pdf}
  \end{subfigure}
  \begin{subfigure}{0.16\linewidth}
    \includegraphics[width=\linewidth]{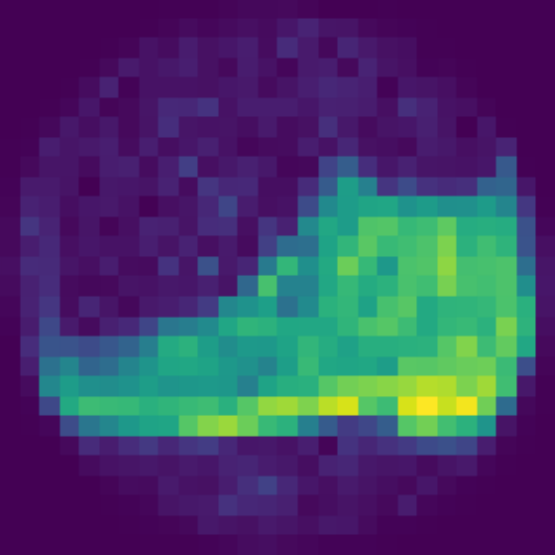}
  \end{subfigure}
  \begin{subfigure}{0.16\linewidth}
    \includegraphics[width=\linewidth]{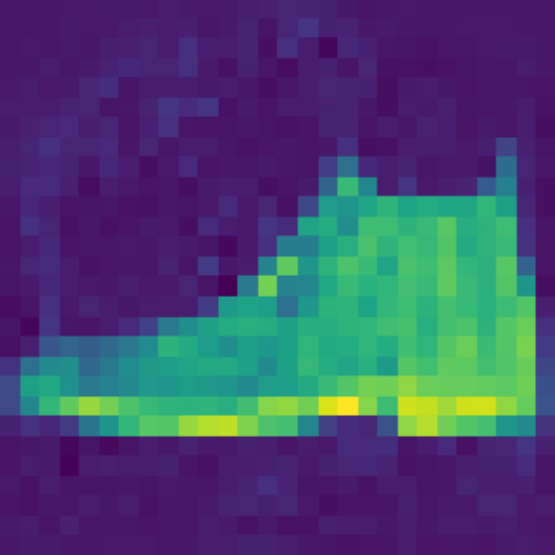}
  \end{subfigure}
  
  \begin{subfigure}{0.16\linewidth}
    \includegraphics[width=\linewidth]{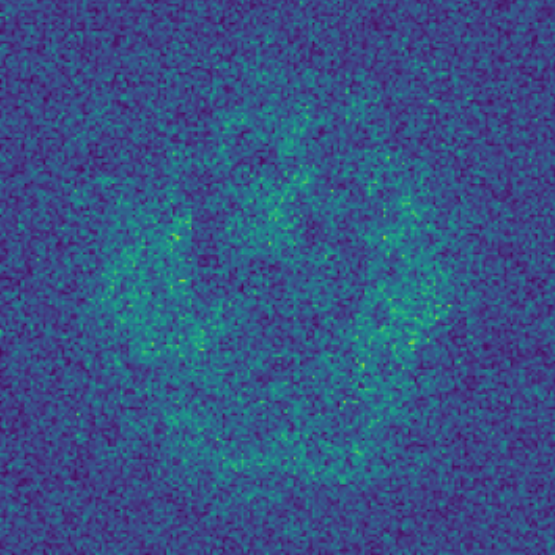}
  \end{subfigure}
  \begin{subfigure}{0.16\linewidth}
    \includegraphics[width=\linewidth]{results/fmnist/noise/original_1.pdf}
  \end{subfigure}
  \begin{subfigure}{0.16\linewidth}
    \includegraphics[width=\linewidth]{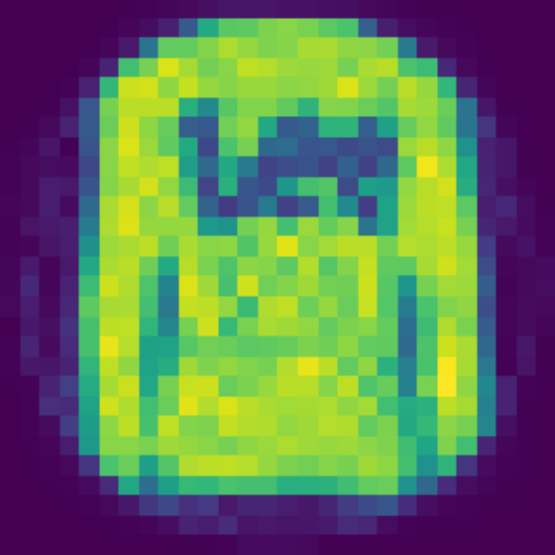}
  \end{subfigure}
  \begin{subfigure}{0.16\linewidth}
    \includegraphics[width=\linewidth]{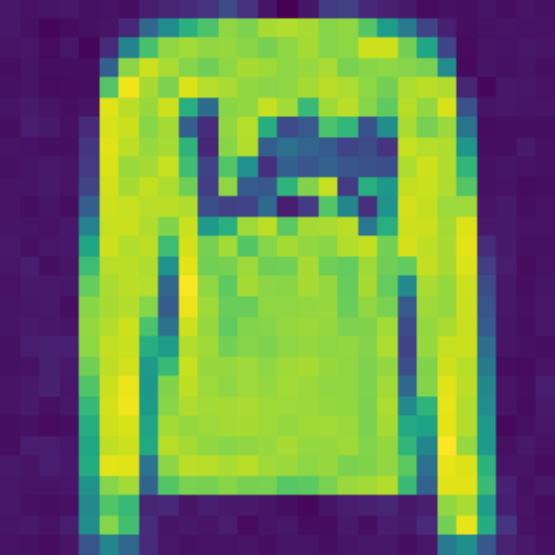}
  \end{subfigure}
  
  \begin{subfigure}{0.16\linewidth}
    \includegraphics[width=\linewidth]{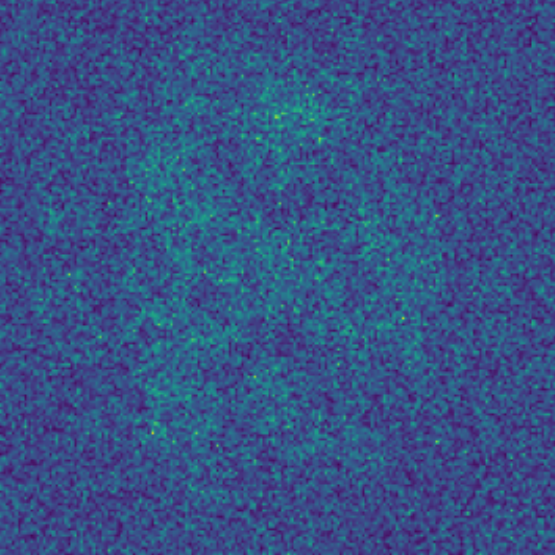}
    \centering
    (a) Input
  \end{subfigure}
  \begin{subfigure}{0.16\linewidth}
    \includegraphics[width=\linewidth]{results/fmnist/noise/original_2.pdf}
    \centering
    (b) Target
  \end{subfigure}
  \begin{subfigure}{0.16\linewidth}
    \includegraphics[width=\linewidth]{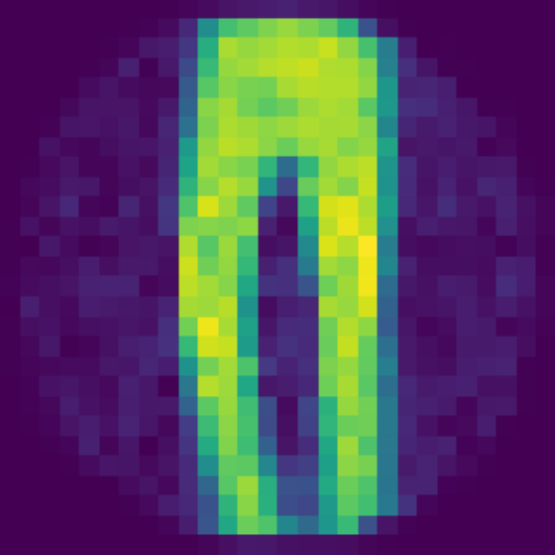}
    \centering
    (c) BEM
  \end{subfigure}
  \begin{subfigure}{0.16\linewidth}
    \includegraphics[width=\linewidth]{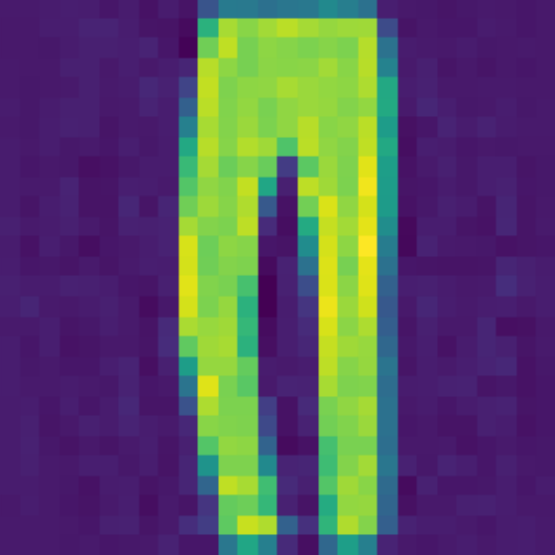}
    \centering
    (d) BEM + PP
  \end{subfigure}
  \caption{Comparison of predicted images from inverting transmission effects of a MMF using FMnist data created with a theoretical TM when the speckled images are saturated at $90\%$ of their maximum value and Gaussian noise with $0.1$ standard deviation added. (a) The input noisy speckled image, (b) the target original image to reconstruct, (c) the output of the Bessel equivariant model, and (d) the output of the combination of Bessel equivariant and post-processing model.}
  \label{fig:fmnist01noise}
\end{figure}

\begin{figure}[htb]
  \centering
  \begin{subfigure}{0.16\linewidth}
    \includegraphics[width=\linewidth]{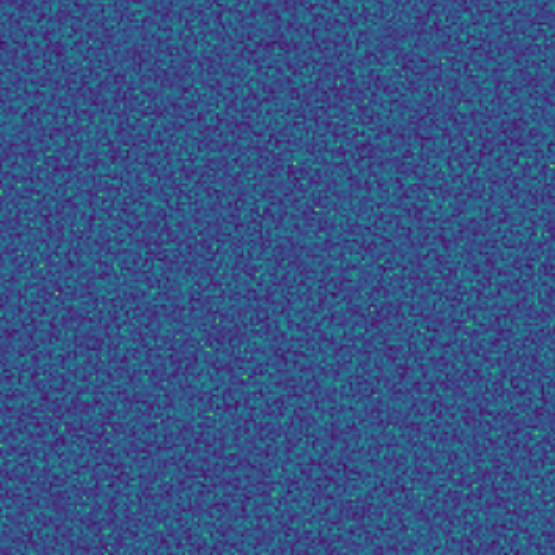}
  \end{subfigure}
  \begin{subfigure}{0.16\linewidth}
    \includegraphics[width=\linewidth]{results/fmnist/noise/original_0.pdf}
  \end{subfigure}
  \begin{subfigure}{0.16\linewidth}
    \includegraphics[width=\linewidth]{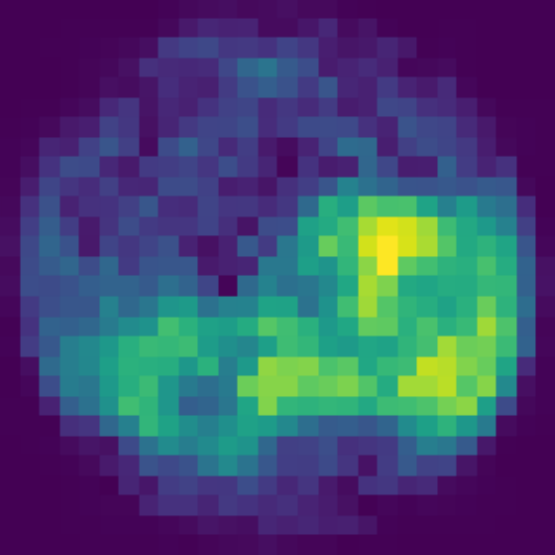}
  \end{subfigure}
  \begin{subfigure}{0.16\linewidth}
    \includegraphics[width=\linewidth]{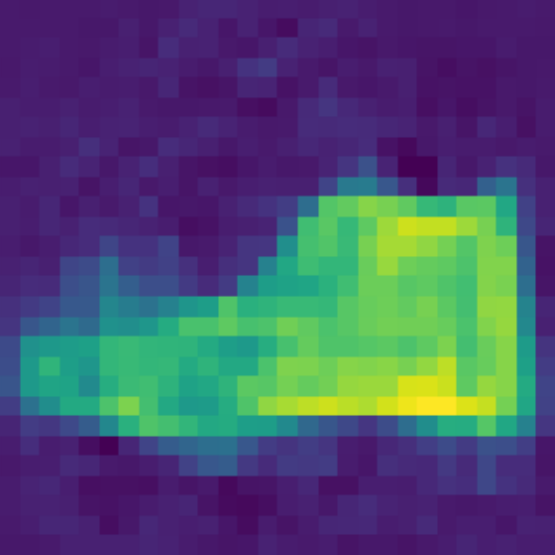}
  \end{subfigure}
  
  \begin{subfigure}{0.16\linewidth}
    \includegraphics[width=\linewidth]{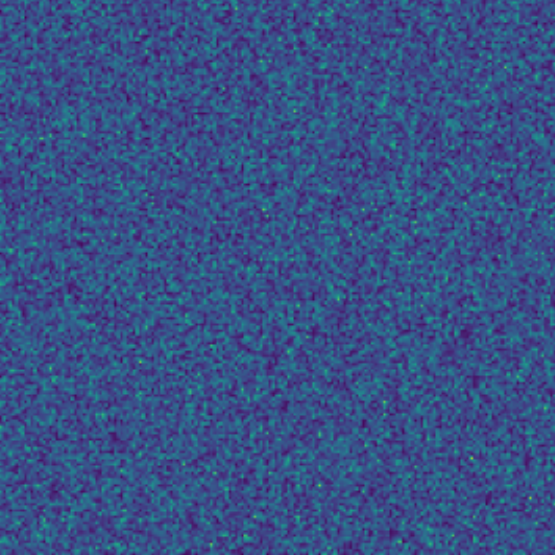}
  \end{subfigure}
  \begin{subfigure}{0.16\linewidth}
    \includegraphics[width=\linewidth]{results/fmnist/noise/original_1.pdf}
  \end{subfigure}
  \begin{subfigure}{0.16\linewidth}
    \includegraphics[width=\linewidth]{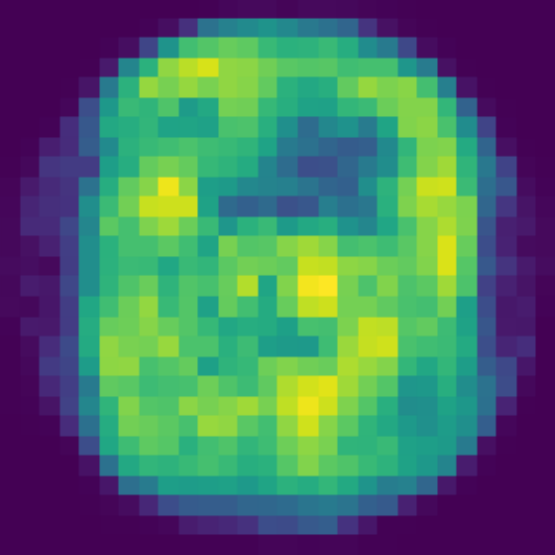}
  \end{subfigure}
  \begin{subfigure}{0.16\linewidth}
    \includegraphics[width=\linewidth]{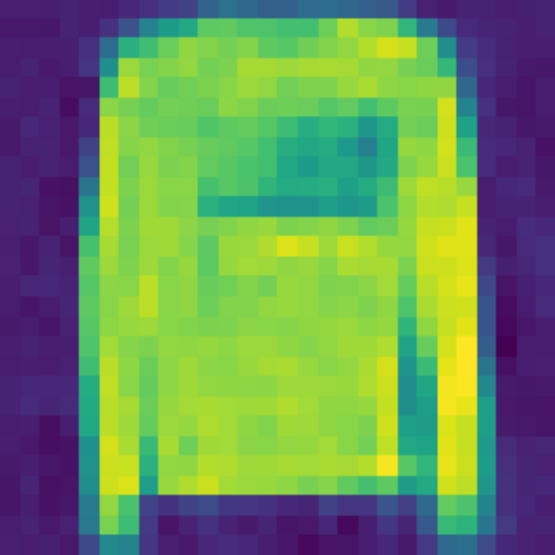}
  \end{subfigure}
  
  \begin{subfigure}{0.16\linewidth}
    \includegraphics[width=\linewidth]{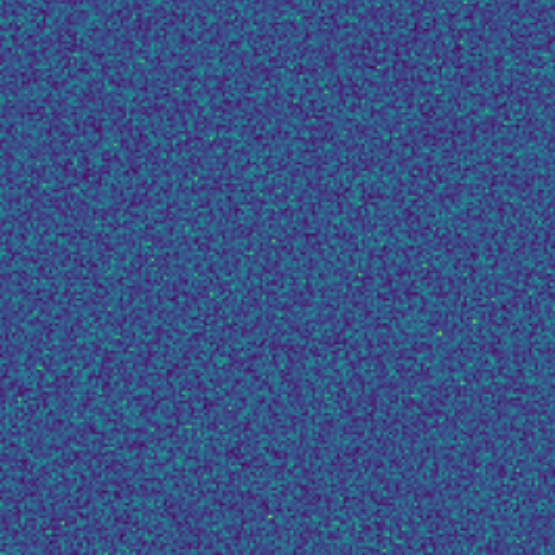}
    \centering
    (a) Input
  \end{subfigure}
  \begin{subfigure}{0.16\linewidth}
    \includegraphics[width=\linewidth]{results/fmnist/noise/original_2.pdf}
    \centering
    (b) Target
  \end{subfigure}
  \begin{subfigure}{0.16\linewidth}
    \includegraphics[width=\linewidth]{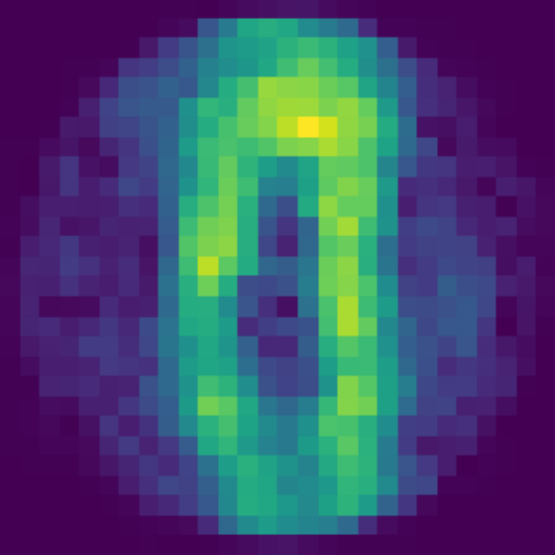}
    \centering
    (c) BEM
  \end{subfigure}
  \begin{subfigure}{0.16\linewidth}
    \includegraphics[width=\linewidth]{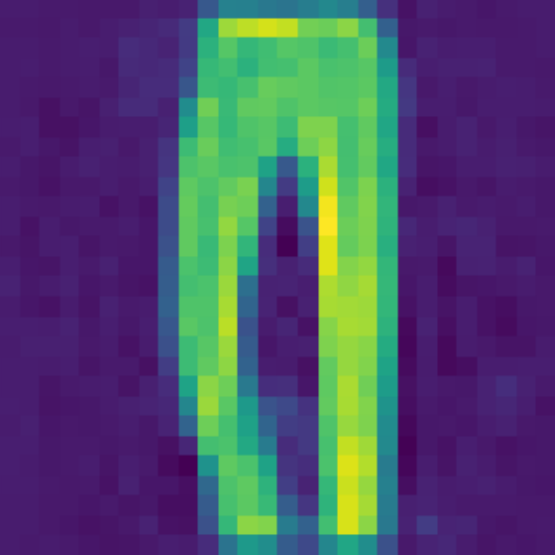}
    \centering
    (d) BEM + PP
  \end{subfigure}
  \caption{Comparison of predicted images from inverting transmission effects of a MMF using FMnist data created with a theoretical TM when the speckled images are saturated at $90\%$ of their maximum value and Gaussian noise with $0.05$ standard deviation added. (a) The input noisy speckled image, (b) the target original image to reconstruct, (c) the output of the Bessel equivariant model, and (d) the output of the combination of Bessel equivariant and post-processing model.}
  \label{fig:fmnist05noise}
\end{figure}

Here we compare the effect of noise on the speckled images when using a theoretical TM. In each case the speckled images are saturated at $90\%$ of there maximum value and Gaussian noise is added with different standard deviation values. We demonstrate the reconstruction ability of our model in Figure~\ref{fig:fmnist001noise} when Gaussian noise with $0.01$ standard deviation, in Figure~\ref{fig:fmnist005noise} when Gaussian noise with $0.05$ standard deviation, in Figure~\ref{fig:fmnist01noise} when Gaussian noise with $0.1$ standard deviation, and in Figure~\ref{fig:fmnist05noise} when Gaussian noise with $0.5$ standard deviation. Further, we provide the loss values in Table~\ref{tab:fmnistlossnoise}. This demonstrates that our approach is robust to noise.

\begin{table}[htb]
  \caption{Comparison of the loss values of each model trained with F-MNIST data when the speckled images are saturated at $90\%$ and Gaussian noise is added with standard deviation given in the column Noise Level.}
  \label{tab:fmnistlossnoise}
  \centering
  \begin{tabular}{clccc}
    \toprule
    Noise Level & Model & Train Loss & Test Loss  \\
    \midrule
    \multirow{2}{*}{0.01} & Bessel Equivariant             & 0.0140 & 0.0142 \\
                          & Bessel Equivariant + Post Proc & 0.0033 & 0.0033 \\
    \multirow{2}{*}{0.05} & Bessel Equivariant             & 0.0147 & 0.0149 \\
                          & Bessel Equivariant + Post Proc & 0.0043 & 0.0043 \\
    \multirow{2}{*}{0.1}  & Bessel Equivariant             & 0.0166 & 0.0168 \\
                          & Bessel Equivariant + Post Proc & 0.0049 & 0.0049 \\
    \multirow{2}{*}{0.5 } & Bessel Equivariant             & 0.0373 & 0.0377 \\
                          & Bessel Equivariant + Post Proc & 0.0164 & 0.0166 \\
    \bottomrule
  \end{tabular}
\end{table}

\FloatBarrier

\newpage
\subsection{Additional Reconstructions - Real Fibre MNIST}

\begin{table}[htb]
  \caption{Comparison of the loss values of each model trained with MNIST data.}
  \label{tab:reallossmnistapp}
  \centering
  \begin{tabular}{lcc}
    \toprule
     & \multicolumn{2}{c}{MNIST} \\
    Model & Train Loss & Test Loss \\
    \midrule
    Real Linear & 0.00466 & 0.00683 \\
    Complex Linear & 0.00396 & 0.00684 \\
    Bessel Equivariant Diag & 0.03004 & 0.03125 \\
    Bessel Equivariant Diag + Post Proc & 0.01317 & 0.01453 \\
    Bessel Equivariant 5 Off Diag & 0.01782 & 0.01931 \\
    Bessel Equivariant 5 Off Diag + Post Proc & 0.00658 & 0.00740 \\
    Bessel Equivariant 10 Off Diag & 0.01362 & 0.01521 \\
    Bessel Equivariant 10 Off Diag + Post Proc & 0.00488 & 0.00576 \\
    Bessel Equivariant Full & 0.00300 & 0.00684 \\
    Bessel Equivariant Full + Post Proc & \textbf{0.00145} & \textbf{0.00378} \\
    \bottomrule
  \end{tabular}
\end{table}

\begin{figure}[htb]
  \centering
  \foreach \n in {3,...,6}
  {
  \begin{subfigure}{0.16\linewidth}
    \includegraphics[width=\linewidth]{results/real/figs1/speckled_\n.pdf}
  \end{subfigure}
  \begin{subfigure}{0.16\linewidth}
    \includegraphics[width=\linewidth]{results/real/figs1/original_\n.pdf}
  \end{subfigure}
  \begin{subfigure}{0.16\linewidth}
    \includegraphics[width=\linewidth]{results/real/figs1/TM_bases_\n.pdf}
  \end{subfigure}
  \begin{subfigure}{0.16\linewidth}
    \includegraphics[width=\linewidth]{results/real/figs1/TM_bases_SR_\n.pdf}
  \end{subfigure}
  \begin{subfigure}{0.16\linewidth}
    \includegraphics[width=\linewidth]{results/real/figs1/MLP_\n.pdf}
  \end{subfigure}
  \begin{subfigure}{0.16\linewidth}
    \includegraphics[width=\linewidth]{results/real/figs1/CMLP_\n.pdf}
  \end{subfigure}
  }

  \begin{subfigure}{0.16\linewidth}
    \includegraphics[width=\linewidth]{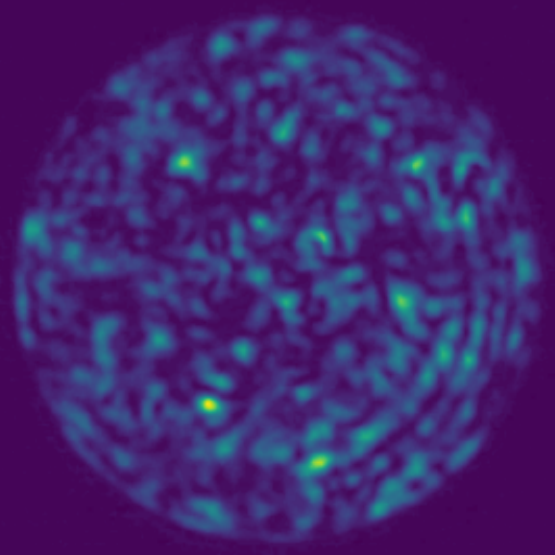}
    \centering
    (a) Input
  \end{subfigure}
  \begin{subfigure}{0.16\linewidth}
    \includegraphics[width=\linewidth]{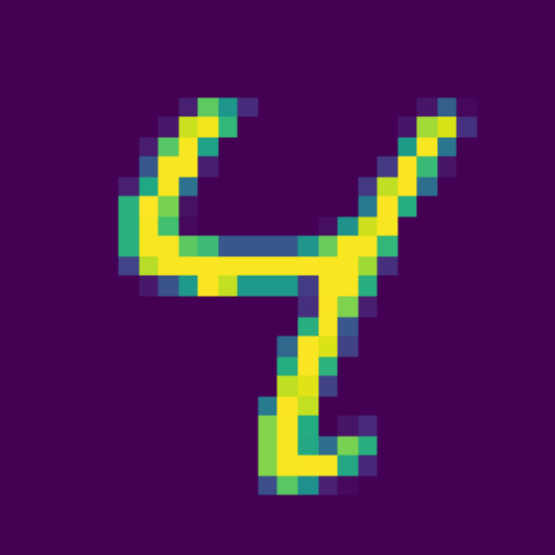}
    \centering
    (b) Target
  \end{subfigure}
  \begin{subfigure}{0.16\linewidth}
    \includegraphics[width=\linewidth]{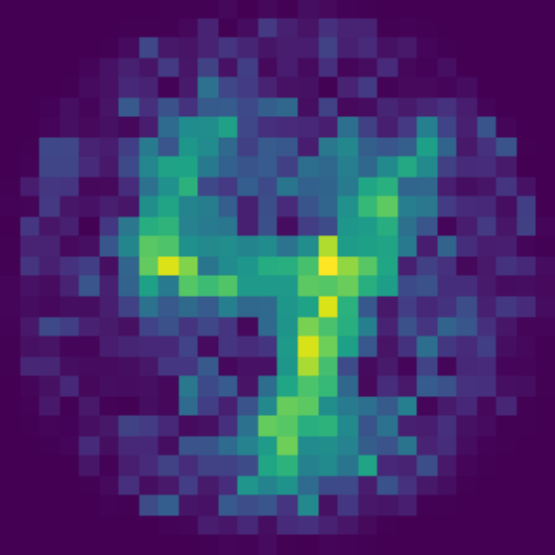}
    \centering
    (c) BEM
  \end{subfigure}
  \begin{subfigure}{0.16\linewidth}
    \includegraphics[width=\linewidth]{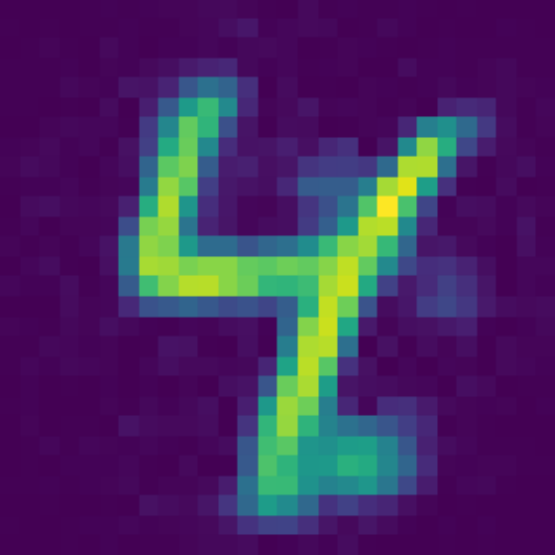}
    \centering
    (d) BEM + PP
  \end{subfigure}
  \begin{subfigure}{0.16\linewidth}
    \includegraphics[width=\linewidth]{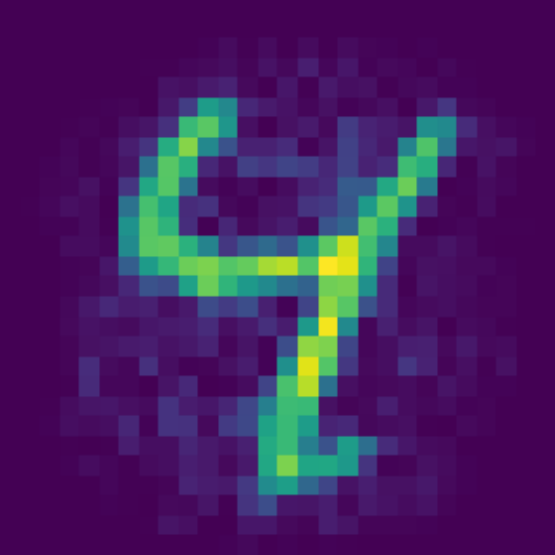}
    \centering
    (e) Real
  \end{subfigure}
  \begin{subfigure}{0.16\linewidth}
    \includegraphics[width=\linewidth]{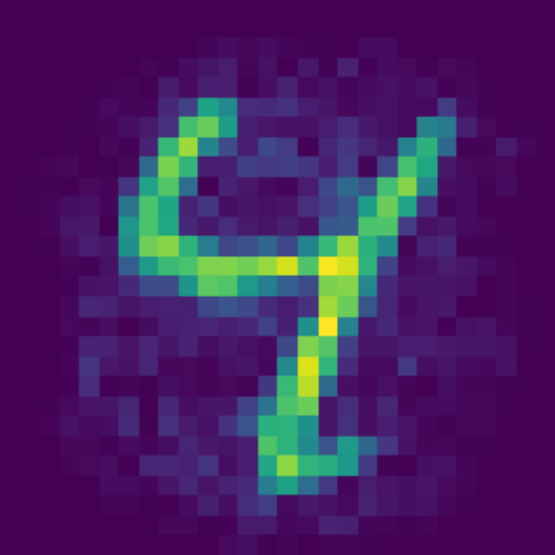}
    \centering
    (f) Complex
  \end{subfigure}
  \caption{Comparison of predicted images from inverting transmission effects of a MMF. (a) The input speckled image, (b) the target original image to reconstruct, (c) the output of the Bessel equivariant model, (d) the output of the combination of Bessel equivariant and post-processing model, (e) the output of the Real valued linear model, and (f) the output of the Complex valued linear model.}
  \label{fig:realmnistapp}
\end{figure}

\newpage
\subsection{Additional Reconstructions - Real Fibre fMNIST}

\begin{table}[htb]
  \caption{Comparison of the loss values of each model trained with fMNIST data.}
  \label{tab:reallossfmnistapp}
  \centering
  \begin{tabular}{lcc}
    \toprule
    & \multicolumn{2}{c}{fMNIST}  \\
    Model & Train Loss & Test Loss  \\
    \midrule
    Real Linear & 0.00586 & 0.01061 \\
    Complex Linear & 0.00509 & 0.01061 \\
    Bessel Equivariant Diag & 0.02903 & 0.03140 \\
    Bessel Equivariant Diag + Post Proc & 0.01609 & 0.01749 \\
    Bessel Equivariant 5 Off Diag & 0.02057 & 0.02354 \\
    Bessel Equivariant 5 Off Diag + Post Proc & 0.01138 & 0.01315 \\
    Bessel Equivariant 10 Off Diag & 0.01788 & 0.02140 \\
    Bessel Equivariant 10 Off Diag + Post Proc & 0.00943 & 0.01162 \\
    Bessel Equivariant Full & 0.00548 & 0.01380 \\
    Bessel Equivariant Full + Post Proc & \textbf{0.00306} & \textbf{0.00964} \\
    \bottomrule
  \end{tabular}
\end{table}

\begin{figure}[htb]
  \centering
  \foreach \n in {3,...,6}
  {
  \begin{subfigure}{0.16\linewidth}
    \includegraphics[width=\linewidth]{results/real/fmnist/speckled_\n.pdf}
  \end{subfigure}
  \begin{subfigure}{0.16\linewidth}
    \includegraphics[width=\linewidth]{results/real/fmnist/original_\n.pdf}
  \end{subfigure}
  \begin{subfigure}{0.16\linewidth}
    \includegraphics[width=\linewidth]{results/real/fmnist/TM_bases_\n.pdf}
  \end{subfigure}
  \begin{subfigure}{0.16\linewidth}
    \includegraphics[width=\linewidth]{results/real/fmnist/TM_bases_SR_\n.pdf}
  \end{subfigure}
  \begin{subfigure}{0.16\linewidth}
    \includegraphics[width=\linewidth]{results/real/fmnist/MLP_\n.pdf}
  \end{subfigure}
  \begin{subfigure}{0.16\linewidth}
    \includegraphics[width=\linewidth]{results/real/fmnist/CMLP_\n.pdf}
  \end{subfigure}
  }

  \begin{subfigure}{0.16\linewidth}
    \includegraphics[width=\linewidth]{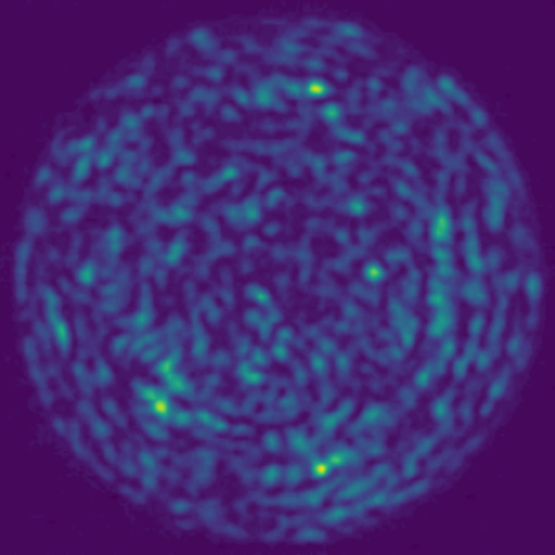}
    \centering
    (a) Input
  \end{subfigure}
  \begin{subfigure}{0.16\linewidth}
    \includegraphics[width=\linewidth]{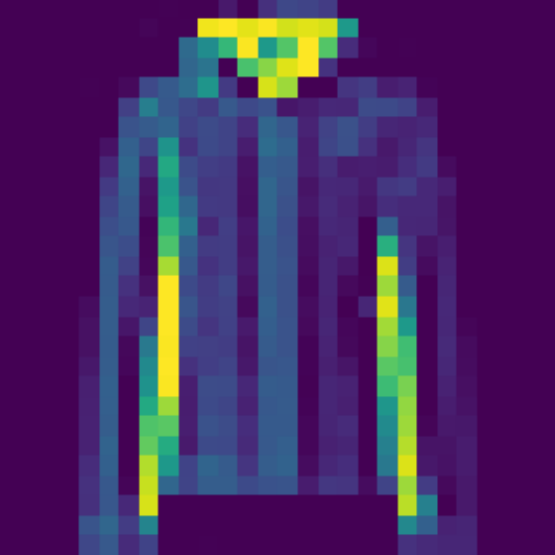}
    \centering
    (b) Target
  \end{subfigure}
  \begin{subfigure}{0.16\linewidth}
    \includegraphics[width=\linewidth]{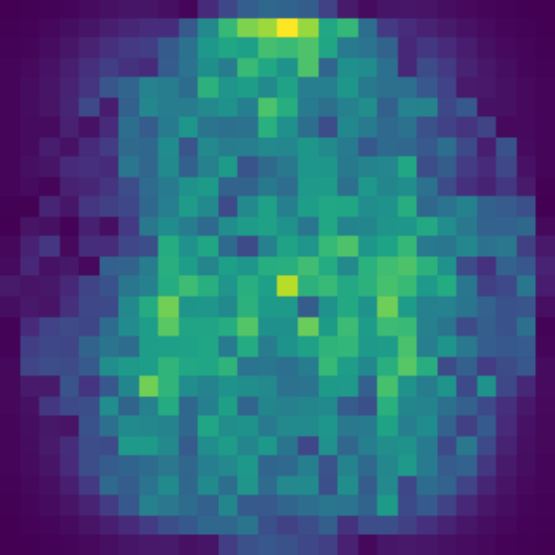}
    \centering
    (c) BEM
  \end{subfigure}
  \begin{subfigure}{0.16\linewidth}
    \includegraphics[width=\linewidth]{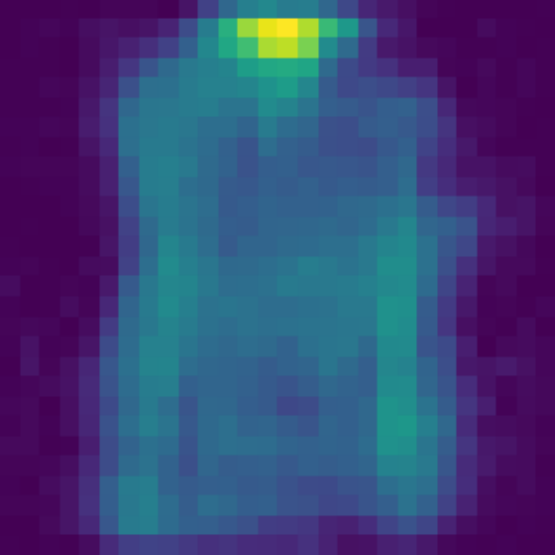}
    \centering
    (d) BEM + PP
  \end{subfigure}
  \begin{subfigure}{0.16\linewidth}
    \includegraphics[width=\linewidth]{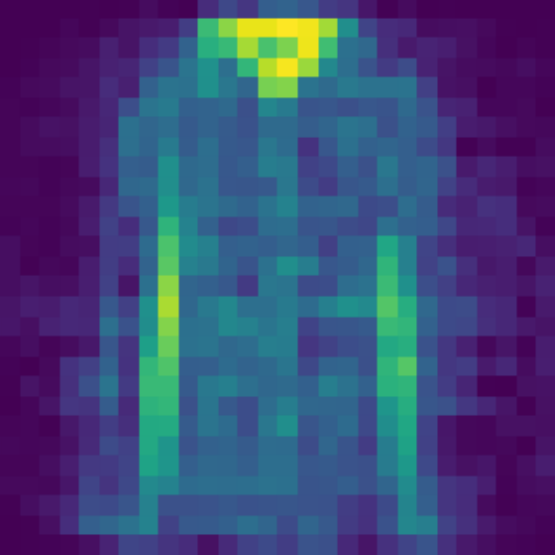}
    \centering
    (e) Real
  \end{subfigure}
  \begin{subfigure}{0.16\linewidth}
    \includegraphics[width=\linewidth]{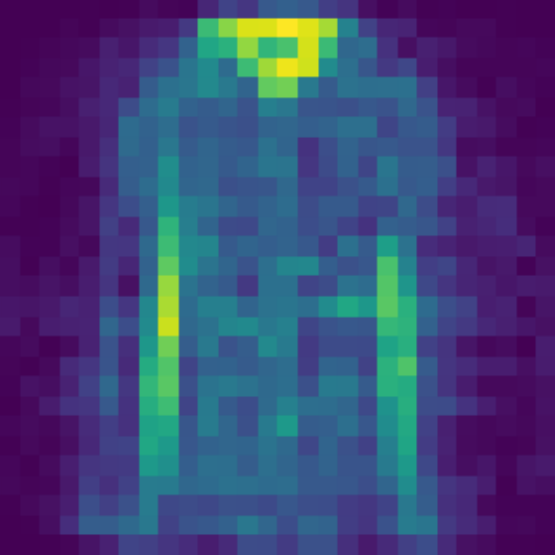}
    \centering
    (f) Complex
  \end{subfigure}
  \caption{Comparison of predicted images from inverting transmission effects of a MMF. (a) The input speckled image, (b) the target original image to reconstruct, (c) the output of the Bessel equivariant model, (d) the output of the combination of Bessel equivariant and post-processing model, (e) the output of the Real valued linear model, and (f) the output of the Complex valued linear model.}
  \label{fig:realfmnistapp}
\end{figure}

\newpage
\subsection{Additional Reconstructions - Theory TM fMNIST}

\begin{table}[htb]
  \caption{Comparison of the loss values of each model trained with fMNIST data.}
  \label{tab:fmnistlossfmnisttestapp}
  \centering
  \begin{tabular}{lcc}
    \toprule
          & \multicolumn{2}{c}{fMNIST} \\
    Model & Train Loss & Test Loss \\
    \midrule
    Real Linear                    & 0.0256 & 0.0250 \\
    Complex Linear                 & 0.0149 & 0.0146 \\
    Bessel Equivariant             & 0.0141 & 0.0139  \\
    Bessel Equivariant + Post Proc & \textbf{0.0032} & \textbf{0.0032} \\
    \bottomrule
  \end{tabular}
\end{table}

\begin{figure}[htb]
  \centering
  \foreach \n in {3,...,7}
  {
  \begin{subfigure}{0.16\linewidth}
    \includegraphics[width=\linewidth]{results/fmnist/figs1/speckled_\n.pdf}
  \end{subfigure}
  \begin{subfigure}{0.16\linewidth}
    \includegraphics[width=\linewidth]{results/fmnist/figs1/original_\n.pdf}
  \end{subfigure}
  \begin{subfigure}{0.16\linewidth}
    \includegraphics[width=\linewidth]{results/fmnist/figs1/TM_bases_\n.pdf}
  \end{subfigure}
  \begin{subfigure}{0.16\linewidth}
    \includegraphics[width=\linewidth]{results/fmnist/figs1/TM_bases_SR_\n.pdf}
  \end{subfigure}
  \begin{subfigure}{0.16\linewidth}
    \includegraphics[width=\linewidth]{results/fmnist/figs1/MLP_\n.pdf}
  \end{subfigure}
  \begin{subfigure}{0.16\linewidth}
    \includegraphics[width=\linewidth]{results/fmnist/figs1/CMLP_\n.pdf}
  \end{subfigure}
  }

  \begin{subfigure}{0.16\linewidth}
    \includegraphics[width=\linewidth]{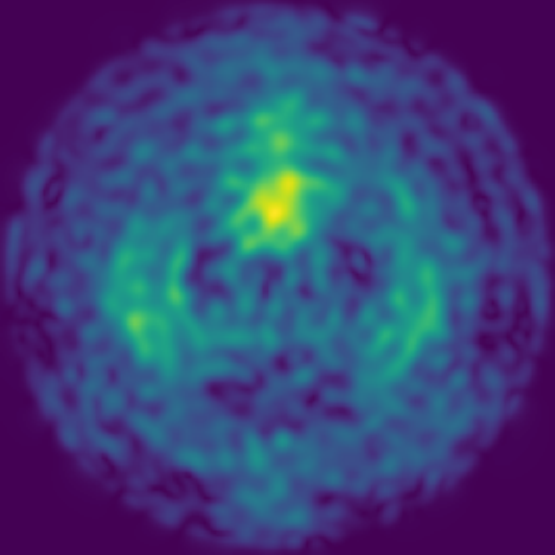}
    \centering
    (a) Input
  \end{subfigure}
  \begin{subfigure}{0.16\linewidth}
    \includegraphics[width=\linewidth]{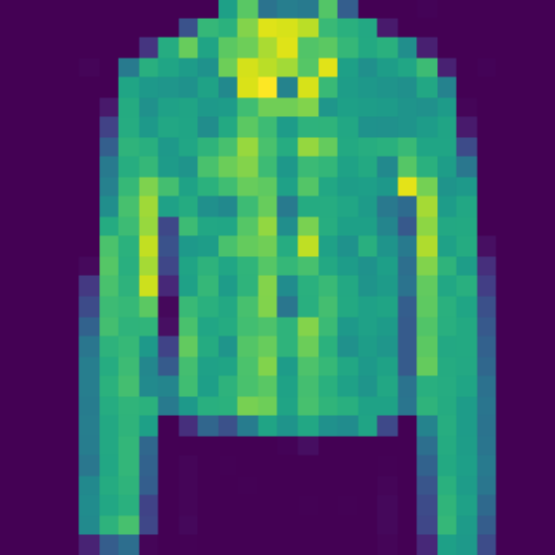}
    \centering
    (b) Target
  \end{subfigure}
  \begin{subfigure}{0.16\linewidth}
    \includegraphics[width=\linewidth]{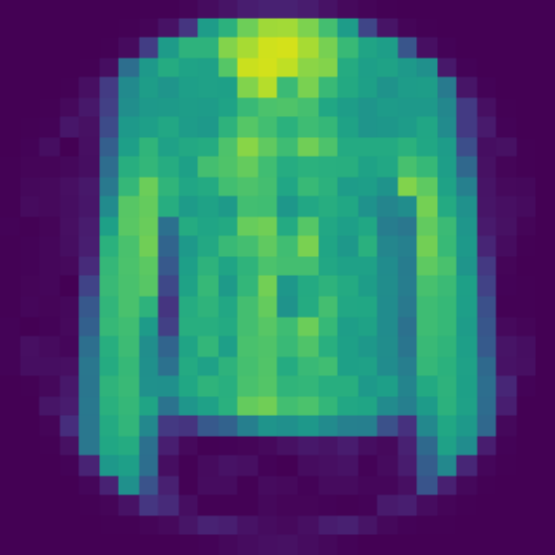}
    \centering
    (c) BEM
  \end{subfigure}
  \begin{subfigure}{0.16\linewidth}
    \includegraphics[width=\linewidth]{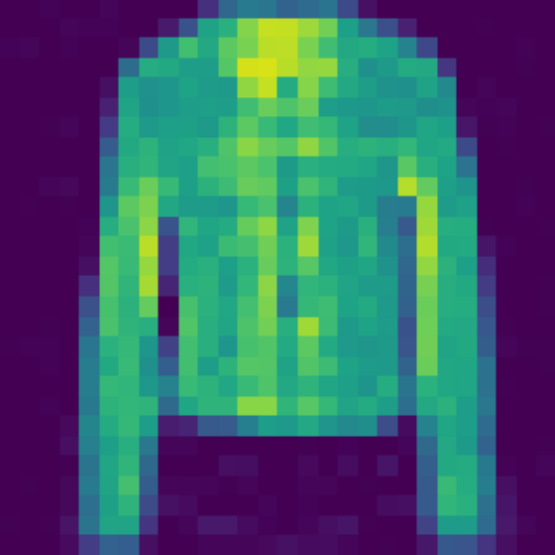}
    \centering
    (d) BEM + PP
  \end{subfigure}
  \begin{subfigure}{0.16\linewidth}
    \includegraphics[width=\linewidth]{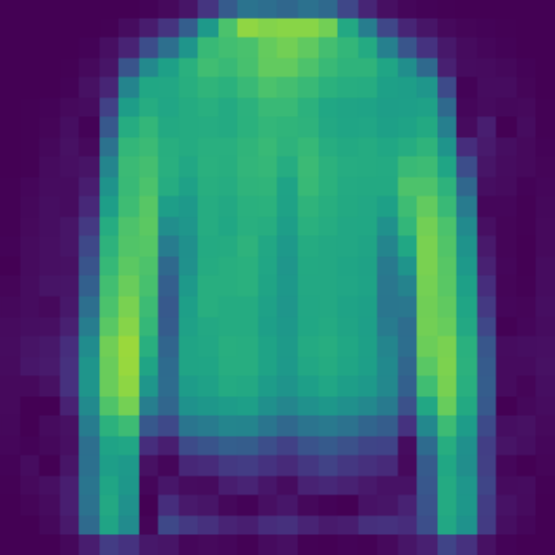}
    \centering
    (e) Real
  \end{subfigure}
  \begin{subfigure}{0.16\linewidth}
    \includegraphics[width=\linewidth]{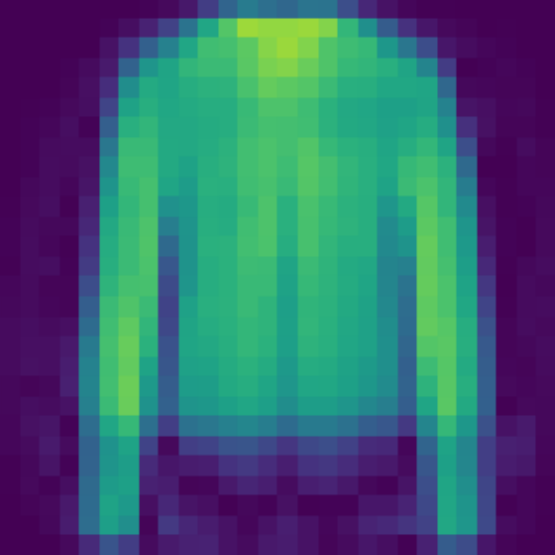}
    \centering
    (f) Complex
  \end{subfigure}
  \caption{Comparison of predicted images from inverting transmission effects of a MMF. (a) The input speckled image, (b) the target original image to reconstruct, (c) the output of the Bessel equivariant model, (d) the output of the combination of Bessel equivariant and post-processing model, (e) the output of the Real valued linear model, and (f) the output of the Complex valued linear model.}
  \label{fig:theoryfmnistapp}
\end{figure}

\newpage
\subsection{Additional Reconstructions - Theory TM MNIST}

\begin{table}[htb]
  \caption{Comparison of the loss values of each model trained with fMNIST data.}
  \label{tab:fmnistlossmnisttestapp}
  \centering
  \begin{tabular}{lc}
    \toprule
          & MNIST \\
    Model & Test Loss  \\
    \midrule
    Real Linear                    & 0.0641 \\
    Complex Linear                 & 0.0363 \\
    Bessel Equivariant             & \textbf{0.0026} \\
    Bessel Equivariant + Post Proc & 0.0028 \\
    \bottomrule
  \end{tabular}
\end{table}

\begin{figure}[htb]
  \centering
  \foreach \n in {3,...,7}
  {
  \begin{subfigure}{0.16\linewidth}
    \includegraphics[width=\linewidth]{results/fmnist/mnist_test/speckled_\n.pdf}
  \end{subfigure}
  \begin{subfigure}{0.16\linewidth}
    \includegraphics[width=\linewidth]{results/fmnist/mnist_test/original_\n.pdf}
  \end{subfigure}
  \begin{subfigure}{0.16\linewidth}
    \includegraphics[width=\linewidth]{results/fmnist/mnist_test/TM_bases_\n.pdf}
  \end{subfigure}
  \begin{subfigure}{0.16\linewidth}
    \includegraphics[width=\linewidth]{results/fmnist/mnist_test/TM_bases_SR_\n.pdf}
  \end{subfigure}
  \begin{subfigure}{0.16\linewidth}
    \includegraphics[width=\linewidth]{results/fmnist/mnist_test/MLP_\n.pdf}
  \end{subfigure}
  \begin{subfigure}{0.16\linewidth}
    \includegraphics[width=\linewidth]{results/fmnist/mnist_test/CMLP_\n.pdf}
  \end{subfigure}
  }

  \begin{subfigure}{0.16\linewidth}
    \includegraphics[width=\linewidth]{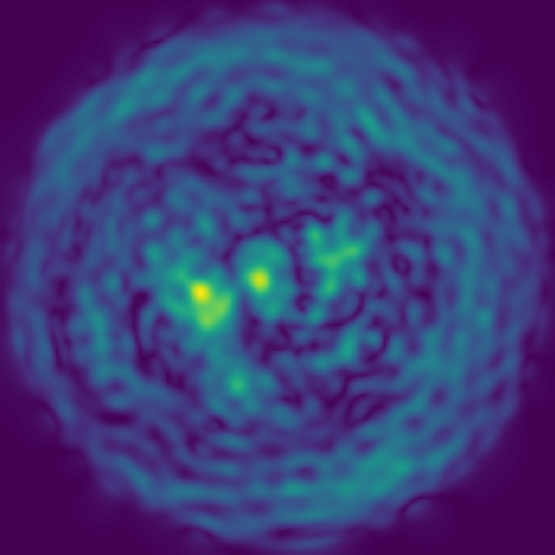}
    \centering
    (a) Input
  \end{subfigure}
  \begin{subfigure}{0.16\linewidth}
    \includegraphics[width=\linewidth]{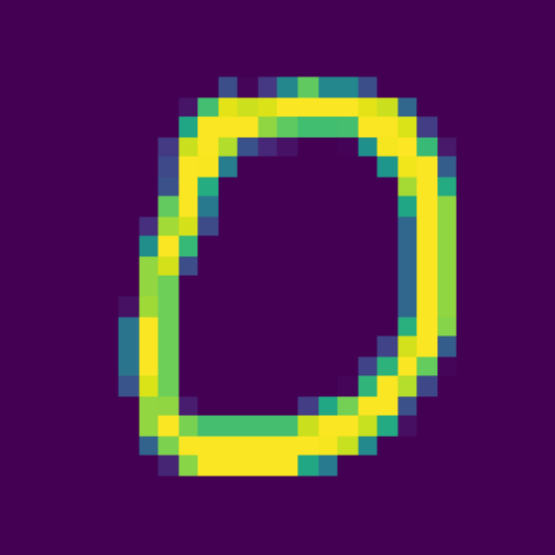}
    \centering
    (b) Target
  \end{subfigure}
  \begin{subfigure}{0.16\linewidth}
    \includegraphics[width=\linewidth]{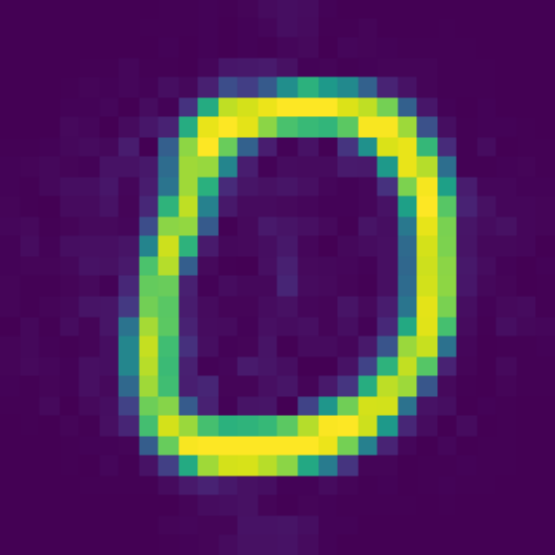}
    \centering
    (c) BEM
  \end{subfigure}
  \begin{subfigure}{0.16\linewidth}
    \includegraphics[width=\linewidth]{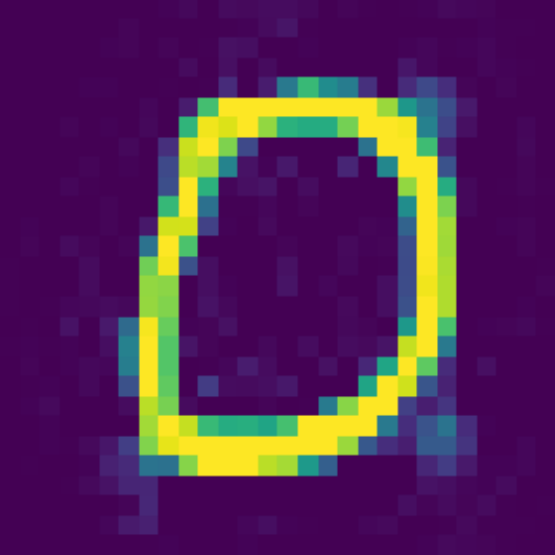}
    \centering
    (d) BEM + PP
  \end{subfigure}
  \begin{subfigure}{0.16\linewidth}
    \includegraphics[width=\linewidth]{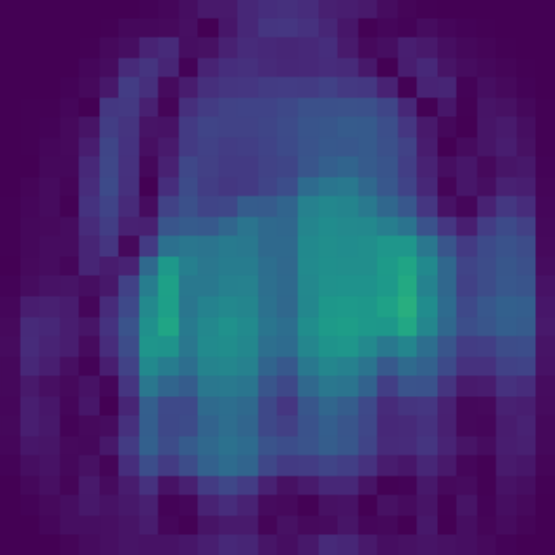}
    \centering
    (e) Real
  \end{subfigure}
  \begin{subfigure}{0.16\linewidth}
    \includegraphics[width=\linewidth]{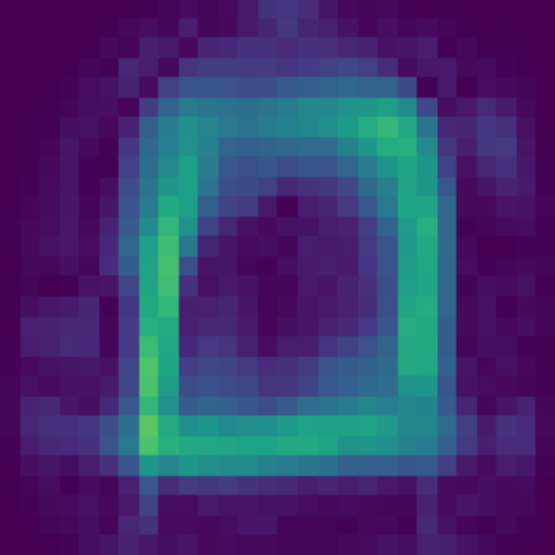}
    \centering
    (f) Complex
  \end{subfigure}
  \caption{Comparison of predicted images from inverting transmission effects of a MMF. (a) The input speckled image, (b) the target original image to reconstruct, (c) the output of the Bessel equivariant model, (d) the output of the combination of Bessel equivariant and post-processing model, (e) the output of the Real valued linear model, and (f) the output of the Complex valued linear model.}
  \label{fig:theorymnistapp}
\end{figure}

\newpage
\subsection{Additional Reconstructions - Theory TM ImageNet}

\begin{figure}[htb]
  \centering
  \foreach \n in {0,...,5}
  {
  \begin{subfigure}{0.16\linewidth}
    \includegraphics[width=\linewidth]{results/imagenet/figs1/speckled_\n.pdf}
  \end{subfigure}
  \begin{subfigure}{0.16\linewidth}
    \includegraphics[width=\linewidth]{results/imagenet/figs1/original_\n.pdf}
  \end{subfigure}
  \begin{subfigure}{0.16\linewidth}
    \includegraphics[width=\linewidth]{results/imagenet/figs1/TM_bases_\n.pdf}
  \end{subfigure}
  \begin{subfigure}{0.16\linewidth}
    \includegraphics[width=\linewidth]{results/imagenet/figs1/TM_bases_SR_\n.pdf}
  \end{subfigure}
  
  }

  \begin{subfigure}{0.16\linewidth}
    \includegraphics[width=\linewidth]{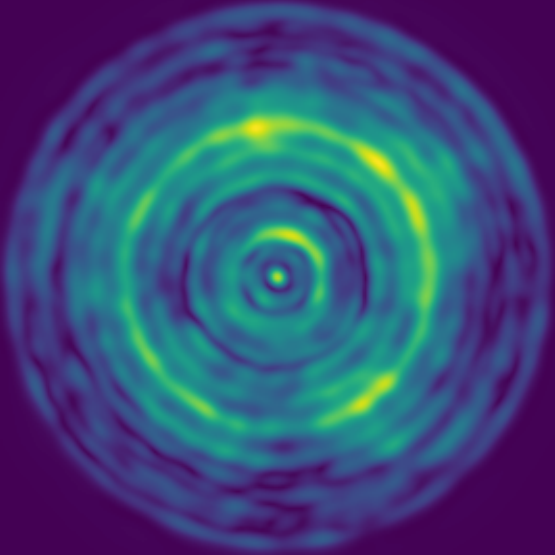}
    \centering
    (a) Input
  \end{subfigure}
  \begin{subfigure}{0.16\linewidth}
    \includegraphics[width=\linewidth]{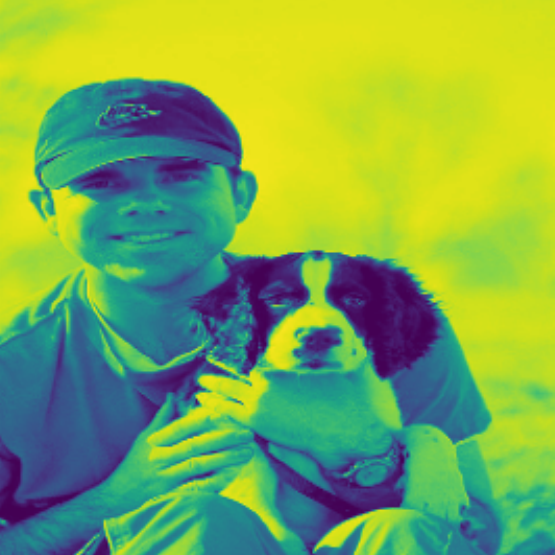}
    \centering
    (b) Target
  \end{subfigure}
  \begin{subfigure}{0.16\linewidth}
    \includegraphics[width=\linewidth]{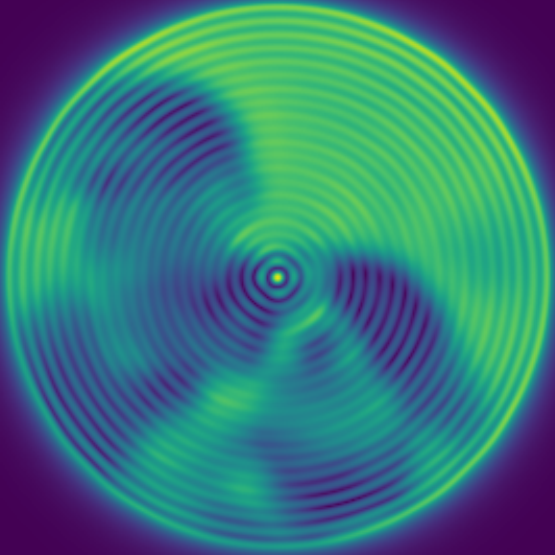}
    \centering
    (c) BEM
  \end{subfigure}
  \begin{subfigure}{0.16\linewidth}
    \includegraphics[width=\linewidth]{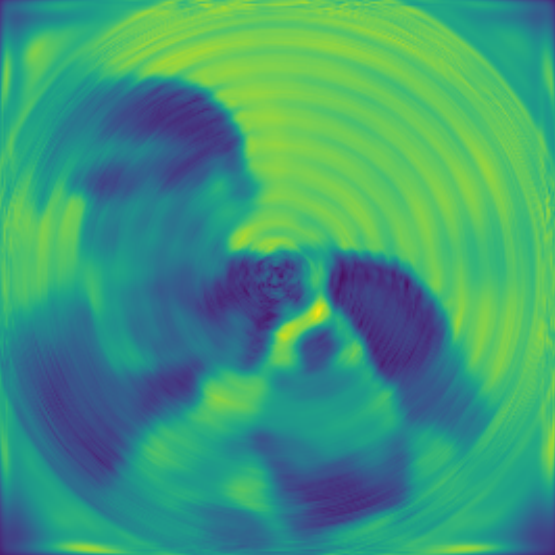}
    \centering
    (d) BEM + PP
  \end{subfigure}
  \caption{Comparison of predicted images from inverting transmission effects of a MMF. (a) The input speckled image, (b) the target original image to reconstruct, (c) the output of the Bessel equivariant model, and (d) the output of the combination of Bessel equivariant and post-processing model.}
  \label{fig:theoryimagenetapp}
\end{figure}

\newpage
\subsection{Impacts of Reducing Training Dataset Size - Theory TM fMNIST}
\label{app:redtrain}

\begin{figure}[htb]
  \centering
  \foreach \n in {3,...,8}
  {
  \begin{subfigure}{0.16\linewidth}
    \includegraphics[width=\linewidth]{results/fmnist/reduce_data/speckled_\n_6000.pdf}
  \end{subfigure}
  \begin{subfigure}{0.16\linewidth}
    \includegraphics[width=\linewidth]{results/fmnist/reduce_data/original_\n_6000.pdf}
  \end{subfigure}
  \begin{subfigure}{0.16\linewidth}
    \includegraphics[width=\linewidth]{results/fmnist/reduce_data/TM_bases_\n_6000.pdf}
  \end{subfigure}
  \begin{subfigure}{0.16\linewidth}
    \includegraphics[width=\linewidth]{results/fmnist/reduce_data/TM_bases_SR_\n_6000.pdf}
  \end{subfigure}
  \begin{subfigure}{0.16\linewidth}
    \includegraphics[width=\linewidth]{results/fmnist/reduce_data/MLP_\n_6000.pdf}
  \end{subfigure}
  \begin{subfigure}{0.16\linewidth}
    \includegraphics[width=\linewidth]{results/fmnist/reduce_data/CMLP_\n_6000.pdf}
  \end{subfigure}
  }

  \begin{subfigure}{0.16\linewidth}
    \includegraphics[width=\linewidth]{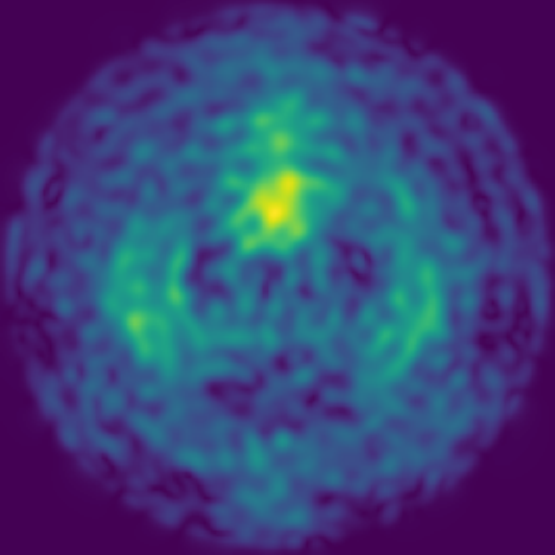}
    \centering
    (a) Input
  \end{subfigure}
  \begin{subfigure}{0.16\linewidth}
    \includegraphics[width=\linewidth]{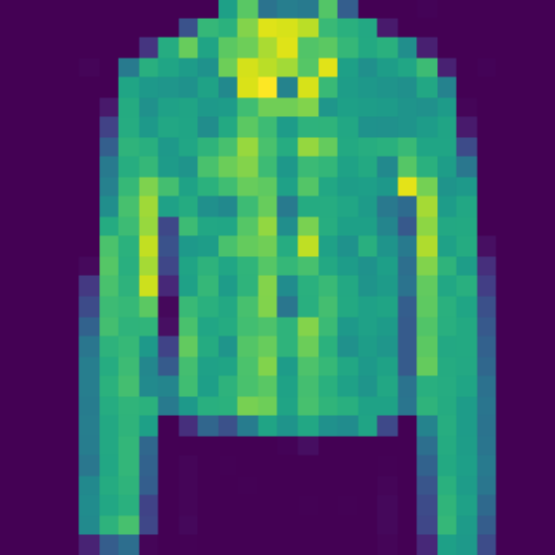}
    \centering
    (b) Target
  \end{subfigure}
  \begin{subfigure}{0.16\linewidth}
    \includegraphics[width=\linewidth]{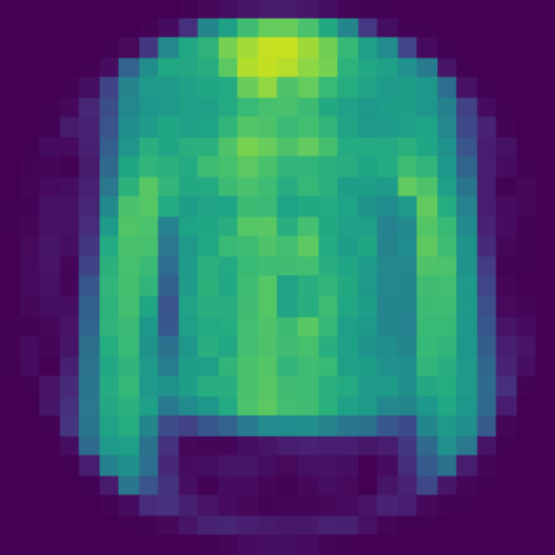}
    \centering
    (c) BEM
  \end{subfigure}
  \begin{subfigure}{0.16\linewidth}
    \includegraphics[width=\linewidth]{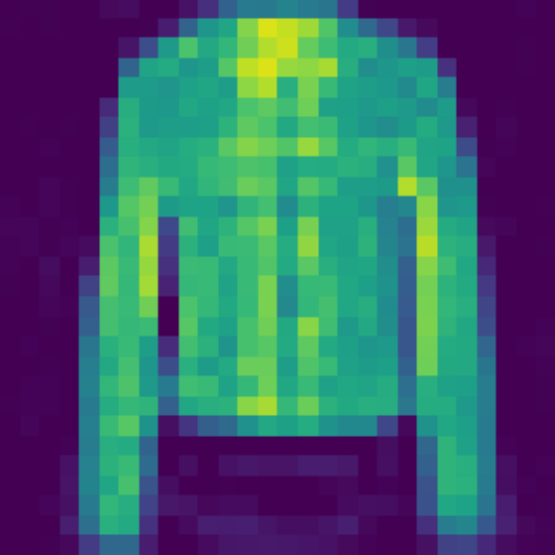}
    \centering
    (d) BEM + PP
  \end{subfigure}
  \begin{subfigure}{0.16\linewidth}
    \includegraphics[width=\linewidth]{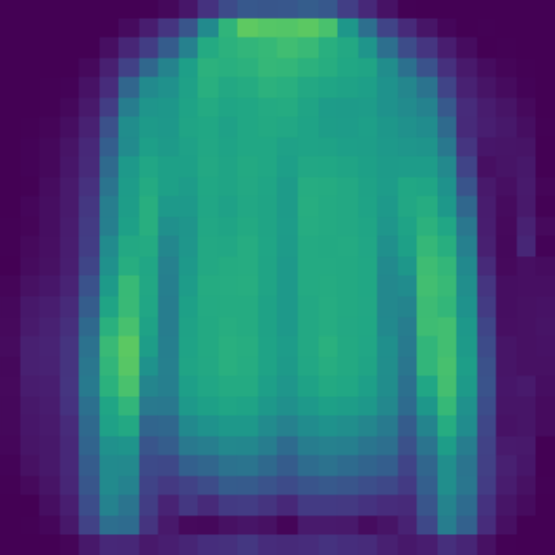}
    \centering
    (e) Real
  \end{subfigure}
  \begin{subfigure}{0.16\linewidth}
    \includegraphics[width=\linewidth]{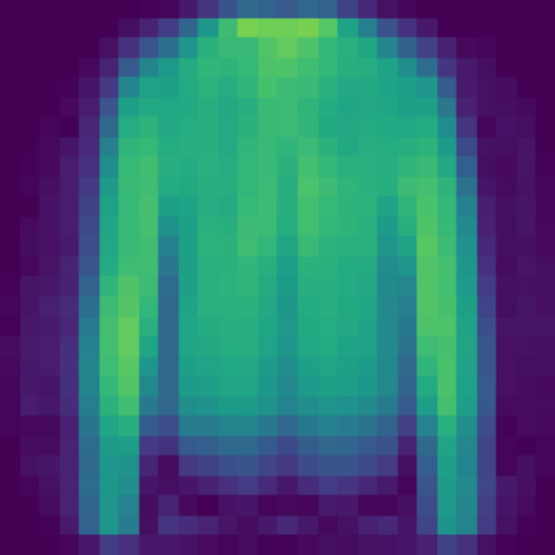}
    \centering
    (f) Complex
  \end{subfigure}
  \caption{Comparison of predicted images from inverting transmission effects of a MMF. The training dataset was reduced from the original size of 12000 to 6000. (a) The input speckled image, (b) the target original image to reconstruct, (c) the output of the Bessel equivariant model, (d) the output of the combination of Bessel equivariant and post-processing model, (e) the output of the Real valued linear model, and (f) the output of the Complex valued linear model. The data used was fMNIST with speckled patterns created with a theoretical TM.}
  \label{fig:dataset6000mnistapp}
\end{figure}

\newpage

\begin{figure}[htb]
  \centering
  \foreach \n in {3,...,8}
  {
  \begin{subfigure}{0.16\linewidth}
    \includegraphics[width=\linewidth]{results/fmnist/reduce_data/speckled_\n_6000.pdf}
  \end{subfigure}
  \begin{subfigure}{0.16\linewidth}
    \includegraphics[width=\linewidth]{results/fmnist/reduce_data/original_\n_2400.pdf}
  \end{subfigure}
  \begin{subfigure}{0.16\linewidth}
    \includegraphics[width=\linewidth]{results/fmnist/reduce_data/TM_bases_\n_2400.pdf}
  \end{subfigure}
  \begin{subfigure}{0.16\linewidth}
    \includegraphics[width=\linewidth]{results/fmnist/reduce_data/TM_bases_SR_\n_2400.pdf}
  \end{subfigure}
  \begin{subfigure}{0.16\linewidth}
    \includegraphics[width=\linewidth]{results/fmnist/reduce_data/MLP_\n_2400.pdf}
  \end{subfigure}
  \begin{subfigure}{0.16\linewidth}
    \includegraphics[width=\linewidth]{results/fmnist/reduce_data/CMLP_\n_2400.pdf}
  \end{subfigure}
  }

  \begin{subfigure}{0.16\linewidth}
    \includegraphics[width=\linewidth]{results/fmnist/reduce_data/speckled_10_6000.pdf}
    \centering
    (a) Input
  \end{subfigure}
  \begin{subfigure}{0.16\linewidth}
    \includegraphics[width=\linewidth]{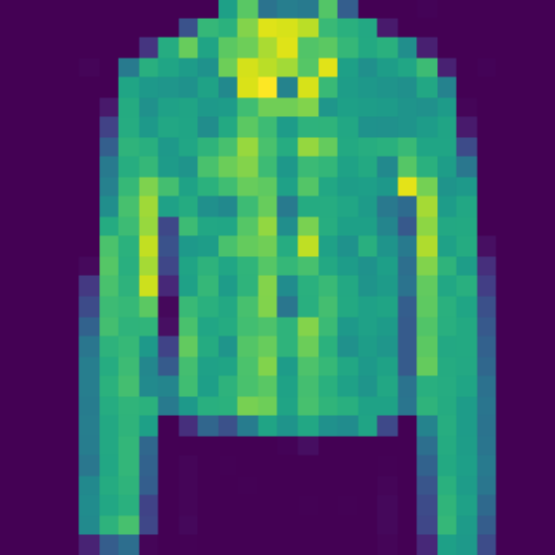}
    \centering
    (b) Target
  \end{subfigure}
  \begin{subfigure}{0.16\linewidth}
    \includegraphics[width=\linewidth]{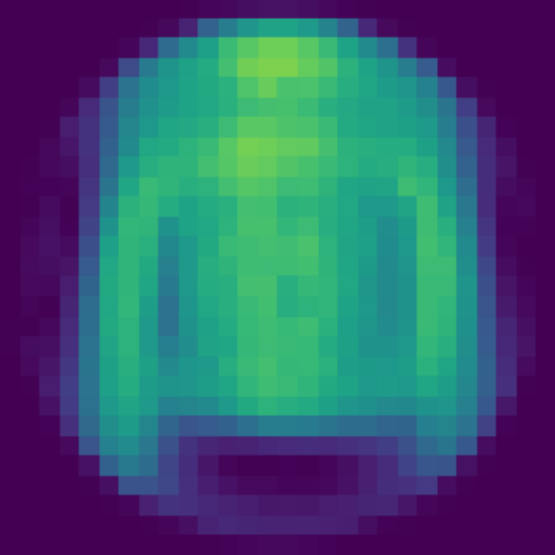}
    \centering
    (c) BEM
  \end{subfigure}
  \begin{subfigure}{0.16\linewidth}
    \includegraphics[width=\linewidth]{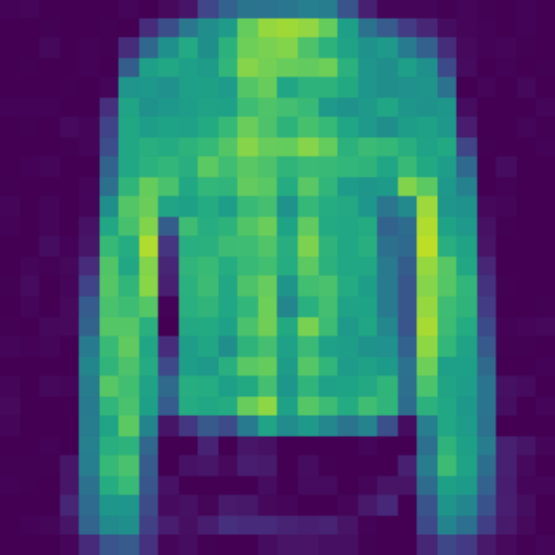}
    \centering
    (d) BEM + PP
  \end{subfigure}
  \begin{subfigure}{0.16\linewidth}
    \includegraphics[width=\linewidth]{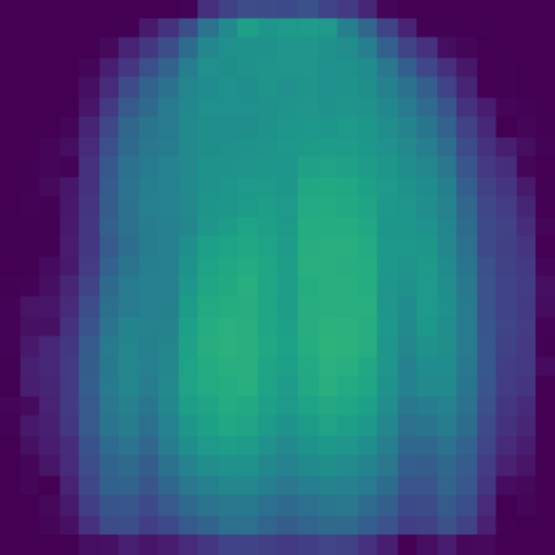}
    \centering
    (e) Real
  \end{subfigure}
  \begin{subfigure}{0.16\linewidth}
    \includegraphics[width=\linewidth]{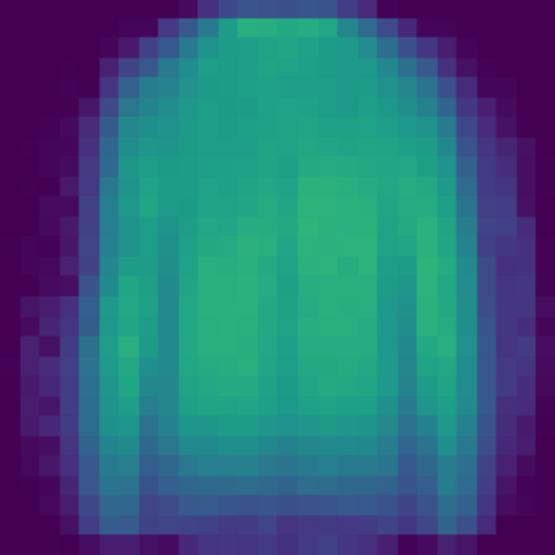}
    \centering
    (f) Complex
  \end{subfigure}
  \caption{Comparison of predicted images from inverting transmission effects of a MMF. The training dataset was reduced from the original size of 12000 to 2400. (a) The input speckled image, (b) the target original image to reconstruct, (c) the output of the Bessel equivariant model, (d) the output of the combination of Bessel equivariant and post-processing model, (e) the output of the Real valued linear model, and (f) the output of the Complex valued linear model. The data used was fMNIST with speckled patterns created with a theoretical TM.}
  \label{fig:dataset2400mnistapp}
\end{figure}

\newpage

\begin{figure}[htb]
  \centering
  \foreach \n in {3,...,8}
  {
  \begin{subfigure}{0.16\linewidth}
    \includegraphics[width=\linewidth]{results/fmnist/reduce_data/speckled_\n_6000.pdf}
  \end{subfigure}
  \begin{subfigure}{0.16\linewidth}
    \includegraphics[width=\linewidth]{results/fmnist/reduce_data/original_\n_1200.pdf}
  \end{subfigure}
  \begin{subfigure}{0.16\linewidth}
    \includegraphics[width=\linewidth]{results/fmnist/reduce_data/TM_bases_\n_1200.pdf}
  \end{subfigure}
  \begin{subfigure}{0.16\linewidth}
    \includegraphics[width=\linewidth]{results/fmnist/reduce_data/TM_bases_SR_\n_1200.pdf}
  \end{subfigure}
  \begin{subfigure}{0.16\linewidth}
    \includegraphics[width=\linewidth]{results/fmnist/reduce_data/MLP_\n_1200.pdf}
  \end{subfigure}
  \begin{subfigure}{0.16\linewidth}
    \includegraphics[width=\linewidth]{results/fmnist/reduce_data/CMLP_\n_1200.pdf}
  \end{subfigure}
  }

  \begin{subfigure}{0.16\linewidth}
    \includegraphics[width=\linewidth]{results/fmnist/reduce_data/speckled_10_6000.pdf}
    \centering
    (a) Input
  \end{subfigure}
  \begin{subfigure}{0.16\linewidth}
    \includegraphics[width=\linewidth]{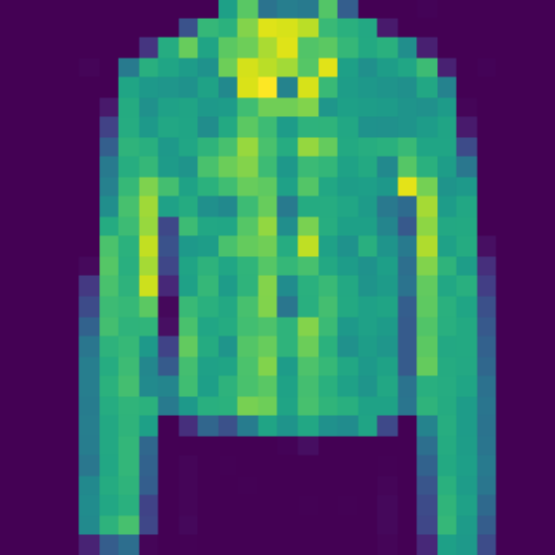}
    \centering
    (b) Target
  \end{subfigure}
  \begin{subfigure}{0.16\linewidth}
    \includegraphics[width=\linewidth]{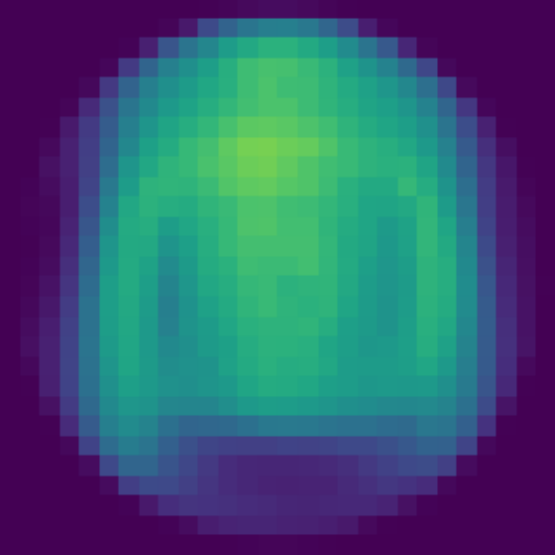}
    \centering
    (c) BEM
  \end{subfigure}
  \begin{subfigure}{0.16\linewidth}
    \includegraphics[width=\linewidth]{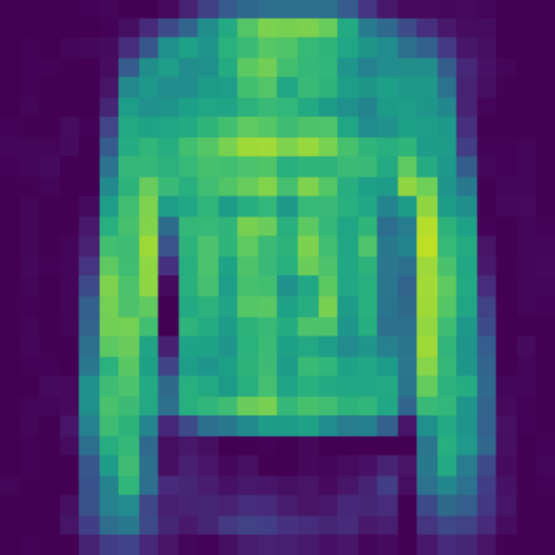}
    \centering
    (d) BEM + PP
  \end{subfigure}
  \begin{subfigure}{0.16\linewidth}
    \includegraphics[width=\linewidth]{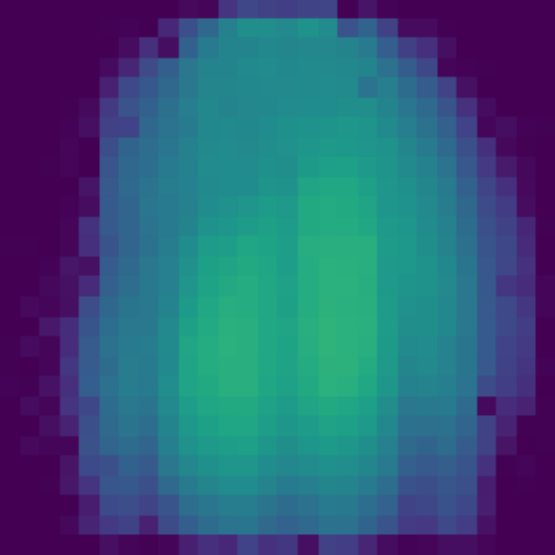}
    \centering
    (e) Real
  \end{subfigure}
  \begin{subfigure}{0.16\linewidth}
    \includegraphics[width=\linewidth]{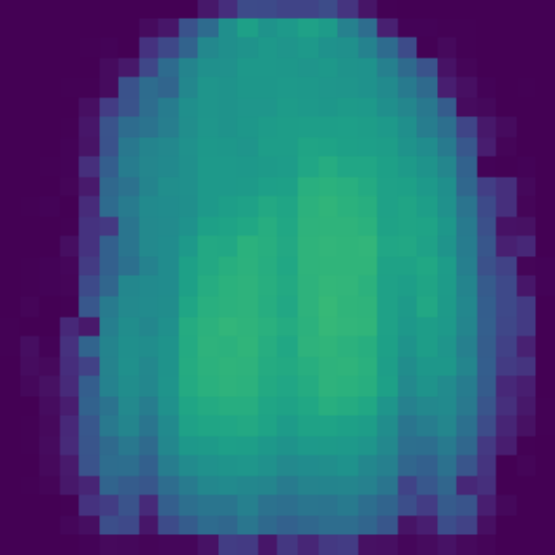}
    \centering
    (f) Complex
  \end{subfigure}
  \caption{Comparison of predicted images from inverting transmission effects of a MMF. The training dataset was reduced from the original size of 12000 to 1200. (a) The input speckled image, (b) the target original image to reconstruct, (c) the output of the Bessel equivariant model, (d) the output of the combination of Bessel equivariant and post-processing model, (e) the output of the Real valued linear model, and (f) the output of the Complex valued linear model. The data used was fMNIST with speckled patterns created with a theoretical TM.}
  \label{fig:dataset1200mnistapp}
\end{figure}

\newpage

\begin{figure}[htb]
  \centering
  \foreach \n in {3,...,8}
  {
  \begin{subfigure}{0.16\linewidth}
    \includegraphics[width=\linewidth]{results/fmnist/reduce_data/speckled_\n_6000.pdf}
  \end{subfigure}
  \begin{subfigure}{0.16\linewidth}
    \includegraphics[width=\linewidth]{results/fmnist/reduce_data/original_\n_120.pdf}
  \end{subfigure}
  \begin{subfigure}{0.16\linewidth}
    \includegraphics[width=\linewidth]{results/fmnist/reduce_data/TM_bases_\n_120.pdf}
  \end{subfigure}
  \begin{subfigure}{0.16\linewidth}
    \includegraphics[width=\linewidth]{results/fmnist/reduce_data/TM_bases_SR_\n_120.pdf}
  \end{subfigure}
  \begin{subfigure}{0.16\linewidth}
    \includegraphics[width=\linewidth]{results/fmnist/reduce_data/MLP_\n_120.pdf}
  \end{subfigure}
  \begin{subfigure}{0.16\linewidth}
    \includegraphics[width=\linewidth]{results/fmnist/reduce_data/CMLP_\n_120.pdf}
  \end{subfigure}
  }

  \begin{subfigure}{0.16\linewidth}
    \includegraphics[width=\linewidth]{results/fmnist/reduce_data/speckled_10_6000.pdf}
    \centering
    (a) Input
  \end{subfigure}
  \begin{subfigure}{0.16\linewidth}
    \includegraphics[width=\linewidth]{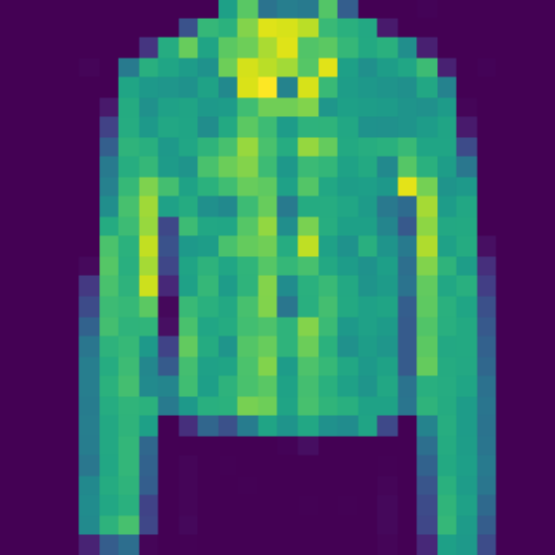}
    \centering
    (b) Target
  \end{subfigure}
  \begin{subfigure}{0.16\linewidth}
    \includegraphics[width=\linewidth]{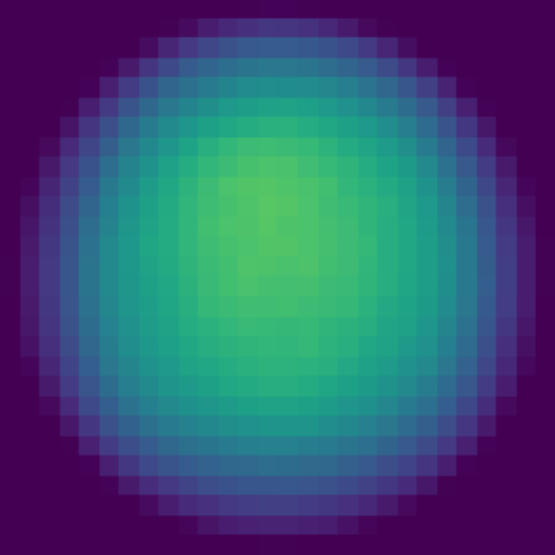}
    \centering
    (c) BEM
  \end{subfigure}
  \begin{subfigure}{0.16\linewidth}
    \includegraphics[width=\linewidth]{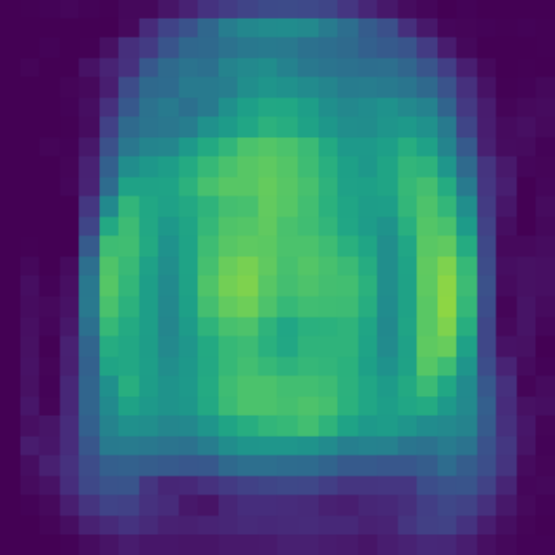}
    \centering
    (d) BEM + PP
  \end{subfigure}
  \begin{subfigure}{0.16\linewidth}
    \includegraphics[width=\linewidth]{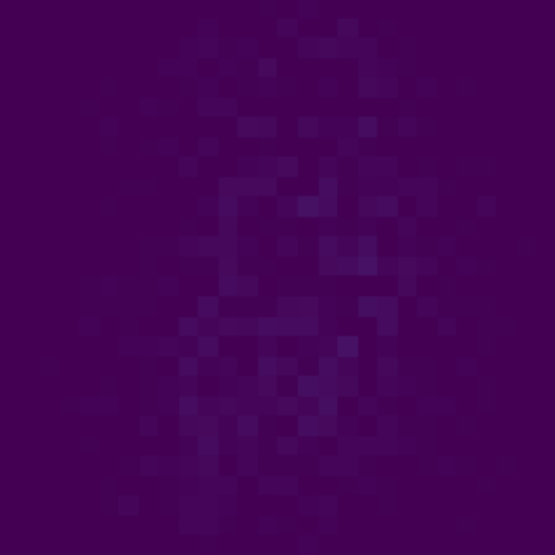}
    \centering
    (e) Real
  \end{subfigure}
  \begin{subfigure}{0.16\linewidth}
    \includegraphics[width=\linewidth]{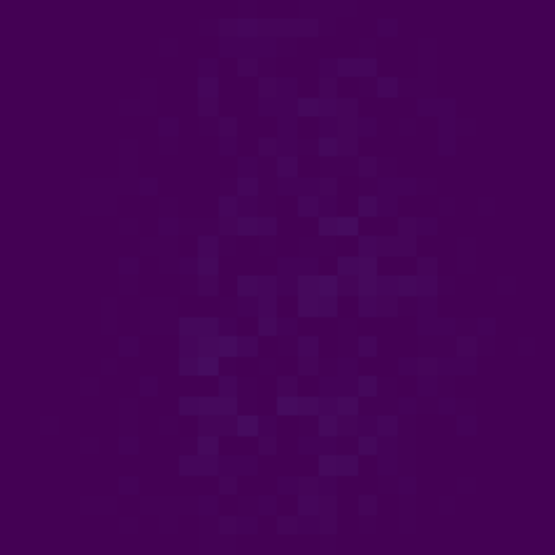}
    \centering
    (f) Complex
  \end{subfigure}
  \caption{Comparison of predicted images from inverting transmission effects of a MMF. The training dataset was reduced from the original size of 12000 to 120. (a) The input speckled image, (b) the target original image to reconstruct, (c) the output of the Bessel equivariant model, (d) the output of the combination of Bessel equivariant and post-processing model, (e) the output of the Real valued linear model, and (f) the output of the Complex valued linear model. The data used was fMNIST with speckled patterns created with a theoretical TM.}
  \label{fig:dataset120mnistapp}
\end{figure}

\newpage
\subsection{Impacts of Under Parameterising the set of Bessel Function Bases - Theory TM fMNIST}
\label{app:redbases}

\begin{figure}[htb]
  \centering
  \foreach \n in {0,...,3}
  {
  \begin{subfigure}{0.16\linewidth}
    \includegraphics[width=\linewidth]{results/fmnist/figs1/original_\n.pdf}
  \end{subfigure}
  \begin{subfigure}{0.16\linewidth}
    \includegraphics[width=\linewidth]{results/fmnist/figs1/TM_bases_\n.pdf}
  \end{subfigure}
  \begin{subfigure}{0.16\linewidth}
    \includegraphics[width=\linewidth]{results/fmnist/reduced_bases/TM_bases_\n_930.pdf}
  \end{subfigure}
  \begin{subfigure}{0.16\linewidth}
    \includegraphics[width=\linewidth]{results/fmnist/reduced_bases/TM_bases_\n_567.pdf}
  \end{subfigure}
  \begin{subfigure}{0.16\linewidth}
    \includegraphics[width=\linewidth]{results/fmnist/reduced_bases/TM_bases_\n_322.pdf}
  \end{subfigure}
  
  }

  \begin{subfigure}{0.16\linewidth}
    \includegraphics[width=\linewidth]{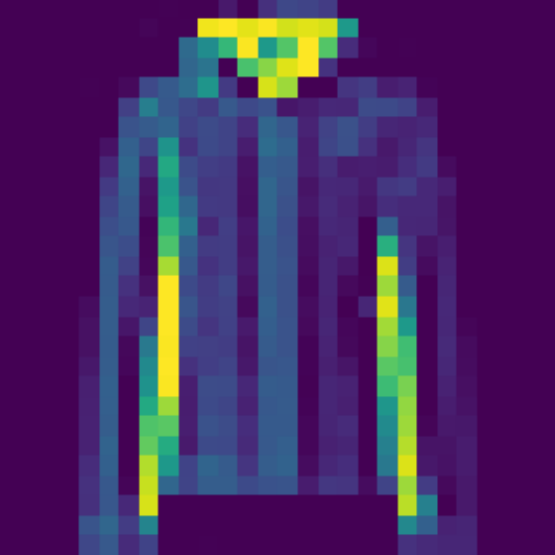}
    \centering
    (a) Target
  \end{subfigure}
  \begin{subfigure}{0.16\linewidth}
    \includegraphics[width=\linewidth]{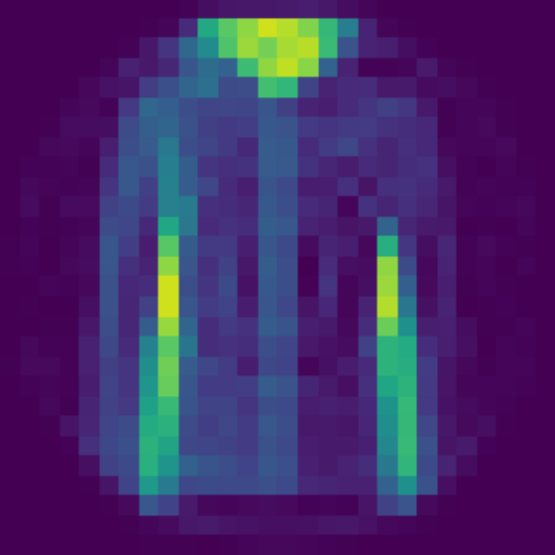}
    \centering
    (b) BEM (21)
  \end{subfigure}
  \begin{subfigure}{0.16\linewidth}
    \includegraphics[width=\linewidth]{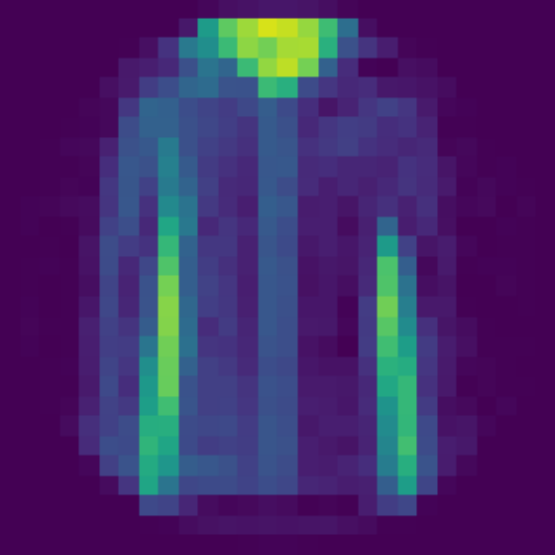}
    \centering
    (c) BEM (14)
  \end{subfigure}
  \begin{subfigure}{0.16\linewidth}
    \includegraphics[width=\linewidth]{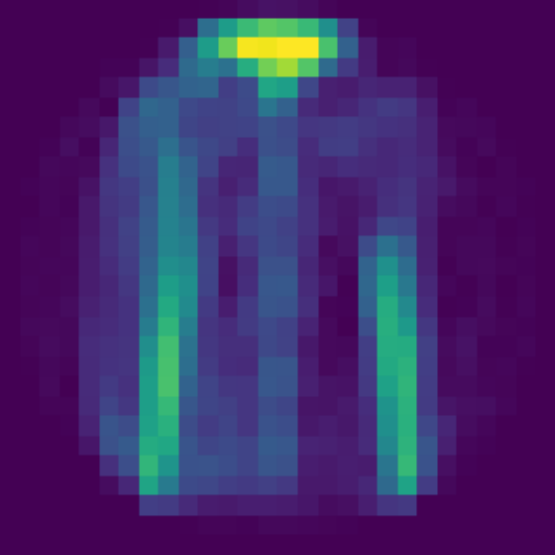}
    \centering
    (d) BEM (7)
  \end{subfigure}
  \begin{subfigure}{0.16\linewidth}
    \includegraphics[width=\linewidth]{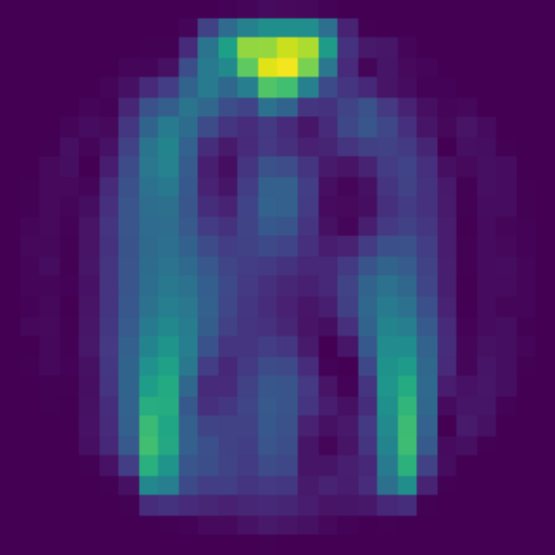}
    \centering
    (e) BEM (4)
  \end{subfigure}

  \caption{Comparison of predicted images from inverting transmission effects of a MMF, when reducing the number of radial frequencies present in the Bessel function basis. (a) The input speckled image. (b) The number of radial frequencies present in the Bessel function baiis was left at the original value of 21, which represents 1061 bases. (c) The number of radial frequencies present in the Bessel function basis was reduced from the original value of 21 to 14, which represents a reduction in the number of bases from the original value of 1061 to 930. (d) The number of radial frequencies present in the Bessel function basis was reduced from the original value of 21 to 7, which represents a reduction in the number of bases from the original value of 1061 to 567. (e) The number of radial frequencies present in the Bessel function basis was reduced from the original value of 21 to 4, which represents a reduction in the number of bases from the original value of 1061 to 322. The data used was fMNIST with speckled patterns created with a theoretical TM.}
  \label{fig:bases1061}
\end{figure}

Here we consider the impact on under parameterising the Bessel function basis. For this we reduce the number of radial frequencies that feature in the set of bases functions. The original basis set we utilised for the data collected using a theoretical fibre for fMNIST comprises of 21 radial frequencies and 1061 bases. Here we show the original results using this full bases set in Figure~\ref{fig:bases1061} column (b). Next we show the results of removing the high frequency bases by only considering the first 14 radial frequencies, which amounts to having 932 bases in Figure~\ref{fig:bases1061} column (c). In addition, we show the results of removing further high frequency bases by only considering the first 7 radial frequencies, which amounts to having 567 bases in Figure~\ref{fig:bases1061} column (d). Finally, we show the results of removing further high frequency bases by only considering the first 4 radial frequencies, which amounts to having 322 bases in Figure~\ref{fig:bases1061} column (e).

\begin{table}[htb]
  \caption{Comparison of the loss values of each model trained with fMNIST data created with a theoretical TM.This shows the impact of under parameterising the Bessel function basis set by removing higher frequency modes. 21 radial frequencies is the natural choice for the fibre considered, hence all values less than 21 give an under parameterised model.}
  \label{tab:fmnist_red_bases}
  \centering
  \begin{tabular}{lcccc}
    \toprule
    Model & Radial Frequencies & \# Modes & Train Loss & Test Loss \\
    \midrule
    Bessel Equivariant             & 21 & 1061 & 0.0141 & 0.0139  \\
    Bessel Equivariant + Post Proc & 21 & 1061 & \textbf{0.0032} & \textbf{0.0032} \\
    Bessel Equivariant             & 14 & 932 & 0.0158 & 0.0156  \\
    Bessel Equivariant + Post Proc & 14 & 932 & 0.0036 & 0.0036 \\
    Bessel Equivariant             & 7 & 567 & 0.0203 & 0.0201  \\
    Bessel Equivariant + Post Proc & 7 & 567 & 0.0053 & 0.0053 \\
    Bessel Equivariant             & 4 & 332 & 0.0299 & 0.0296  \\
    Bessel Equivariant + Post Proc & 4 & 332 & 0.0090 & 0.0090 \\
    \bottomrule
  \end{tabular}
\end{table}

\begin{figure}[htb]
    \centering
    \includegraphics[width=\linewidth]{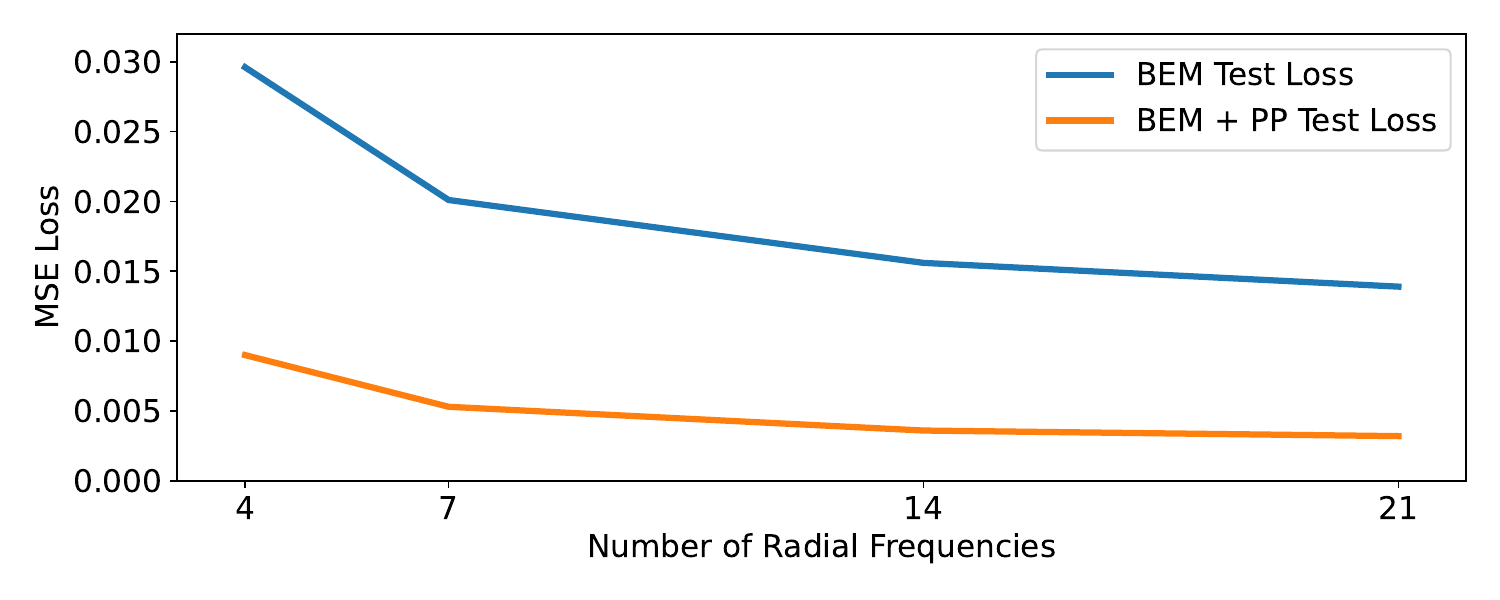}
    \caption{Comparison of the loss values of each model trained with fMNIST data created with a theoretical TM. This shows the impact of under parameterising the Bessel function basis set by removing higher frequency modes. 21 radial frequencies is the natural choice for the fibre considered, hence all values less than 21 give an under parameterised model.}
    \label{fig:fmnist_red_bases}
\end{figure}

\FloatBarrier
\newpage

\subsection{Negative Societal Impacts}
\label{sec:negsocimpact}

Our work enables imaging using a multi-mode optical fibre at higher resolutions than has been previously been achievable with a machine learning based approach. In addition, our new method generalises better to new data classes out of the training data classes. Therefore, it better enables imaging using multi-mode fibres to be used practically. Thus, this technology could potentially be used in a way that negatively impacts society through any negative use case of multi-mode optical fibre imaging (e.g. possible uses in espionage). Never-the-less the technology developed during this paper does not have a direct negative impact.

\end{document}